\documentclass[useAMS,usenatbib]{mnras}
\usepackage{natbib}
\usepackage{amsmath}
\usepackage{url}
\usepackage{longtable}
\usepackage{aas_macros}
\usepackage{amssymb}
\usepackage{graphicx}
\usepackage{deluxetable}
\usepackage{pdflscape}

\newcommand{\about}{$\sim\!\!$~}
\newcommand{\kms}{\,km\,s$^{-1}$}

\def\lsim{\hbox{\rlap{\raise 0.425ex\hbox{$<$}}\lower 0.65ex\hbox{$\sim$}}}
\def\gsim{\hbox{\rlap{\raise 0.425ex\hbox{$>$}}\lower 0.65ex\hbox{$\sim$}}}

\title[Late-time Spectra of SNe~Iax]{Late-time Spectroscopy of Type
  Iax Supernovae}

\def\illast{1}
\def\illphys{2}
\def\rut{3}
\def\berk{4}
\def\kitp{5}
\def\berkp{6}
\def\tac{7}
\def\lbl{8}

\begin{document}

\author[Foley et~al.]{Ryan~J.~Foley$^{\illast,\illphys}$\thanks{E-mail:rfoley@illinois.edu},
Saurabh~W.~Jha$^{\rut}$,
Yen-Chen~Pan$^{\illast}$,
WeiKang~Zheng$^{\berk}$,
\newauthor
Lars~Bildsten$^{\kitp}$,
Alexei~V.~Filippenko$^{\berk}$,
Daniel~Kasen$^{\berkp,\tac,\lbl}$\\
$^{\illast}$Astronomy Department, University of Illinois at Urbana-Champaign, 1002 W.\ Green Street, Urbana, IL 61801, USA\\
$^{\illphys}$Department of Physics, University of Illinois at Urbana-Champaign, 1110 W.\ Green Street, Urbana, IL 61801, USA\\
$^{\rut}$Department of Physics and Astronomy, Rutgers, The State University of New Jersey, 136 Frelinghuysen Road, Piscataway, NJ 08854, USA\\
$^{\berk}$Department of Astronomy, University of California, Berkeley, CA 94720-3411, USA\\
$^{\kitp}$Kavli Institute for Theoretical Physics and Department of Physics Kohn Hall, University of California, Santa Barbara, CA 93106, USA\\
$^{\berkp}$Department of Physics, University of California, Berkeley, CA
94720, USA\\
$^{\tac}$Department of Astronomy and Theoretical Astrophysics Center, University of California, Berkeley, CA 94720, USA\\
$^{\lbl}$Nuclear Science Division, Lawrence Berkeley National Laboratory, Berkeley, CA 94720, USA}

\date{Accepted  . Received   ; in original form  }
\pagerange{\pageref{firstpage}--\pageref{lastpage}} \pubyear{2015}
\maketitle
\label{firstpage}

\begin{abstract}
  We examine the late-time ($t \gtrsim 200$~days after peak
  brightness) spectra of Type Iax supernovae (SNe~Iax), a
  low-luminosity, low-energy class of thermonuclear stellar explosions
  observationally similar to, but distinct from, Type Ia supernovae.
  We present new spectra of SN~2014dt, resulting in the most complete
  published late-time spectral sequence of a SN~Iax.  At late times,
  SNe~Iax have generally similar spectra, all with a similar continuum
  shape and strong forbidden-line emission.  However, there is also
  significant diversity where some SN~Iax spectra display narrow
  P-Cygni features from permitted lines and a continuum indicative of
  a photosphere at late times in addition to strong narrow forbidden
  lines, while others have no obvious P-Cygni features, strong broad
  forbidden lines, and weak narrow forbidden lines.  Finally, some
  SNe~Iax have spectra intermediate to these two varieties with weak
  P-Cygni features and broad/narrow forbidden lines of similar
  strength.  We find that SNe~Iax with strong broad forbidden lines
  also tend to be more luminous and have higher-velocity ejecta at
  peak brightness.  We find no evidence for dust formation in the SN
  ejecta or the presence of circumstellar dust, including for the
  infrared-bright SN~2014dt.  Late-time SN~Iax spectra have strong
  [\ion{Ni}{II}] emission, which must come from stable Ni, requiring
  electron captures that can only occur at the high densities of a
  (nearly) Chandrasekhar-mass WD.  Therefore, such a star is the
  likely progenitor of SNe~Iax.  We estimate blackbody and kinematic
  radii of the late-time photosphere, finding the latter an order of
  magnitude larger than the former for at least one SN~Iax.  We
  propose a two-component model that solves this discrepancy and
  explains the diversity of the late-time spectra of SNe~Iax.  In this
  model, the broad forbidden lines originate from the SN ejecta,
  similar to the spectra of all other types of SNe, while the
  photosphere, P-Cygni lines, and narrow forbidden lines originate
  from a wind launched from the remnant of the progenitor white dwarf
  and is driven by the radioactive decay of newly synthesised material
  left in the remnant.  The relative strength of the two components
  accounts for the diversity of late-time SN~Iax spectra.  This model
  also solves the puzzle of a long-lived photosphere and slow
  late-time decline of SNe~Iax.
\end{abstract}

\begin{keywords}
  {supernovae---general, supernovae---individual (PTF09ego, PTF09eiy,
    PTF10bvr, SN~2002cx, SN~2004cs, SN~2005P, SN~2005hk, SN~2007J,
    SN~2008A, SN~2008ge, SN~2008ha, SN~2010ae, SN~2011ay, SN~2011ce,
    SN~2012Z, SN~2014dt)}
\end{keywords}


\defcitealias{Foley13:iax}{F13}
\defcitealias{Jha06:02cx}{J06}

\section{Introduction}\label{s:intro}

Type Iax supernovae (SNe~Iax) are a newly defined class of stellar
death \citep[][hereafter \citetalias{Foley13:iax}]{Foley13:iax}.
These thermonuclear explosions are observationally similar to, but
distinct from, SNe~Ia.  The main observational differences between the
two classes are related to energetics: SNe~Iax have peak luminosities,
integrated luminosity, and near-maximum ejecta velocities that are
substantially lower than that of SNe~Ia
\citep[e.g.,][]{Filippenko03:02cx, Li03:02cx, Jha06:02cx}, with the
most extreme members of the class having peak luminosities and ejecta
velocities 1\% and 20\% those of typical SNe~Ia, respectively
\citep{Foley09:08ha, Foley10:08ha, Stritzinger14:10ae}.

While SNe~Ia and Iax have somewhat similar spectra near maximum
brightness \citep[e.g.,][]{Li03:02cx, Branch04, Chornock06,
  Jha06:02cx, Phillips07, Sahu08, Foley10:08ha, Foley13:iax,
  Stritzinger14:10ae, Stritzinger15}, the late-time ($t \gtrsim
200$~d) spectra of SNe~Iax are more distinct from SNe~Ia and SNe of
all other classes (\citealt{Jha06:02cx}; \citetalias{Foley13:iax};
\citealt{McCully14:iax}).  Specifically, even a year after explosion,
SNe~Iax lack the strong forbidden Fe lines at blue optical wavelengths
([\ion{Fe}{II}] $\lambda 4200$, [\ion{Fe}{III}] $\lambda 4700$, and
[\ion{Fe}{II}] $\lambda 5270$) and still have a continuum and P-Cygni
profiles with very low velocities \citep[\about 500~\kms;][hereafter
\citetalias{Jha06:02cx}]{Jha06:02cx}.

The large differences at late times likely point to different
explosion mechanisms and progenitors for SNe~Ia and Iax.  Since the
probable progenitor system of one SN~Iax (SN~2012Z) has been detected
in pre-explosion images \citep{McCully14:12z}, while no progenitor
system has yet been detected for SNe~Ia even in deep pre-explosion
images \citep[e.g.,][]{Li11:11fe, Kelly14}, there is additional
evidence that SNe~Ia and Iax have different progenitor systems,
although this difference may be primarily constrained to the companion
stars.

Currently, the leading progenitor model for SNe~Iax is a C/O white
dwarf (WD) accreting material from a He-star donor \citep[although see
\citealt{Kromer15}]{Foley09:08ha, Foley13:iax, Liu15}.  This model is
consistent with all current observational data
\citepalias{Foley13:iax} including the probable progenitor detection
of SN~2012Z \citep[][and the nondetection of the progenitor system for
SN~2014dt; \citealt{Foley15:14dt}]{McCully14:12z}.

Because of the low ejecta masses required for some SNe~Iax
\citep[perhaps as low as 0.1~M$_{\sun}$; e.g.,][]{Foley09:08ha,
  Foley10:08ha, McCully14:iax, Valenti09}, there is indirect evidence
that the progenitor star is not completely disrupted.  Models of a C/O
WD undergoing a deflagration that does not fully disrupt the
progenitor WD \citep[e.g.,][]{Jordan12, Kromer13, Kromer15} can
explain most of the observations including the low luminosity, low
ejecta velocities, and slow late-time luminosity decline.  However,
additional constraints on the explosion mechanism are required for
further progress.  The potential detection of the remnant WD years
after SN~2008ha exploded \citep{Foley14:08ha} would be the most direct
indication that some SNe~Iax do not completely disrupt their
progenitor stars.

Here, we examine the late-time spectra of a sample of 10 SNe~Iax to
further understand the physical mechanisms of this class of SNe.  The
diverse spectra at $t > 200$~d after peak brightness provide multiple
clues about the explosion and the final fate of the progenitor star.

We describe our sample and data, which includes new observations of
SN~2014dt, in Section~\ref{s:sample}. Section~\ref{s:measure} presents
various physical quantities for the late-time spectra of SNe~Iax and
the measurements are analysed in Section~\ref{s:analysis}.  We discuss
our findings in Section~\ref{s:disc} and conclude in
Section~\ref{s:conc}.


\section{Sample}\label{s:sample}

For our sample, we begin with the data presented by
\citetalias{Foley13:iax}, which represents the largest sample of
SNe~Iax to date.  This sample contains 25 SNe~Iax, of which 7 have
late-time ($t \gtrsim 200$~d) spectra.  In addition to the data
presented by \citetalias{Foley13:iax}, \citet{Sahu08},
\citet{Foley10:08ge}, \citet{Stritzinger14:10ae}, and
\citet{Stritzinger15} present late-time spectra for SNe~2005hk,
2008ge, 2010ae, and 2012Z, which we include here.  In addition, we use
the updated light-curve parameters for SNe~2010ae and 2012Z
\citep[respectively]{Stritzinger14:10ae, Stritzinger15}.

We add to this sample SN~2014dt, the closest SN~Iax yet discovered
\citep{Foley15:14dt}.  Below, we present late-time spectra of
SN~2014dt.

We also examined the sample of \citet{White15}, which includes a
compilation of six SNe identified as SNe~Iax that are not in the
\citetalias{Foley13:iax} sample.  In Appendix~\ref{a:ptf}, we
determine that while four are genuine SNe~Iax, two are most likely not
SNe~Iax.  Of the genuine \citet{White15} SNe~Iax, two have spectra at
$t > 100$~d.  However, none is at $t > 125$~d nor has sufficiently
high quality for inclusion in this analysis.

The combined sample has 10 SNe~Iax with late-time spectra.  We give
light-curve parameters and maximum-light photospheric velocity
measurements for these objects in Table~\ref{t:max}.  We present the
phases of our primarily examined spectra in Table~\ref{t:neb_fit}.

\begin{deluxetable}{llll}
\tabletypesize{\footnotesize}
\tablewidth{0pt}
\hspace{-3in}\tablecaption{SN~Iax Maximum-light Parameters\label{t:max}}
\tablehead{
\colhead{SN} &
\colhead{$M_{V, {\rm peak}}$ (mag)} &
\colhead{$\Delta m_{15}(V)$ (mag)} &
\colhead{$v_{\rm ph}$ (\kms)}}

\startdata

2002cx & $-17.52$ (0.18) & 0.84 (0.09) & $-5550$ (20) \\
2005P  & \nodata         & \nodata     & \nodata \\
2005hk & $-18.07$ (0.25) & 0.92 (0.01) & $-4490$ (430) \\
2008A  & $-18.16$ (0.15) & 0.82 (0.06) & $-6350$ (160) \\
2008ge & $-17.60$ (0.25) & 0.34 (0.24) & \nodata \\
2010ae & $-15.33$ (0.54) & 1.15 (0.04) & $-4390$ (60) \\
2011ay & $-18.40$ (0.16) & 0.75 (0.12) & $-5560$ (80) \\
2011ce & \nodata         & \nodata     & \nodata \\
2012Z  & $-18.50$ (0.09) & 0.89 (0.01) & $-6030$ (180) \\
2014dt & $-17.40$ (0.50) & \nodata     & \nodata

\enddata

\vspace{-0.7cm}
\tablecomments{Uncertainties listed in parentheses.}

\end{deluxetable}

\subsection{SN~2014dt}

The newest addition to our sample is SN~2014dt, which was detected in
M61 on 2014 October 29.8 (all dates are UT) at $V = 13.6$~mag by
\citet{Nakano14} and promptly classified as a SN~Iax by
\citet{Ochner14} from a spectrum obtained 2014 October 31.2.  The SN
was past peak at discovery and there are no recent nondetections which
constrain the date of explosion.

\citet{Foley15:14dt} present a spectrum from 2014 November 18.6, 19.6
rest-frame days after discovery.  Using SNID \citep{Blondin07}, we
determine that SN~2014dt was at a phase of $+23 \pm 7$~d for that
spectrum.  The classification spectrum, taken at 1.4 rest-frame days
after discovery, yields a phase of $+15 \pm 19$~d.  Using both
constraints, we estimate that SN~2014dt was discovered $+4 \pm 7$~d
after maximum brightness, consistent with the photometry.  This puts
maximum light for SN~2014dt on 2014 October 25 ($\pm$7~d).

At discovery, SN~2014dt had an absolute magnitude $M_{V} = -16.9 \pm
0.3$~mag, where we use a distance modulus to M61 of $30.45 \pm
0.24$~mag\footnote{\citet{Fox15} use a distance modulus of 31.43~mag,
  which is inconsistent with the Tully-Fisher distance \citep[$\mu =
  30.21 \pm 0.70$~mag;][]{Schoeniger97}, the redshift-derived distance
  (corrected for Virgo infall; $\mu = 30.59 \pm 0.16$~mag), and an
  expanding photosphere method distance using the SN~II~2008in
  \citep[$\mu = 30.45 \pm 0.10$~mag or $\mu = 30.81 \pm 0.20$~mag,
  with the difference resulting from different prescriptions and the
  former being more consistent with external distances for a large
  sample]{Bose14}.  Their assumed distance comes from a separate
  analysis of SN~2008in \citep{Rodriguez14}.  While that distance may
  be correct, the authors specifically point out that their analysis
  yields a significant negative extinction for SN~2008in, the only
  such outlier of their sample.} \citep{Foley15:14dt}.

Since SN~2014dt was discovered close to peak brightness, the discovery
magnitude is a reasonable upper limit on the peak magnitude.  For the
lower limit, we examine the light curves of other SNe~Iax, which have
a maximum $\Delta m_{15} (V) = 1$~mag \citepalias{Foley13:iax}.  Since
the SN was discovered before +15~d, a reasonable lower limit is $M_{V}
= -17.9$~mag.  We use these limits to set the range of peak absolute
magnitudes, $M_{V} = -17.4 \pm 0.5$~mag.

We obtained a series of low-resolution spectra of SN~2014dt.  Here we
focus on the late-time spectra obtained from 2015 April 10 through
July 24, corresponding to phases of 172 to 270~d after $B$-band
maximum brightness.  The remainder of our dataset will be presented by
Jha et~al. (in prep.).  The data were obtained with the Goodman
spectrograph \citep{Clemens04} on the 4~m SOAR telescope, the Robert
Stobie spectrograph \citep{Smith06} on the 10~m SALT telescope, the
Kast double spectrograph \citep{Miller93} on the Shane 3~m telescope
at Lick Observatory, and the Low Resolution Imaging Spectrometer
\citep[LRIS;][]{Oke95} on the 10~m Keck~I telescope.

For most data, standard CCD processing and spectrum extraction were
accomplished with IRAF\footnote{IRAF: the Image Reduction and Analysis
  Facility is distributed by the National Optical Astronomy
  Observatory, which is operated by the Association of Universities
  for Research in Astronomy, Inc.\ (AURA) under cooperative agreement
  with the National Science Foundation (NSF).}.  The SALT spectra were
partially reduced with PySALT \citep{Crawford10}. The data were
extracted using the optimal algorithm of \citet{Horne86}.  Low-order
polynomial fits to calibration-lamp spectra were used to establish the
wavelength scale, and small adjustments derived from night-sky lines
in the object frames were applied.  We employed our own IDL routines
to flux calibrate the data and remove telluric lines using the
well-exposed continua of spectrophotometric standards \citep{Wade88,
  Foley03}.  Details of our spectroscopic reduction techniques are
described by \citet{Silverman12:bsnip}.

A log of spectral observations is presented in Table~\ref{t:spec}, and
the spectra are shown in Figure~\ref{f:14dt_spec}.

\setcounter{figure}{0}
\begin{figure}
\begin{center}
  \includegraphics[angle=0,width=3.2in]{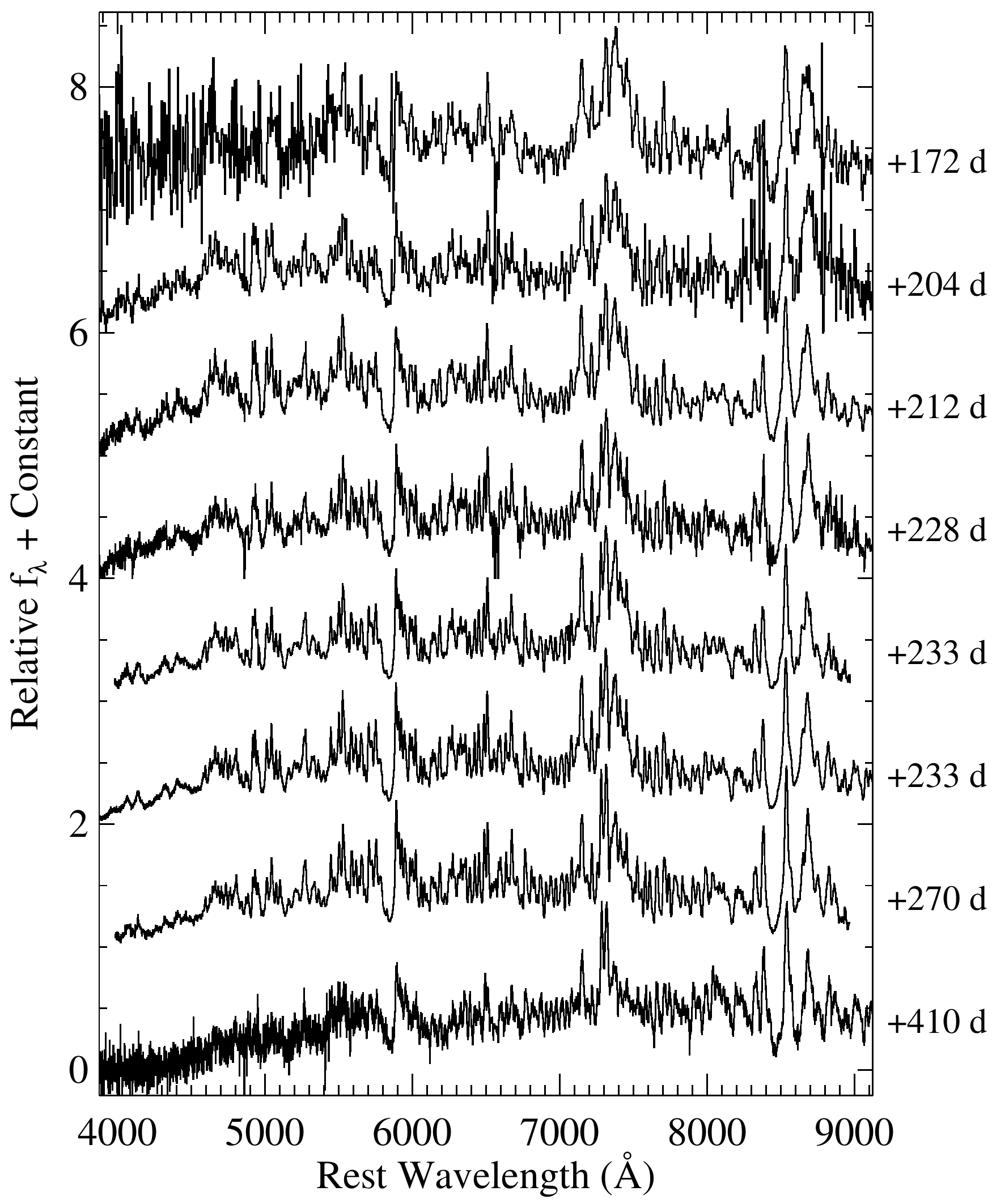}
  \caption{Late-time spectra of SN~2014dt.  Each spectrum is labeled
    by its phase relative to $B$-band maximum brightness.  All spectra
    have a continuum, permitted P-Cygni features, and forbidden
    lines.}\label{f:14dt_spec}
\end{center}
\end{figure}


\section{Properties of Late-time SN~I\lowercase{ax} Spectra}\label{s:measure}

\subsection{Spectral Variations Among SNe~Iax}

The primary difference between maximum-light spectra of different
SNe~Iax is their ejecta velocity \citepalias{Foley13:iax}.  If a
low-velocity SN~Iax spectrum is shifted and smoothed to mimic having a
higher ejecta velocity, the result will resemble that of a
higher-velocity SN~Iax spectrum.

At late times, all SN~Iax spectra share certain characteristics.
There is always a continuum, and the general shapes of the spectra are
similar.  The spectra all have similar permitted features such as the
\ion{Ca}{II} near-infrared (NIR) triplet and \ion{Na}{I}~D.
Similarly, every late-time spectrum has at least some indication of
[\ion{Ca}{II}] emission.

However, the late-time spectra of SNe~Iax show significant diversity,
and variance beyond that seen near peak brightness.  While some
late-time spectra have obvious low-velocity (\about 500~\kms) P-Cygni
profiles \citepalias[e.g., SN~2002cx;][]{Jha06:02cx}, others have
higher velocities blending these lines \citep[e.g.,
SN~2008ge;][]{Foley10:08ge}.  In addition to the difference in
velocities, there are differences in the strength of forbidden lines.
In particular, the [\ion{Fe}{II}] $\lambda 7155$, [\ion{Ca}{II}]
$\lambda \lambda 7291$, 7324, and [\ion{Ni}{II}] $\lambda 7378$
features have significantly different line strengths and widths.

\begin{figure*}
\begin{center}
  \includegraphics[angle=0,width=6.8in]{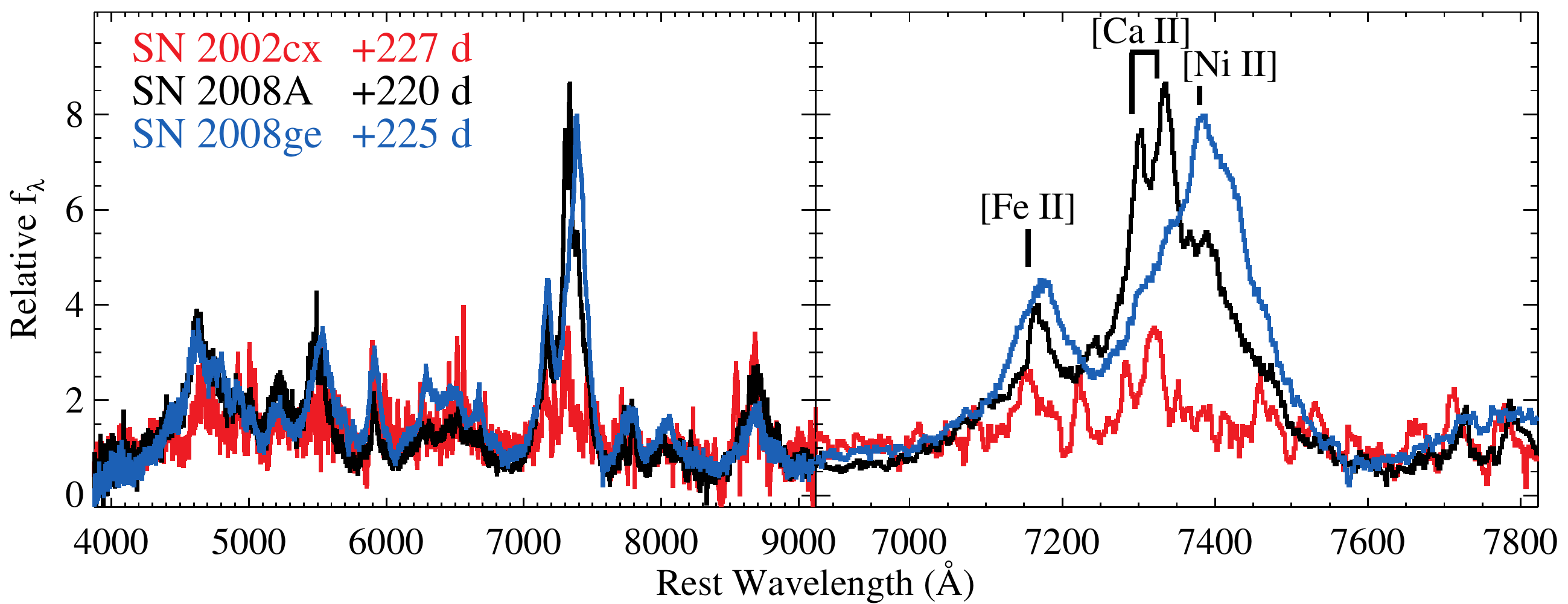}
  \caption{Late-time spectra of SNe~2002cx at a phase of +227~d, (red
    curve), 2008A at a phase of +220~d (black curve), and 2008ge at a
    phase of +225~d (blue curve).  The left panel shows the entire
    optical region, while the right panel displays the region
    containing the [\ion{Fe}{II}] $\lambda 7155$, [\ion{Ca}{II}]
    $\lambda \lambda 7291$, 7324, and [\ion{Ni}{II}] $\lambda 7378$
    features (all labeled).  The SN~2002cx spectrum has a relatively
    high signal-to-noise ratio (S/N), and the small-amplitude features
    in the SN~2002cx spectrum are mostly real \citepalias{Jha06:02cx}.
    This figure displays the heterogeneous late-time spectra of
    SNe~Iax.}\label{f:example}
\end{center}
\end{figure*}

Example spectra of objects having (1) high velocity, strong
[\ion{Ni}{II}], and weak [\ion{Ca}{II}] (SN~2008ge), (2) low velocity,
weak [\ion{Ni}{II}], and strong [\ion{Ca}{II}] (SN~2002cx), and (3)
intermediate properties (SN~2008A) are displayed in
Figure~\ref{f:example}.  These three spectra are indicative of the
main differences among late-time SN~Iax spectra.

SN~2002cx has low velocities at late times resulting in numerous
P-Cygni features being visible at all optical wavelengths.  It has no
obvious [\ion{Ni}{II}] $\lambda 7378$ emission, but relatively strong
[\ion{Fe}{II}] $\lambda 7155$ and [\ion{Ca}{II}] $\lambda \lambda
7291$, 7324.  SN~2008ge has lines broad enough such that individual
P-Cygni profiles are not obvious except for the strongest lines (e.g.,
Na~D and the Ca NIR triplet).  None the less, its continuum is
consistent with that of SN~2002cx, perhaps indicating that the main
difference between SNe~2002cx and 2008ge at late phases is that the
latter has higher-velocity material.  The [\ion{Ni}{II}] $\lambda
7378$ and [\ion{Fe}{II}] $\lambda 7155$ features for SN~2008ge are
very strong and somewhat strong, respectively, while its
[\ion{Ca}{II}] $\lambda \lambda 7291$, 7324 emission is barely
noticeable as small notches on the wings of the [\ion{Ni}{II}]
$\lambda 7378$ profile.

SN~2008A is intermediate to SNe~2002cx and 2008ge.  It has broad
features similar to SN~2008ge, but there are weak, low-velocity
P-Cygni profiles superimposed on the broader features.  Its
[\ion{Fe}{II}] $\lambda 7155$ emission is similar to that of both
SNe~2002cx and 2008ge, but noticeably narrower than that of SN~2008ge.
Its [\ion{Ca}{II}] $\lambda \lambda 7291$, 7324 emission is relatively
strong.  The [\ion{Ni}{II}] $\lambda 7378$ emission is sufficiently
strong to produce a pronounced ``shoulder'' on the [\ion{Ca}{II}]
profile, but is not strong enough to have a defined peak.

Furthermore, there are obvious line shifts between the different
spectra.  The peaks of the forbidden lines are progressively shifted
further to the blue from SN~2008ge to SN~2008A to SN~2002cx.

While there are additional differences between these spectra, as well
as for other spectra in our sample, these are the most obvious.  They
shape the initial investigations discussed below.

\subsection{Forbidden-Line Diversity}\label{ss:for}

As noted above, the [\ion{Fe}{II}] $\lambda 7155$, [\ion{Ca}{II}]
$\lambda \lambda 7291$, 7324, and [\ion{Ni}{II}] $\lambda 7378$
forbidden lines show significant diversity in the late-time spectra of
SNe~Iax.  Here we fit these features to measure line strengths,
velocity shifts, and velocity widths.

We fit multiple Gaussian profiles to all late-time SN~Iax spectra in
the region 6900 -- 7700~\AA.  Although this ignores other spectral
features in this region, the emission in this region is typically well
described by emission from only the four features listed above.  For
some spectra, it was obvious that two components (a ``broad''
component with a velocity width of \about 8000~\kms\ full width at
half-maximum intensity (FWHM), and a ``narrow'' component with a
velocity width of \about 1000~\kms\ FWHM) were necessary, with each
narrow/broad feature having the same kinematic properties (velocity
shift and velocity width) as the other narrow/broad features.  No
spectrum has obvious broad [\ion{Ca}{II}] emission.

\begin{figure*}
\begin{center}
  \includegraphics[angle=0,width=6.8in]{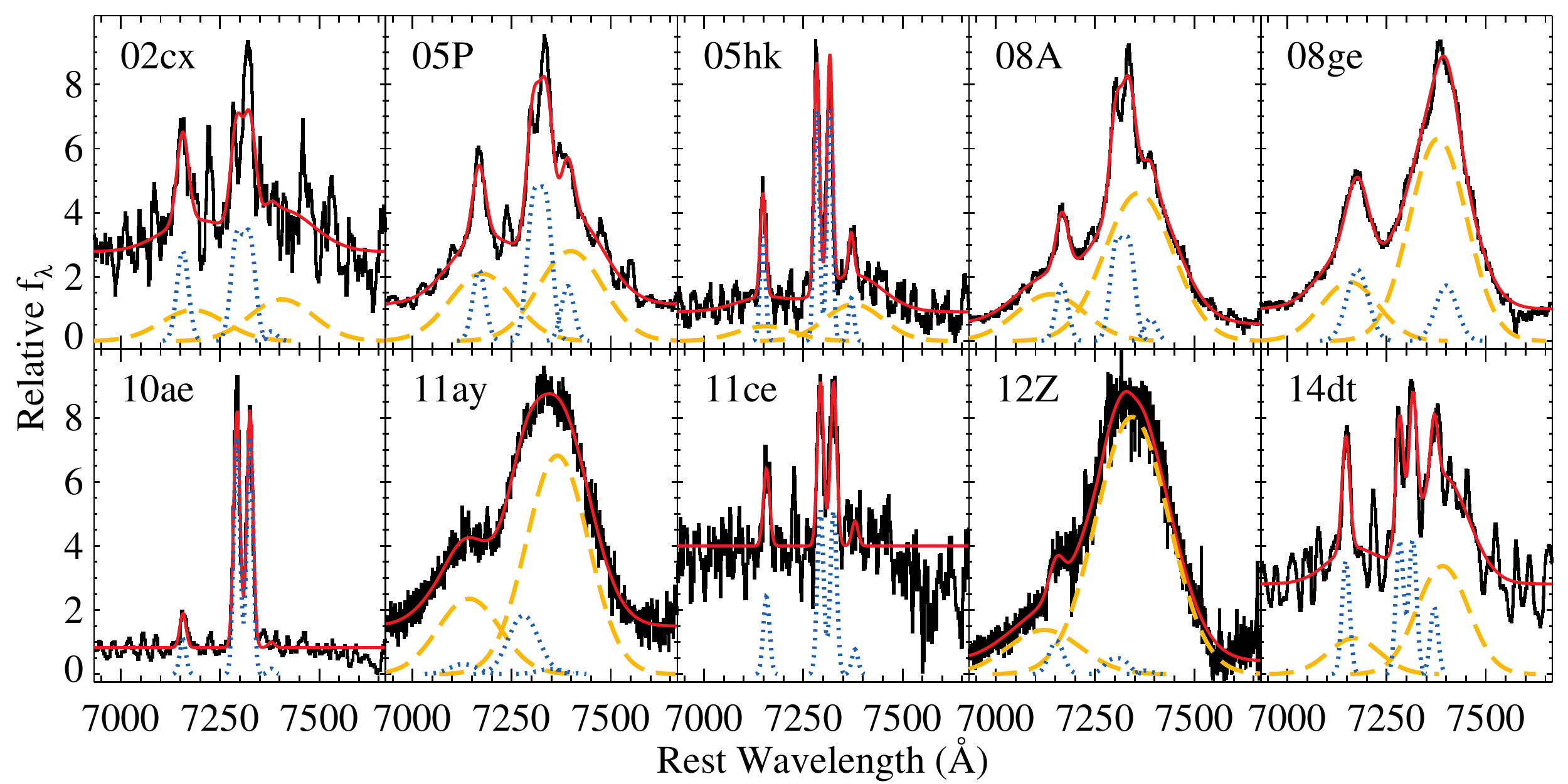}
  \caption{Late-time spectra of SNe~Iax (black).  Each panel displays
    the spectrum of a different SN.  The red curve corresponds to the
    best-fitting 10-parameter model of the forbidden lines.  The blue
    dotted curves and the gold dashed curves correspond to the
    individual narrow and broad components,
    respectively.}\label{f:neb_fit}
\end{center}
\end{figure*}

We fit the spectra with 4 kinematic parameters (2 each for the narrow
and broad components), 5 parameters to describe the line strengths
(fixing each [\ion{Ca}{II}] line to have the same flux), and a
constant flux offset, for a total of 10 parameters.  For a subset, the
fitting procedure could not distinguish between a constant flux offset
and low-flux, extremely broad, often extremely offset emission
features; in such cases (SNe~2002cx, 2010ae, and 2011ce), we fixed
these broad features to have zero flux, effectively removing the broad
components from the fit.  We also tried fitting each feature
separately, but found the parameters for the features from each
kinematic component to be essentially identical.  The best-fitting
models are shown in Figure~\ref{f:neb_fit} and the parameters are
listed in Table~\ref{t:neb_fit}.

In each case, the 10-parameter fit is generally a good description of
the data.  In some cases (particularly SNe~2002cx, 2005hk, 2010ae,
2011ce, and 2014dt), there are additional features, mostly
corresponding to permitted \ion{Fe}{II} lines \citepalias{Jha06:02cx},
which are not well fit by this model.  We do not attempt to account
for these features.  In particular, we note that as seen in the
spectral sequence of SN~2014dt (Section~\ref{ss:evol};
Figure~\ref{f:14dt_diff}), there appears to be a feature at roughly
the position of [\ion{Ca}{II}] $\lambda 7324$ which is unlikely to be
that line.  This feature is present in the 172-day spectrum of
SN~2014dt, but there is no similar line at 7291~\AA.  In all later
epochs (from +203~d onward), the [\ion{Ca}{II}] $\lambda 7291$ line is
present and of similar strength to [\ion{Ca}{II}] $\lambda 7324$,
though we caution that the other, contaminating line may result in
suboptimal fitting of these features, but should not significantly
affect our results for the spectra we examine.  None the less, future
investigations may employ a more detailed analysis where other lines,
including permitted features, are also fitted.

From the fitting, we can see at least three types of behaviour.  There
are SNe~Iax where the narrow components dominate, corresponding to
SNe~2002cx, 2005hk, 2010ae, and 2011ce; SNe~Iax where the broad
components dominate, corresponding to SNe~2008ge, 2011ay, and 2012Z;
and SNe~Iax where the narrow and broad components are roughly similar
in strength, corresponding to SNe~2005P, 2008A, and 2014dt.  These
correspond to the rough characterisation made at the beginning of
Section~\ref{s:measure} and in Figure~\ref{f:example}.

While several SNe have no discernible broad components, all SNe have
at least some narrow emission.  We can remove the need for narrow
components in SN~2011ay if we do not require that the broad components
have the same velocity shifts and velocity widths.  However, the broad
components appear to have the same widths and shifts for all other
SNe~Iax, and all other SNe~Iax require at least some narrow emission
for a reasonable fit.  As such, we include the narrow lines in its
fit, but caution overinterpretation of the strength of these features.

Below, we analyze the correlations between these parameters.

\subsection{Spectral Evolution with Time}\label{ss:evol}

Only a few SNe~Iax have multiple late-time spectra.  Of these objects,
SN~2002cx has two spectra separated by only 50~d \citepalias[+227 and
+277~d;][]{Jha06:02cx}.  SN~2005hk has at least 4 late-time spectra,
spanning a period of +230~d to +455~d \citep[although the last
spectrum with a detected continuum is at +403~d;][]{McCully14:iax}.
SN~2008A has four late-time spectra spanning +200~d to +283~d
\citep{McCully14:iax}.  SN~2012Z has two spectra at +215 and +248~d
\citep[a difference of only 33~d;][]{Stritzinger15}.  Finally,
SN~2014dt has multiple late-time spectra spanning +172 to +410~d.  For
SNe~2002cx and 2012Z, the time spans are relatively short, and there
is no obvious difference in the spectra at our disposal.  Therefore,
there are three SNe worth further investigation: SNe~2005hk, 2008A,
and 2014dt.

\begin{figure}
\begin{center}
  \includegraphics[angle=0,width=3.2in]{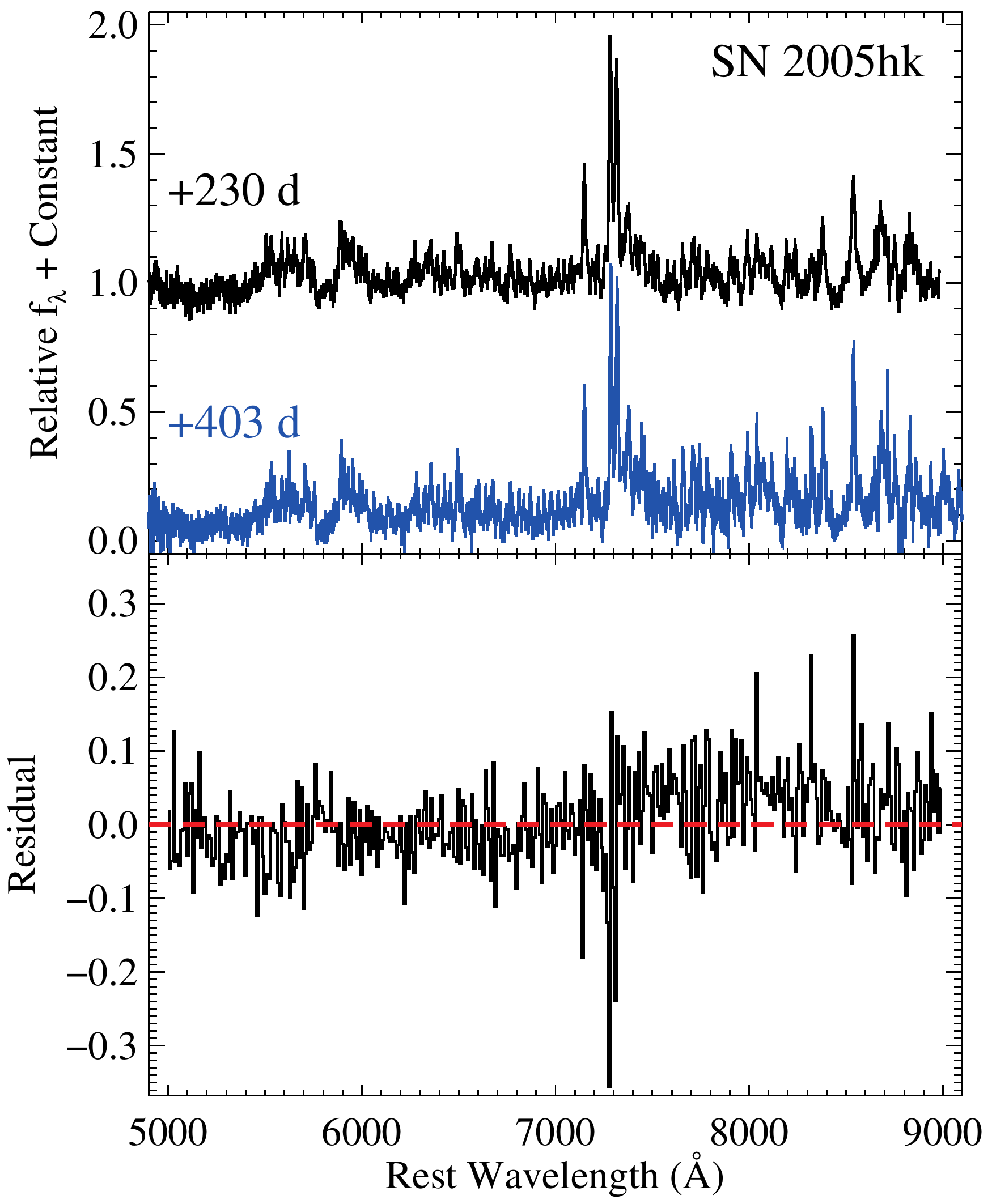}
  \caption{{\it Top panel}: Late-time spectra of SN~2005hk at phases
    of roughly $+230$ (top, black curve) and $+403$~d (bottom, blue
    curve), respectively.  The spectra are nearly identical in
    appearance despite the SN fading significantly between these
    epochs.  {\it Bottom panel}: Residual spectrum for these two
    spectra where the earlier spectrum is subtracted from the later
    spectrum.  The main difference is in the [\ion{Ca}{II}] $\lambda
    \lambda 7291$, 7324 feature.  This difference is the result of the
    lines becoming somewhat narrower with time (see
    Figure~\ref{f:sn2005hk2}).}\label{f:sn2005hk}
\end{center}
\end{figure}

For SN~2005hk, there is very little difference in the spectral
appearance between +230 and +403~d (Figure~\ref{f:sn2005hk}).
Although roughly 6 months has passed between these epochs, and the SN
is nearly twice as old in the second epoch as in the first and has
faded significantly, the spectra are nearly identical.

\begin{figure}
\begin{center}
  \includegraphics[angle=0,width=3.2in]{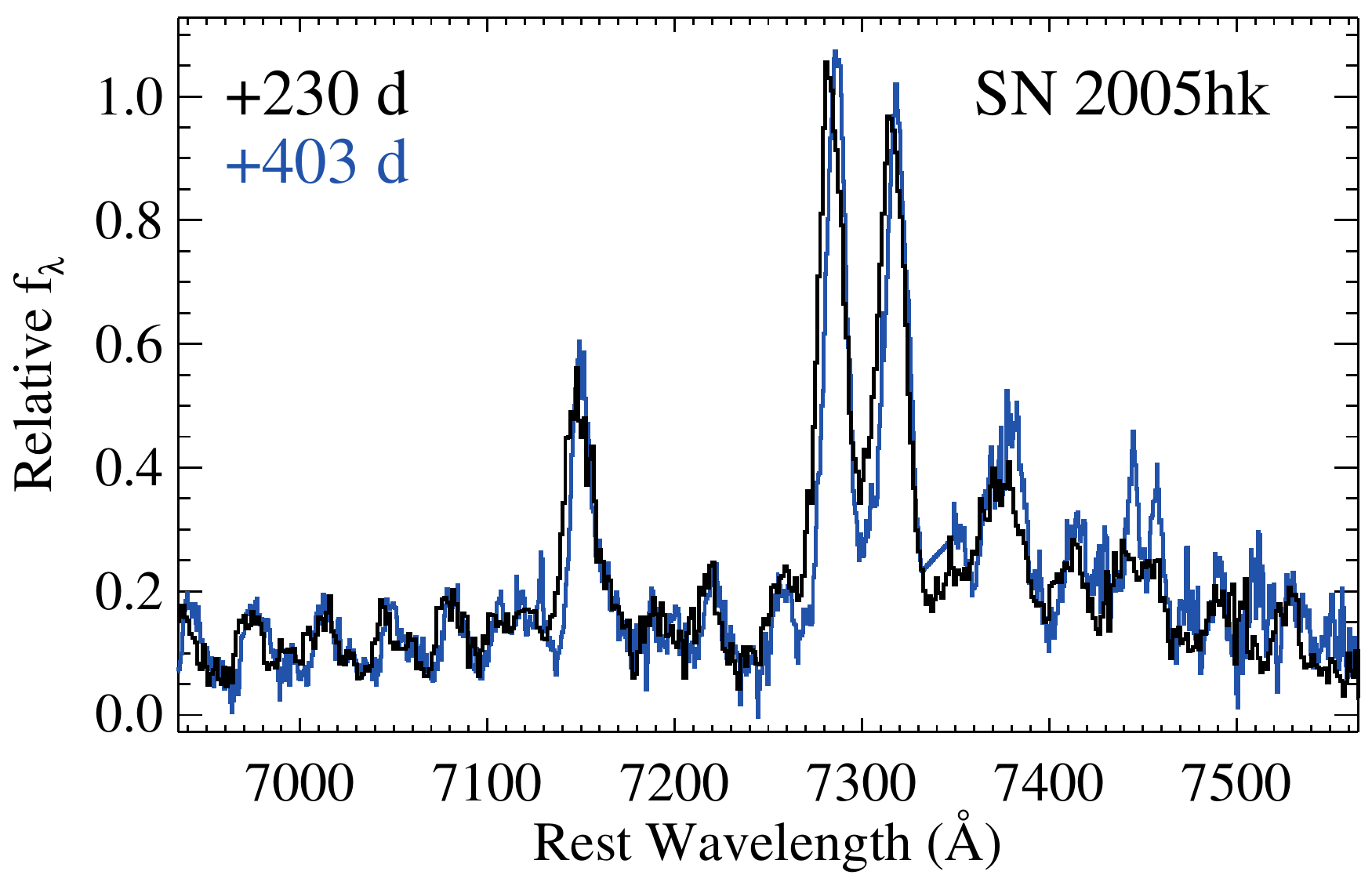}
  \caption{Late-time spectra of SN~2005hk at phases of roughly $+230$
    (black curve) and $+403$~d (blue curve), respectively.  The later
    spectrum has narrower and more blueshifted features for both the
    permitted and forbidden lines.}\label{f:sn2005hk2}
\end{center}
\end{figure}

Examining the differences between the two spectra
(Figure~\ref{f:sn2005hk}), we note that there is a slight difference
in the continuum strength, which may be the result of small errors in
flux calibration or a slight change to the temperature of the
photosphere.  Additionally, the [\ion{Ca}{II}] $\lambda \lambda 7291$,
7324 lines have a smaller equivalent width (EW) in the later spectrum
(Figure~\ref{f:sn2005hk}).  This difference is caused by the
[\ion{Ca}{II}] lines becoming narrower, with the FWHM decreasing from
290~\kms\ to 230~\kms, and moving slightly to the red (as determined
by simultaneously fitting both lines with Gaussians), with the
velocity shift increasing from $-360$~\kms\ to $-180$~\kms\ (where a
negative velocity indicates a blueshifted feature;
Figure~\ref{f:sn2005hk2}).  Similar behaviour is seen in the permitted
lines.  The decrease in velocity for both the permitted and narrow
forbidden lines suggests a physical connection.  We note that these
velocity shifts are unlikely to be caused by reddening from newly
formed dust; in that case, we would expect the lines to shift to the
blue \citep[e.g.,][]{Smith08:06jc}.

\begin{figure}
\begin{center}
  \includegraphics[angle=0,width=3.2in]{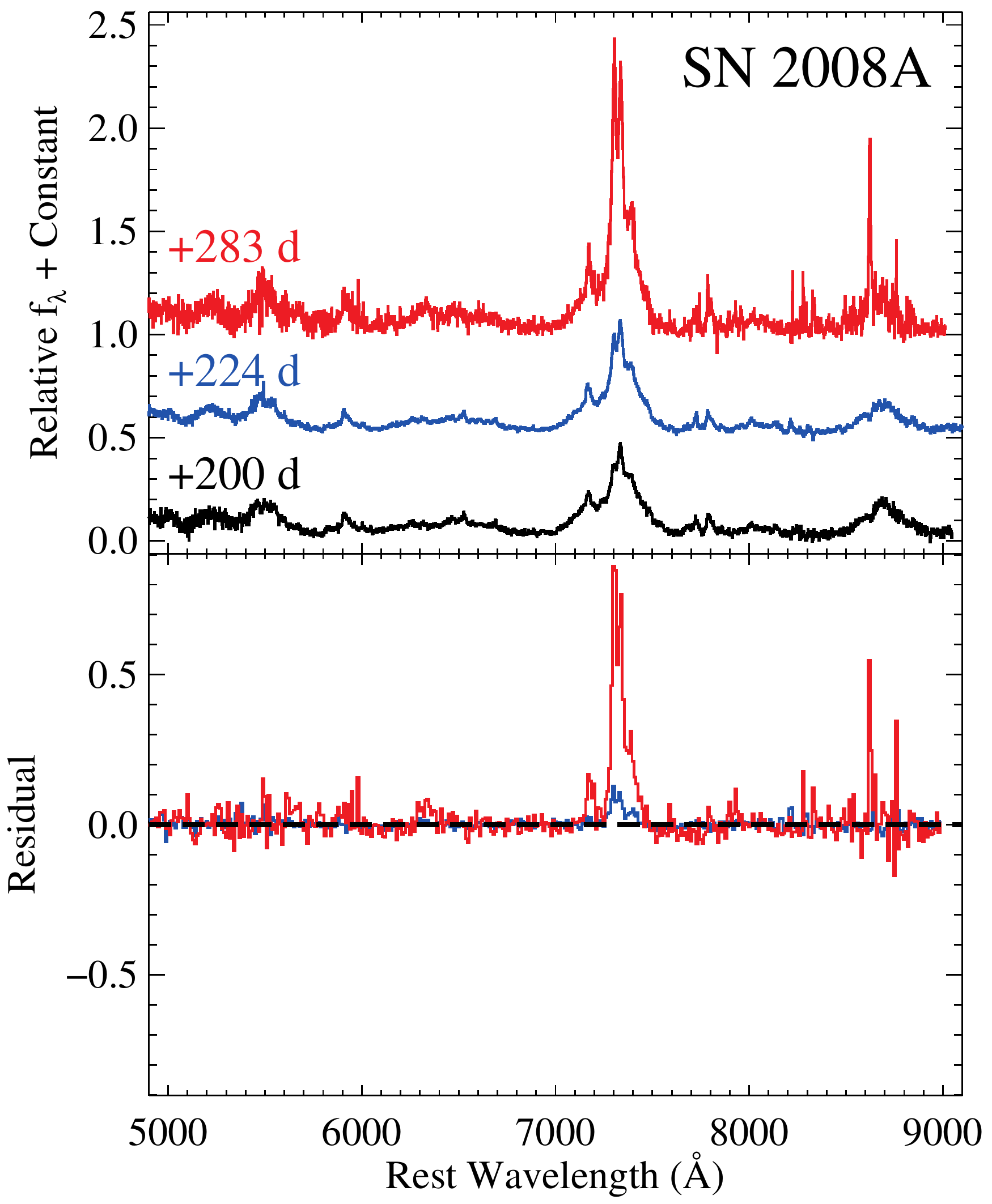}
  \caption{{\it Top panel}: Late-time spectra of SN~2008A at phases of
    roughly $+200$ (bottom, black curve), $+224$~d (middle, blue
    curve), and $+283$~d (top, red curve), respectively.  The spectra
    are nearly identical, except for at wavelengths of 7000 --
    7600~\AA.  {\it Bottom panel}: Residual spectra for these spectra
    where the $+200$~d spectrum is subtracted from the later spectra.
    The main differences are in [\ion{Fe}{II}] $\lambda 7155$,
    [\ion{Ca}{II}] $\lambda \lambda 7291$, 7324, and [\ion{Ni}{II}]
    $\lambda 7378$, with the later spectra having generally stronger
    lines.}\label{f:sn2008a}
\end{center}
\end{figure}

SN~2008A, unlike SN~2005hk, has significant spectral evolution between
+200 and +283~d.  Again, SN~2008A has faded significantly between
these epochs.  While most of the spectrum is nearly identical during
this time (Figure~\ref{f:sn2008a}), the strengths of the forbidden
lines ([\ion{Fe}{II}] $\lambda 7155$, [\ion{Ca}{II}] $\lambda \lambda
7291$, 7324, and [\ion{Ni}{II}] $\lambda 7378$) change dramatically
between the three epochs (at +200, +224, and +283~d).  Most of this
evolution occurs between +224 and +283~d, with only minor changes to
the features between +200 and +224~d.  While the forbidden-line
strengths change, the SN does not transition to (or from) a spectrum
more similar to SN~2002cx or SN~2008ge; SN~2008A always has relatively
strong narrow and forbidden lines.

\begin{figure}
\begin{center}
  \includegraphics[angle=0,width=3.2in]{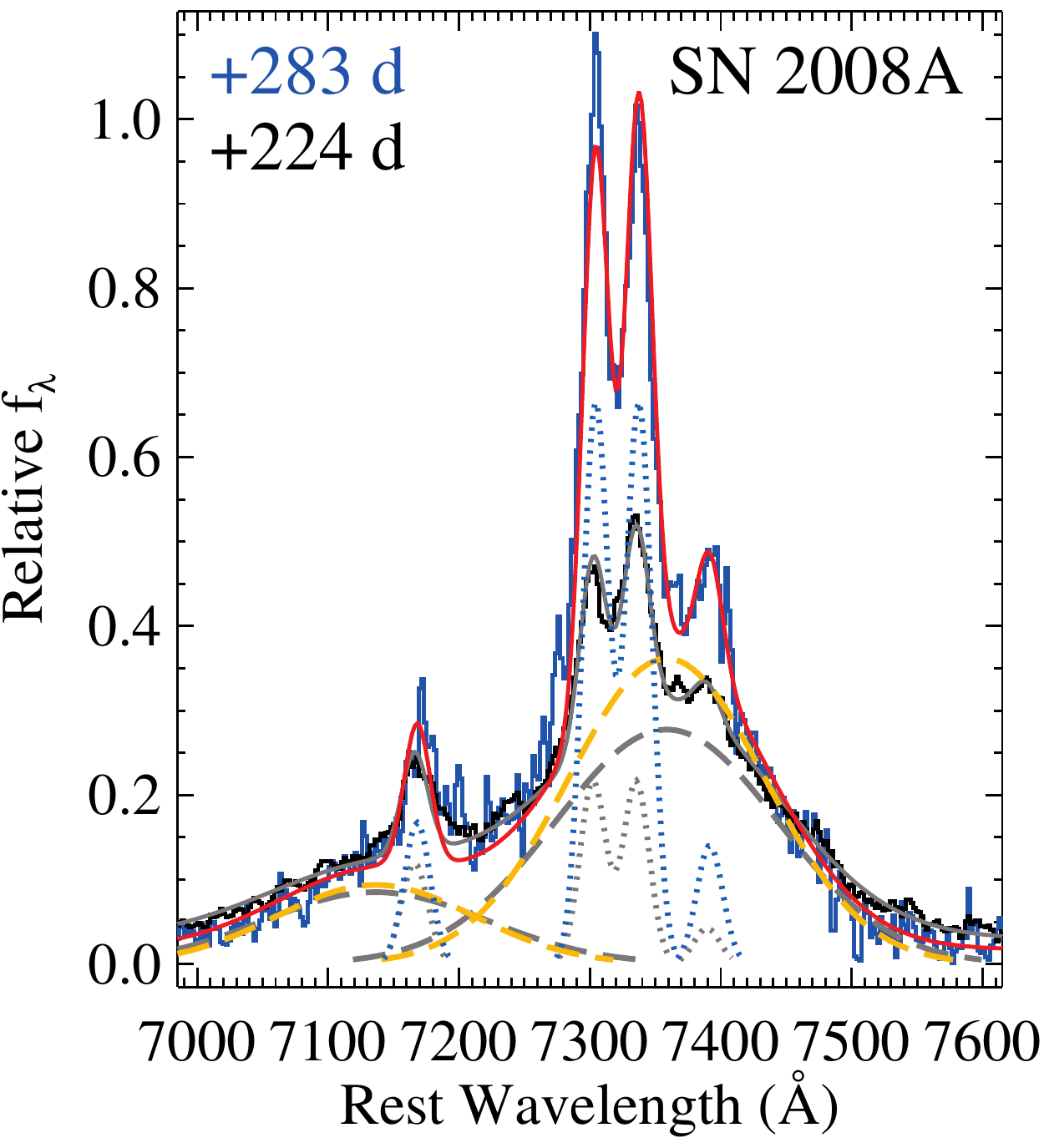}
  \caption{Late-time spectra of SN~2008A at phases of roughly $+200$
    (black curve) and $+283$~d (dark-blue curve), respectively.  Also
    shown are 10-parameter model spectra (see Section~\ref{ss:for}) in
    solid grey and red, respectively.  The components corresponding to
    broad [\ion{Fe}{II}] $\lambda 7155$ and [\ion{Ni}{II}] $\lambda
    7378$ are shown with long-dashed lines, with the grey and gold
    curves corresponding to the $+200$ and $+283$~d spectra,
    respectively.  Similarly, the narrow [\ion{Ca}{II}] $\lambda
    \lambda 7291$, 7324 and [\ion{Ni}{II}] $\lambda 7378$ are shown as
    dotted lines, with the light-blue and grey curves corresponding to
    the $+200$ $+283$~d spectra, respectively.}\label{f:sn2008a2}
\end{center}
\end{figure}

Examining the forbidden lines in detail (Figure~\ref{f:sn2008a2}), we
see that the narrow components (see Section~\ref{ss:for}) of the lines
all get stronger (in EW) by factors of \about 2--5 between +224 and
+283~d.  This is most obvious in the [\ion{Ca}{II}] doublet, which
increases in strength by a factor of \about 4 and is clearly the
dominant feature in the +283~d spectrum.  The broad [\ion{Fe}{II}]
feature is roughly the same strength in both spectra, but the broad
[\ion{Ni}{II}] feature is \about 50\% stronger in the later spectrum.
This behaviour may be the result of the continuum fading while the
narrow features stay relatively constant in flux.

Similar to SN~2005hk, the narrow forbidden lines become slightly more
redshifted with time (from +470~\kms\ to +550~\kms), but the line
widths do not significantly change.  The velocity widths of the broad
components do not significantly change either.  However, detailed
modeling of the full complex, including permitted-line emission, may
reveal subtle shifts.

\begin{figure*}
\begin{center}
  \includegraphics[angle=0,width=6.8in]{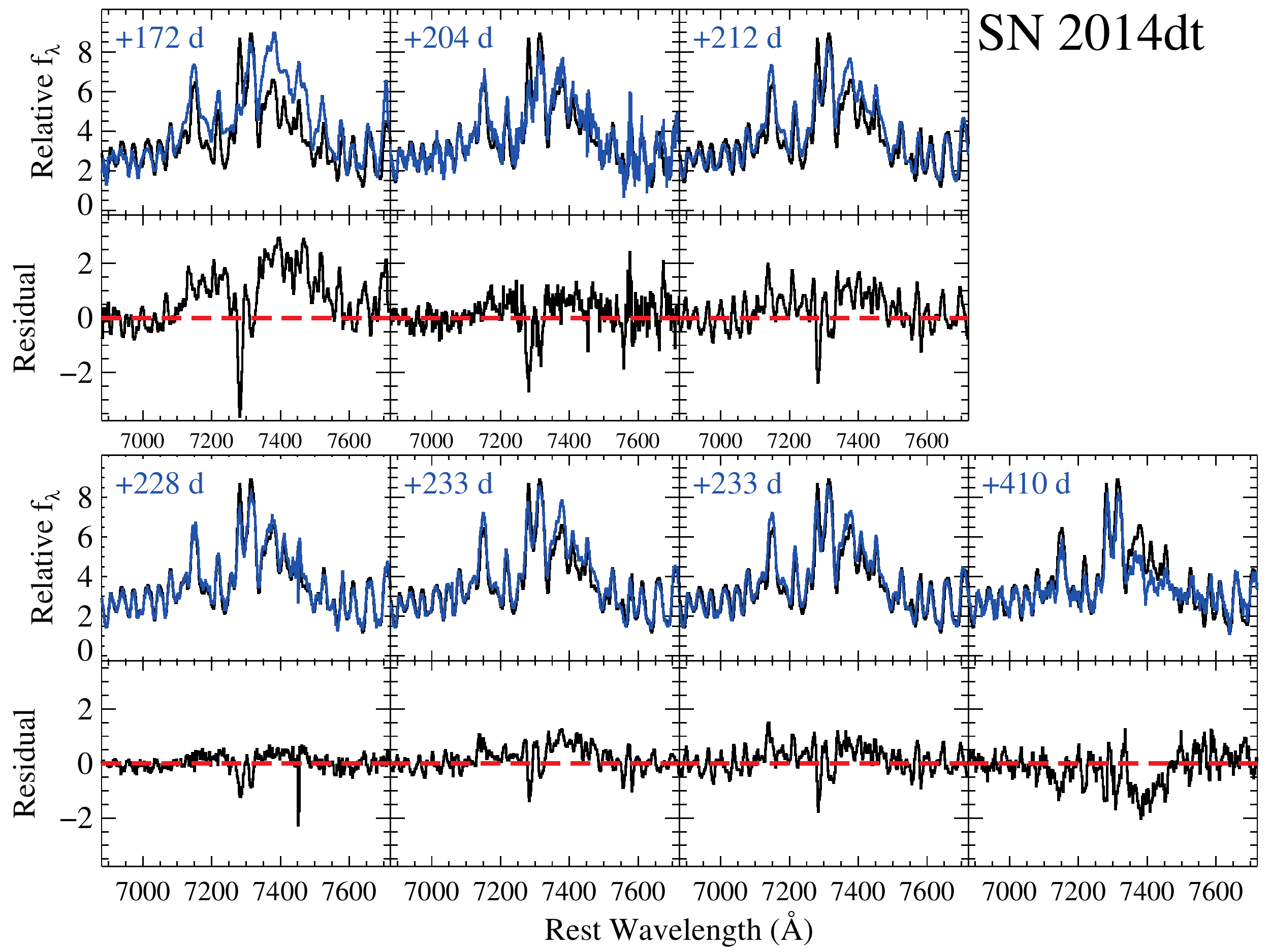}
  \caption{Late-time spectra of SN~2014dt from 172 to 270~d after
    maximum brightness, focusing on the forbidden-line region.  The
    top panel of each row shows a different spectrum in blue, with its
    phase labeled.  The +270~d spectrum is given in black for
    comparison in each subplot.  The bottom panels of each row show
    the residual spectrum relative to the +270~d spectrum after they
    have been arbitrarily scaled to have their continua match just
    blueward and redward of the forbidden-line complex.  The red
    dashed lines represent zero residual flux.}\label{f:14dt_diff}
\end{center}
\end{figure*}

Finally, SN~2014dt has the best spectral sequence of any SN~Iax at
late times.  The details of the spectral evolution will be presented
by Jha et~al.\ (in prep.); here we focus on the region around the
forbidden lines already identified.  We display that spectral region
in Figure~\ref{f:14dt_diff}.  In Figure~\ref{f:14dt_diff}, we also
show the residual spectra compared to the +270~d spectrum.

Notably, the spectra do not evolve from being similar to SN~2002cx
into being similar to SN~2008ge (or vice versa; see
Figure~\ref{f:example}).  The main changes are the continued decrease
in a broad feature that is presumably [\ion{Ni}{II}] $\lambda 7378$
with perhaps some contribution from [\ion{Fe}{II}] $\lambda 7155$, and
the strengthening of narrow [\ion{Ca}{II}] $\lambda \lambda 7291$,
7324.  Despite these noticeable differences between different phases,
there is very little spectral evolution between +203 and +410~d.  The
172-day spectrum is less similar to the other spectra and likely is
still transitioning into being a true ``late-time'' spectrum.  This
relative stability implies that a single spectrum taken after about
200~d relative to maximum brightness is sufficient to characterise the
late-time spectrum of a SN~Iax.  While this statement is consistent
with our findings for SNe~2005hk and 2014dt, more data will be
necessary to determine if the evolution seen in SN~2008A typically
occurs primarily around +270~d or continues steadily between +230 and
+270~d.

Despite the evolutionary changes seen in SNe~2005hk, 2008A, and
2014dt, they are all relatively small and any such late-time evolution
should not significantly affect our results below.

\subsection{Velocity Shifts}

As is evident from Figures~\ref{f:example} and \ref{f:neb_fit}, as
well as Table~\ref{t:neb_fit}, there are large differences in the
forbidden-line shifts in late-time SN~Iax spectra.  These shifts can
be caused by the motion of the progenitor system or asymmetries in the
explosion.

In addition to forbidden-line shifts, the permitted features are at
different velocities for different SNe.  To determine the relative
velocity shifts between spectra, we cross-correlated the SN~2005hk
spectrum and other spectra.  From the measured lag, we can directly
measure the velocity shift.  SN~2005hk was used since it has (1) a
very high-S/N spectrum; (2) relatively low-velocity features, allowing
for precise measurements of any shifts; and (3) both narrow and broad
forbidden lines.

Performing the cross-correlation, we decided to examine different
wavelength ranges.  We measured cross correlations using essentially
all data (4600 -- 9000~\AA), a blue region (4600 -- 6500~\AA, limited
on the red side to avoid any possible correlation with galactic
H$\alpha$ emission), a red region (7600 -- 9000~\AA, bounded on the
blue side to avoid the strong forbidden lines discussed above), a
forbidden-line region (6900 -- 7600~\AA), as well as disjoint
1000~\AA\ regions starting at 5000 -- 6000~\AA\ and ending at 8000 --
9000~\AA.

Unsurprisingly, many of the derived cross-correlation velocities are
strongly correlated with each other.  Interestingly, the
forbidden-line region is uncorrelated with all nonoverlapping regions.
The highest correlation is with the red region: a correlation
coefficient of 0.14.  The forbidden-line region has a higher
anticorrelation with the 5000 -- 6000~\AA\ region (correlation
coefficient of $-0.51$).

However, half of the SN~Iax sample (SNe~2002cx, 2005P, 2005hk, 2010ae,
2011ce, and 2014dt) have forbidden-line shifts similar to that of the
permitted lines.  Notably, these are the SNe~Iax with the weakest
broad emission lines and their forbidden-line shifts are primarily
determined from the narrow forbidden lines.  The remaining SNe have
forbidden-line shifts that are significantly offset from their
permitted-line shifts, as determined by cross correlation.

Examining the velocity shifts for the narrow forbidden lines as
determined in Section~\ref{ss:for}, the permitted line shifts are now
relatively correlated with a correlation coefficient of 0.47.
Comparing these values, the outliers are SNe~2008A, 2008ge, and 2012Z.
Unsurprisingly, these are 3/4 of the SNe~Iax with the weakest narrow
forbidden lines.  Although a possible interpretation is that the
narrow-component forbidden-line velocity shifts are poorly measured
for these SNe, the narrow lines are clearly seen in SN~2008A.  Another
interpretation is that the physical regions producing the permitted
and forbidden lines are essentially independent of each other for the
SNe with strong broad forbidden lines.

Since the velocity shifts for the broad forbidden lines and permitted
lines, even when there are no narrow P-Cygni features visible, are
uncorrelated, it is likely that the material from which the broad
forbidden lines are formed and the photosphere, which is where the
continuum originates, are physically distinct.  However, the
correlation with the narrow forbidden lines and permitted lines
suggests that those components {\it do} originate from the same
material.  These results favour the idea that late-time SNe~Iax are
composed of two physically distinct regions.

None the less, the photosphere and the material generating the broad
forbidden lines are somehow connected.  The SNe~Iax with the broadest
forbidden lines also lack distinct low-velocity P-Cygni features,
suggesting that SNe~Iax with higher-velocity photospheres also have
higher-velocity, and more blueshifted, broad forbidden-line-forming
regions.

\subsection{Principal-component Analysis}\label{ss:pca}

To investigate the possibility that late-time SN~Iax spectra have
distinct physical components and to further examine correlations
between spectral features, we perform a principal-component analysis
(PCA) of the spectra.  To do this, we subtract the average flux from
each spectrum and scale each spectrum to have a similar flux.  We also
shift the spectra in velocity space by their narrow forbidden emission
line velocity shift.  This last step reduces differences from small
velocity shifts and focuses the analysis on differences in
emission-line strengths and widths.

\begin{figure}
\begin{center}
  \includegraphics[angle=0,width=3.2in]{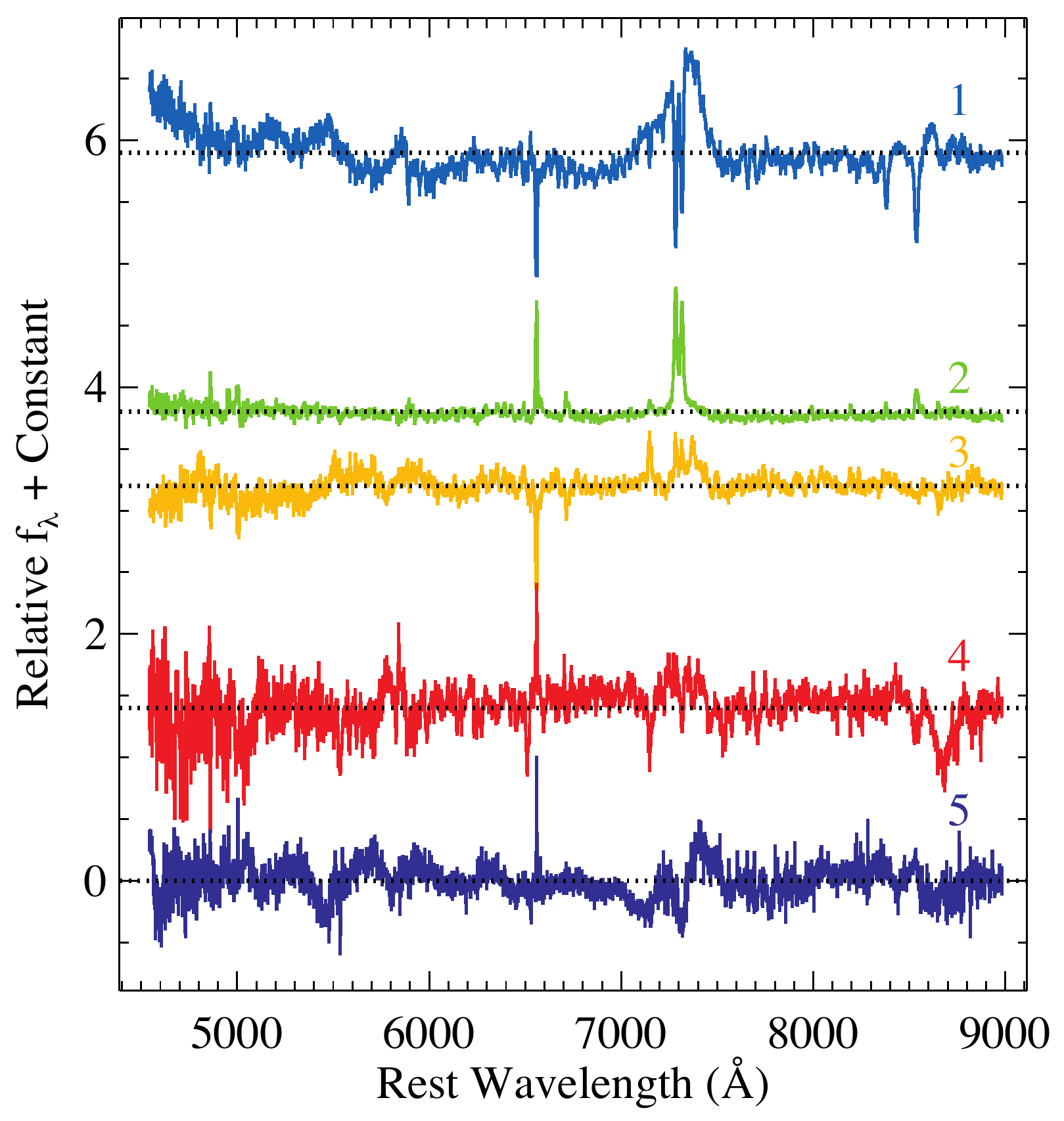}
  \caption{First five eigenspectra for late-time SN~Iax spectra.  The
    ``zero flux'' is annotated as a dotted line for each
    eigenspectrum.}\label{f:pca}
\end{center}
\end{figure}

We present the first 5 eigenvalues for each SN in Table~\ref{t:eigen}.
Figure~\ref{f:pca} displays the first 5 eigenspectra for our sample
(all normalised to have the same maximum amplitude).  The first 5
eigenspectra represent 40.8, 26.5, 11.0, 7.7, and 4.6\% of the total
variance between spectra, respectively.  Cumulatively, this
corresponds to 40.8, 67.3, 78.3, 85.9, and 90.5\% of the total
variance.

\begin{deluxetable}{lrrrrr}
\centering
\tabletypesize{\footnotesize}
\tablewidth{0pt}
\tablecaption{Eigenvalues for Late-time SN~Iax Spectra\label{t:eigen}}
\tablehead{
\colhead{SN} &
\colhead{1st} &
\colhead{2nd} &
\colhead{3rd} &
\colhead{4th} &
\colhead{5th}}

\startdata

2002cx &    1.1  &  10.4  &  1.6 & $-11.2$ &  $-1.2$\\
2005P  &   15.7  &  18.3  & 12.9 &  $-3.6$ &  $-9.9$\\
2005hk &  $-9.2$ &  23.0  & 33.1 &  $-0.6$ &  $-1.2$\\
2008A  &   32.2  &  31.2  & 15.2 &    1.4  & $-12.9$\\
2008ge &   27.9  &  15.0  & 16.7 &    6.8  &  $-4.2$\\
2010ae & $-14.8$ &  51.8  &  2.9 &    5.8  &  $-4.4$\\
2011ay &   22.8  &  15.3  &  3.6 &    2.0  &  $-6.5$\\
2011ce & $-10.8$ & $-0.8$ &  8.2 &   16.3  &  $-8.0$\\
2012Z  &   32.9  &  25.7  &  9.6 &    8.4  &    8.9\\
2014dt &    0.4  &   8.7  & 11.0 &  $-1.7$ &    0.3

\enddata

\end{deluxetable}

The eigenspectra show interesting correlations between features,
including some correlations not identified in the previous sections.
In the first eigenspectrum, the main features are broad components to
the forbidden lines with anticorrelated narrow components.  That is,
the first eigenspectrum describes the relative strengths of the broad
and narrow forbidden lines.  Additionally, the first eigenspectrum
also has a blue continuum correlated with stronger broad components.
It is unclear if the colour difference is the result of additional
emission at these wavelengths or caused by (uncorrected) dust
reddening (the latter is unlikely since the bluer continuum is
correlated with narrow Na~D absorption that is likely ISM absorption;
intriguingly, we are unable to detect any narrow Na~D within the broad
Na~D associated with the SN, but the eigenspectra are able to isolate
this feature).  Finally, the first eigenspectrum shows a correlation
between high-frequency permitted lines and the strength of the narrow
forbidden lines.  Therefore, the first eigenspectrum suggests that
SNe~Iax with relatively strong narrow forbidden lines (and weaker
broad forbidden lines) have more distinct permitted features.

The second eigenspectrum is basically a flat spectrum with mostly
[\ion{Ca}{II}] $\lambda \lambda 7291$, 7324 emission.  This component
is essentially uncorrelated with any other feature, although there is
weak, narrow [\ion{Fe}{II}] emission correlated with the
[\ion{Ca}{II}] emission.  The continuum is slightly negative at nearly
every wavelength, indicating that the overall continuum strength is
anticorrelated with the strength of the [\ion{Ca}{II}] feature.

The third eigenspectrum has correlated narrow and broad forbidden
lines that are anticorrelated with a blue continuum.  This both
confirms the necessity of broad and narrow forbidden lines and is a
key discriminant for ``transition'' objects.  The fourth eigenspectrum
shows a correlation between narrow [\ion{Ni}{II}] $\lambda 7155$ and
broad \ion{Ca}{II} NIR emission.  The fifth eigenspectrum exhibits
``P-Cygni-like'' features for the broad emission lines, and is likely
related to velocity shifts for the broad emission relative to the
narrow emission.

\begin{figure}
\begin{center}
  \includegraphics[angle=0,width=3.2in]{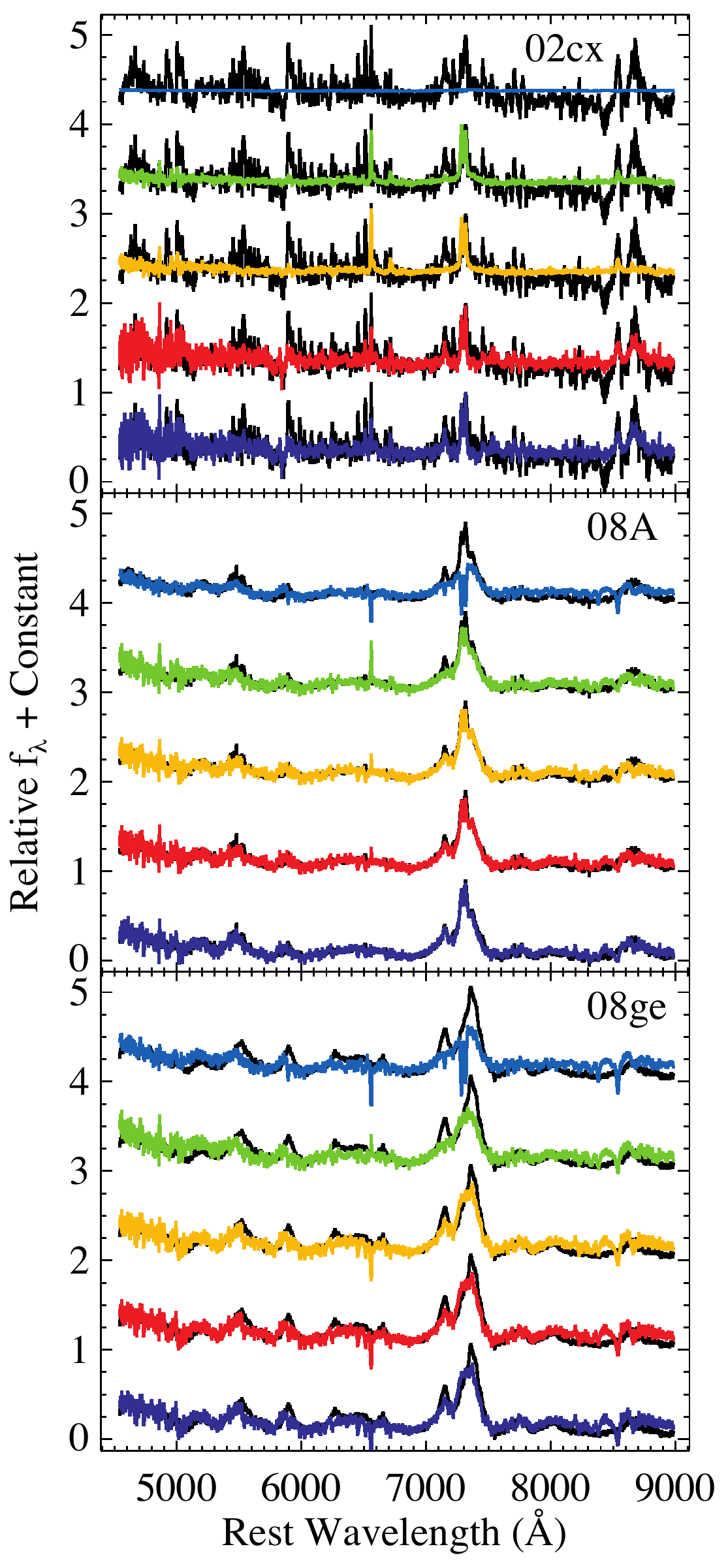}
  \caption{Late-time spectra of SNe~2002cx (top panel), 2008A (middle
    panel), and 2008ge (bottom panel) repeated in black.  The
    successive (from top to bottom) coloured curves correspond to the
    reconstructed spectra using the first $N$ eigenspectra, where $N$
    corresponds to the spectrum's position from the top of the
    panel.}\label{f:pca_example}
\end{center}
\end{figure}

Figure~\ref{f:pca_example} displays the SNe~2002cx, 2008A, and 2008ge
spectra (the same as in Figure~\ref{f:example}) compared to their
progressively reconstructed spectra.  That is, the first comparison
shows the first eigenspectrum multiplied by the first eigenvalue for
that spectrum, while the second comparison shows that same projected
spectrum added to the second eigenspectrum multiplied by the second
eigenvalue for that spectrum.  If there were zero variance beyond the
fifth eigenspectrum, the final comparison would be equivalent to both
the reconstructed spectrum and the true spectrum.  For these spectra,
a reconstruction using the first 5 eigenspectra results in reasonable
reproductions of the data.

While the eigenvalues are representative of the projection of spectra
onto the eigenvectors, the relative eigenvalues are more illustrative
than their absolute values.  Examining the eigenvalues for each
spectrum, it is clear that the first eigenvalue is highly correlated
with the strength of the broad forbidden lines, with SNe~2008A,
2008ge, 2011ay, and 2012Z having the largest (positive) eigenvalues
and SNe~2005hk, 2010ae, and 2011ce having the smallest (negative)
eigenvalues.

The second eigenvalues are positive for all SNe except for SN~2011ce.
While one might naively think that the second eigenvalue dictates the
strength of the observed [\ion{Ca}{II}] emission, this is only
partially correct.  Spectra having large (positive) first eigenvalues
also need large second eigenvalues to ``fill in'' the ``absorption,''
while negative first eigenvalues result in some [\ion{Ca}{II}]
emission, and so the size of the second eigenvalue is not perfectly
correlated with the observed [\ion{Ca}{II}] emission.

The second eigenvalue more directly tracks the continuum strength.
For instance, the SNe with the largest second eigenvalue are
SNe~2005hk, 2008A, 2010ae, and 2012Z, all of which have a small
continuum level relative to their emission lines (see
Figure~\ref{f:example}).  However, SNe~2002cx, 2011ce, and 2014dt,
which have small second eigenvalues, all have relatively high continua
relative to their emission lines.

The third eigenvalue provides some indication if a SN has a
``transition'' spectrum with both narrow and broad components.  The
SNe with the largest third eigenvalue, from strongest to weakest, are
SNe~2005hk, 2008ge, 2008A, 2005P, and 2014dt.  While not a direct
correspondence, this group does include our previously identified
transition objects and excludes the most extreme members of the class
on both ends (e.g., SNe~2010ae and 2011ay).

Additional eigenspectra have more complicated interpretations.
However, we caution against overinterpretation of the eigenspectra.
The correlations, especially for less-significant eigenspectra, do not
necessarily correspond to a physical cause and effect rather than
simply correlation.


\section{Analysis}\label{s:analysis}

In Section~\ref{s:measure}, we described three different methods to
examine the late-time (($t \gtrsim 200$~d) spectra of SNe~Iax:
model-fitting of forbidden lines, cross-correlation to determine
velocity shifts, and a PCA.  Here, we combine measurements from these
methods along with other extant data to examine the causes of the
spectroscopic diversity of SNe~Iax at late times.

In addition to the velocity shifts, velocity widths, line strengths,
line ratios, and eigenvalues derived above, we examine other SN
properties as reported in other studies.  In particular, we
investigate the peak luminosity, the light-curve shape, and the
photospheric velocity at maximum brightness.

\subsection{Spectral Comparisons}

\begin{figure}
\begin{center}
  \includegraphics[angle=0,width=3.2in]{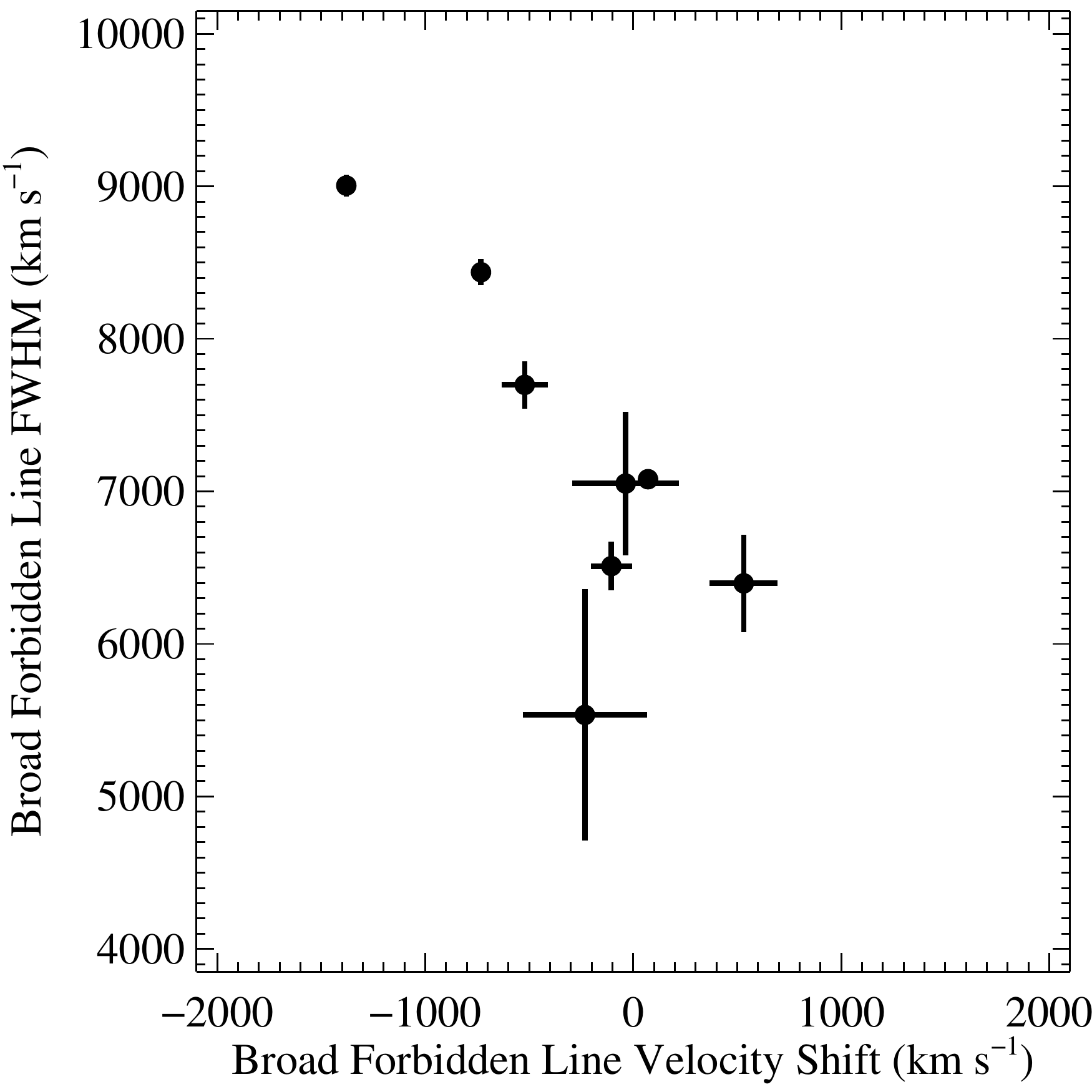}
  \caption{Line width as a function of velocity shifts for the broad
    forbidden-line components as fitted in Section~\ref{ss:for}.  The
    correlation coefficient is $-0.54$.}\label{f:broad}
\end{center}
\end{figure}

We first examine the broad and narrow components of the forbidden
lines individually.  For the broad components, there is a strong
correlation between the velocity shift and the velocity width
\citepalias[Figure~\ref{f:broad}; see also][]{Foley13:iax}.
Specifically, SNe~Iax with blueshifted broad components tend to be
broader than SNe with broad components that have no velocity shift or
are redshifted.  The correlation coefficient for this relation is
$-0.54$; however, the true relation appears to be stronger than this
number suggests.  Performing a Bayesian Monte-Carlo linear regression
on the data \citep{Kelly07}, we exclusively found non-negative slopes
for the fitted lines in all of 200,000 trials, making the results
significant at $>$5.5~$\sigma$.

\begin{figure*}
\begin{center}
  \includegraphics[angle=0,width=3.8in]{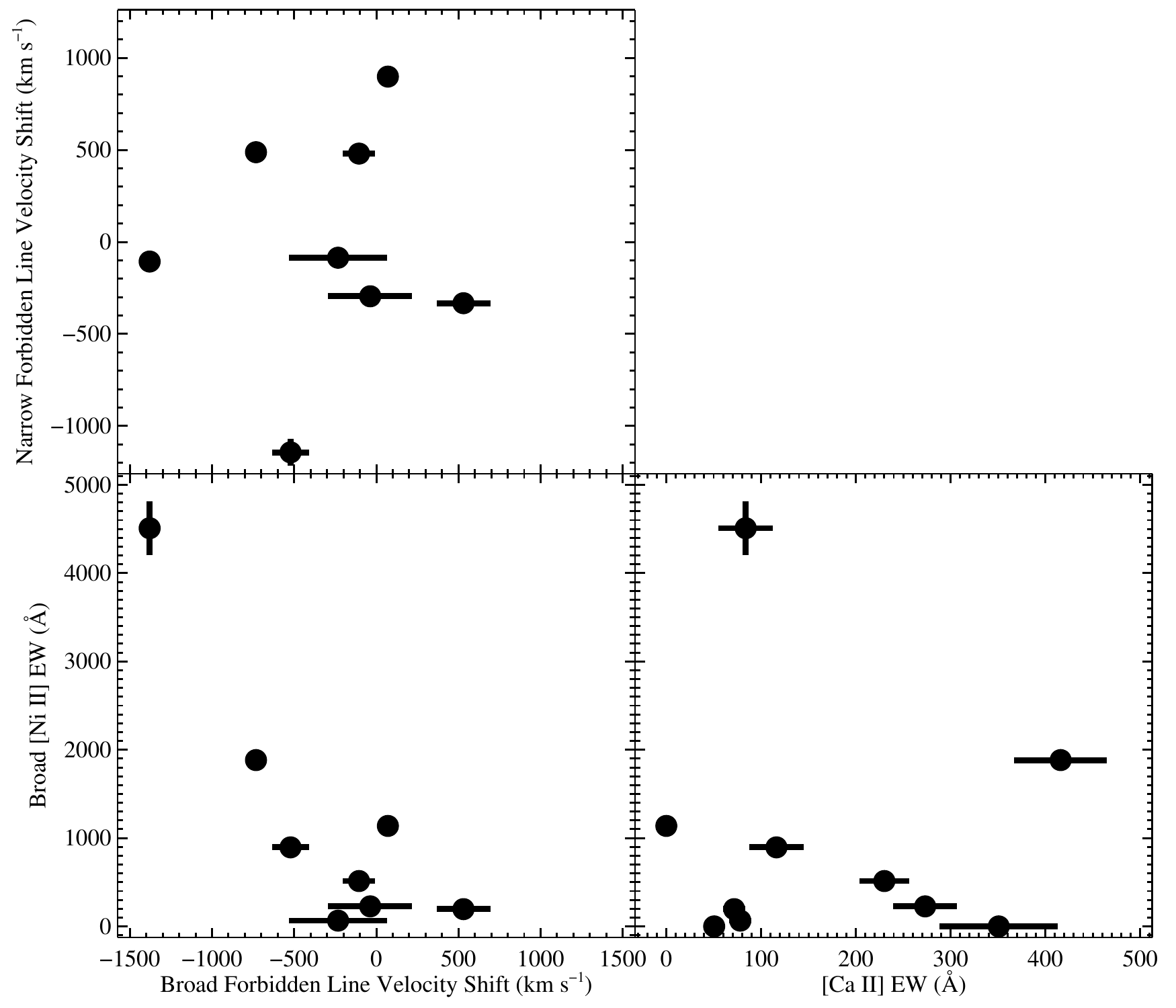}
  \caption{{\it Top-left}: Velocity shifts for the broad and narrow
    components of the forbidden lines as fitted in
    Section~\ref{ss:for}.  The correlation coefficient is 0.20.  {\it
      Bottom-left}: Broad [\ion{Ni}{II}] $\lambda 7378$ EW as a
    function of its velocity shift as fitted in Section~\ref{ss:for}.
    The correlation coefficient is $-0.83$.  {\it Bottom-right}: Line
    strengths, relative to the continuum, for [\ion{Ca}{II}] $\lambda
    \lambda 7291$, 7324 and broad [\ion{Ni}{II}] $\lambda 7378$ as
    fitted in Section~\ref{ss:for}.  The correlation coefficient is
    $-0.06$.}\label{f:comparison}
\end{center}
\end{figure*}

More impressive is the relation between the EW of the broad
[\ion{Ni}{II}] $\lambda 7378$ emission and its velocity shift.  These
parameters are highly correlated: stronger lines correspond to more
blueshifted lines.  Figure~\ref{f:comparison} displays these two
parameters, which have a correlation coefficient of $-0.86$.  A
similar correlation is found with the broad [\ion{Fe}{II}] $\lambda
7155$ emission, where the EW of that feature and its velocity shift
have a correlation coefficient of $-0.85$.

Six SNe~Iax have blueshifted broad forbidden lines, while only two
have redshifted lines (and two with no discernible broad component).
Moreover, the redshifted objects are consistent with being at zero
velocity (shifts of $70 \pm 22$ and $530 \pm 160$~\kms, respectively,
with the uncertainties not including typical galactic rotation of
200--300~\kms).  However, some SNe~Iax appear to have truly
blueshifted features (SNe~2008A and 2012Z).  While this may be caused
by chance (such a distribution has a \about 13\% chance of occurring),
it is also possible that SNe~Iax tend to have their broad nebular
emission blueshifted or do not have broad emission {\it at all}.

This latter explanation is consistent with the correlation between
width/strength and shift for these features.  In this scenario, a
weaker ``broad'' component will be narrower and less blueshifted.  The
extreme of this situation would be a ``broad'' component which is
either too weak to be detected or too narrow to be distinguished from
a separate ``narrow'' component.

We also examined the similar measurements for the narrow components.
Here the correlation between line shift and width was not strong ($r =
-0.17$).  There may be some correlation for the narrow lines, but
SNe~2008ge and 2011ay, which have the most blueshifted and redshifted
lines (respectively), and thus highly influence any relation, pull the
result in opposite directions.  As both have weak narrow lines, either
could be a systemic outlier, but it is currently impossible to
determine if either is.  Alternatively, the underlying physical
relation may be between the magnitude of the shift (i.e., the absolute
value) and the width of the line, for which there is a strong
correlation ($r = 0.88$).  With additional data, this relation should
be re-examined.

Unsurprisingly, the strength of each individual broad/narrow component
is (in general) highly correlated with each other.  The strength of
the two broad features have a correlation coefficient of 0.93, while
the narrow [\ion{Fe}{II}] and [\ion{Ni}{II}] ([\ion{Ca}{II}]) features
have a correlation coefficient of 0.78 (0.46).  The [\ion{Ca}{II}] and
narrow [\ion{Ni}{II}] have a correlation coefficient of 0.61.

Next, we compare the properties of the broad and narrow components of
the forbidden lines.  As seen above, and especially as determined from
the first eigenspectrum, the height of the (narrow) [\ion{Ca}{II}]
$\lambda \lambda 7291$, 7324 emission is anticorrelated with the
height of the broad [\ion{Ni}{II}] $\lambda 7378$ emission.  While the
heights of these features are moderately anticorrelated ($r = -0.52$),
the EWs are uncorrelated ($r = -0.06$; Figure~\ref{f:comparison}).

There is also no correlation between the velocity shifts of the narrow
and broad components.  Figure~\ref{f:comparison} compares these two
values, showing no trend.

The broad and narrow components are generally uncorrelated.  While
there is some trend that SNe~Iax with ``stronger'' [\ion{Ca}{II}]
emission have ``weaker'' broad [\ion{Ni}{II}] emission, this is not
seen in the EW measurements for these features.  This may be partially
caused by the anticorrelation between the continuum flux and
[\ion{Ca}{II}] emission (Section~\ref{ss:pca}).  Again, the lack of a
strong connection between the broad and weak components indicates that
they come from physically distinct components and perhaps physically
distinct mechanisms.

\begin{figure*}
\begin{center}
  \includegraphics[angle=0,width=3.3in]{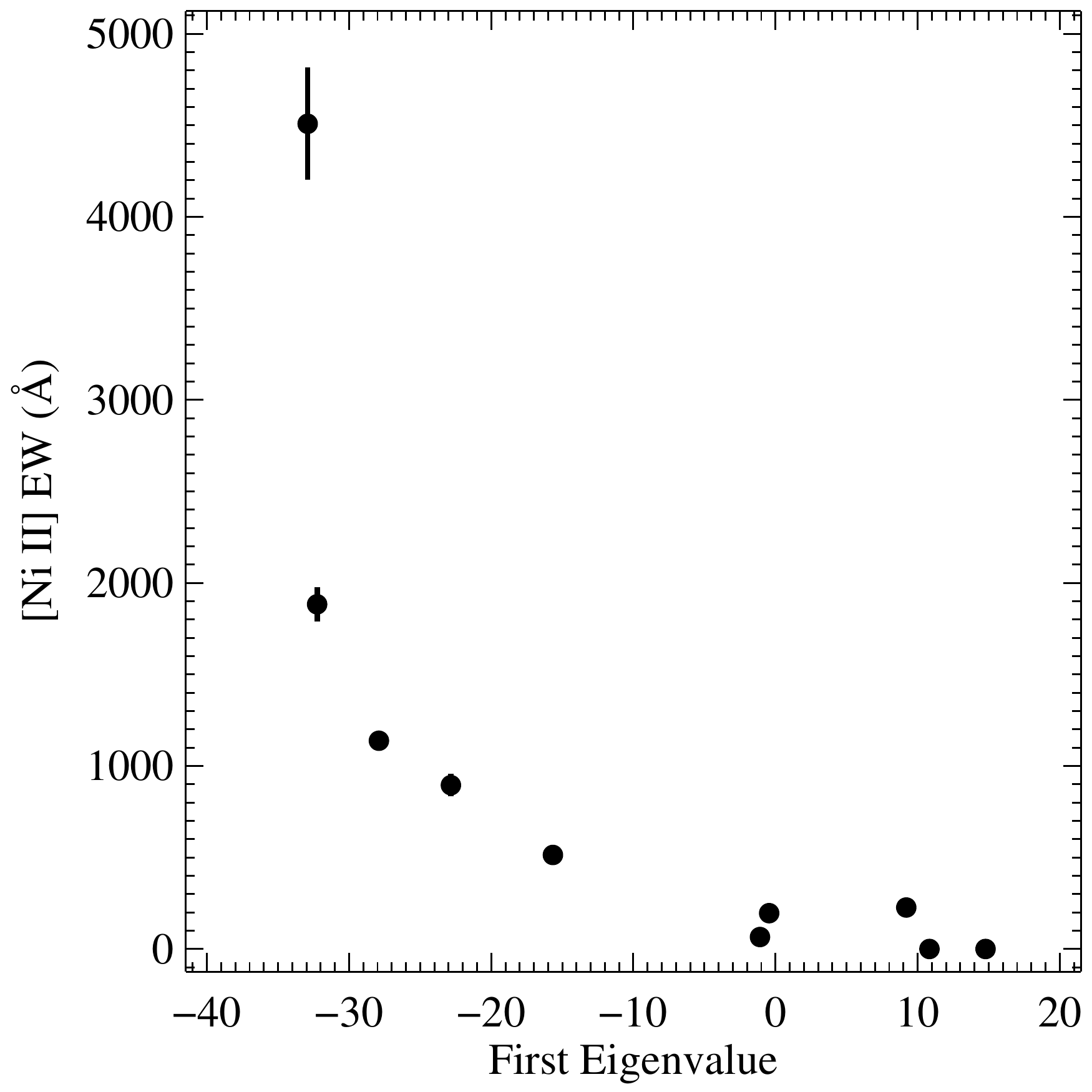}
  \includegraphics[angle=0,width=3.3in]{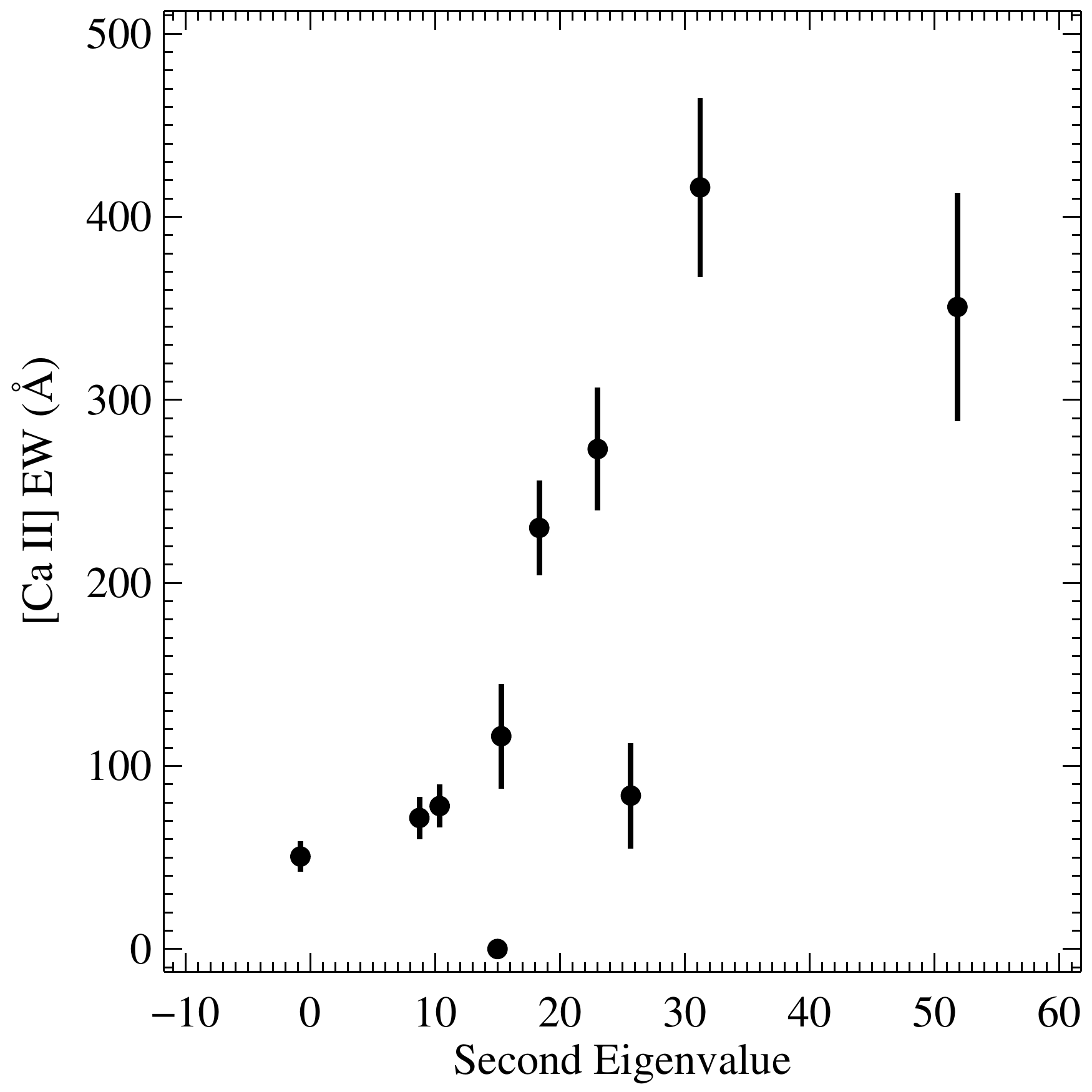}
  \caption{Broad [\ion{Ni}{II}] $\lambda 7378$ (left) and narrow
    [\ion{Ca}{II}] $\lambda \lambda 7291$, 7324 (right) EWs as a
    function of the first and second eigenvalues,
    respectively.}\label{f:eigen_ew}
\end{center}
\end{figure*}

Our physical interpretation of the eigenspectra in
Section~\ref{ss:pca} is confirmed by comparing the eigenvalues of each
spectrum to spectral parameters.  For instance, the first and second
eigenvalues predict the strength of the broad [\ion{Ni}{II}] and
narrow [\ion{Ca}{II}] emission, respectively
(Figure~\ref{f:eigen_ew}).  As such, the eigenvalues can be used as a
proxy for these values when it is difficult to measure them directly.

Finally, we compare the late-time spectral properties to those at
maximum brightness.  \citetalias{Foley13:iax} presented maximum-light
\ion{Si}{II} $\lambda 6355$ velocities for SNe~2002cx, 2005hk, 2008A,
2011ay, and 2012Z.  In addition, we use the spectrum of SN~2010ae
presented by \citet{Foley13:ca}, which was obtained at $+0.8$~d, to
measure a maximum-light velocity of $-4390 \pm 60$~\kms.  Therefore,
six members of our sample have maximum-brightness measurements of
their ejecta velocity.

While there is some correlation between maximum-brightness ejecta
velocities and properties of the broad forbidden lines, the
statistical significance (partially because SN~2010ae does not have a
measured broad component) for any potential correlation is low.

\begin{figure}
\begin{center}
  \includegraphics[angle=0,width=3.2in]{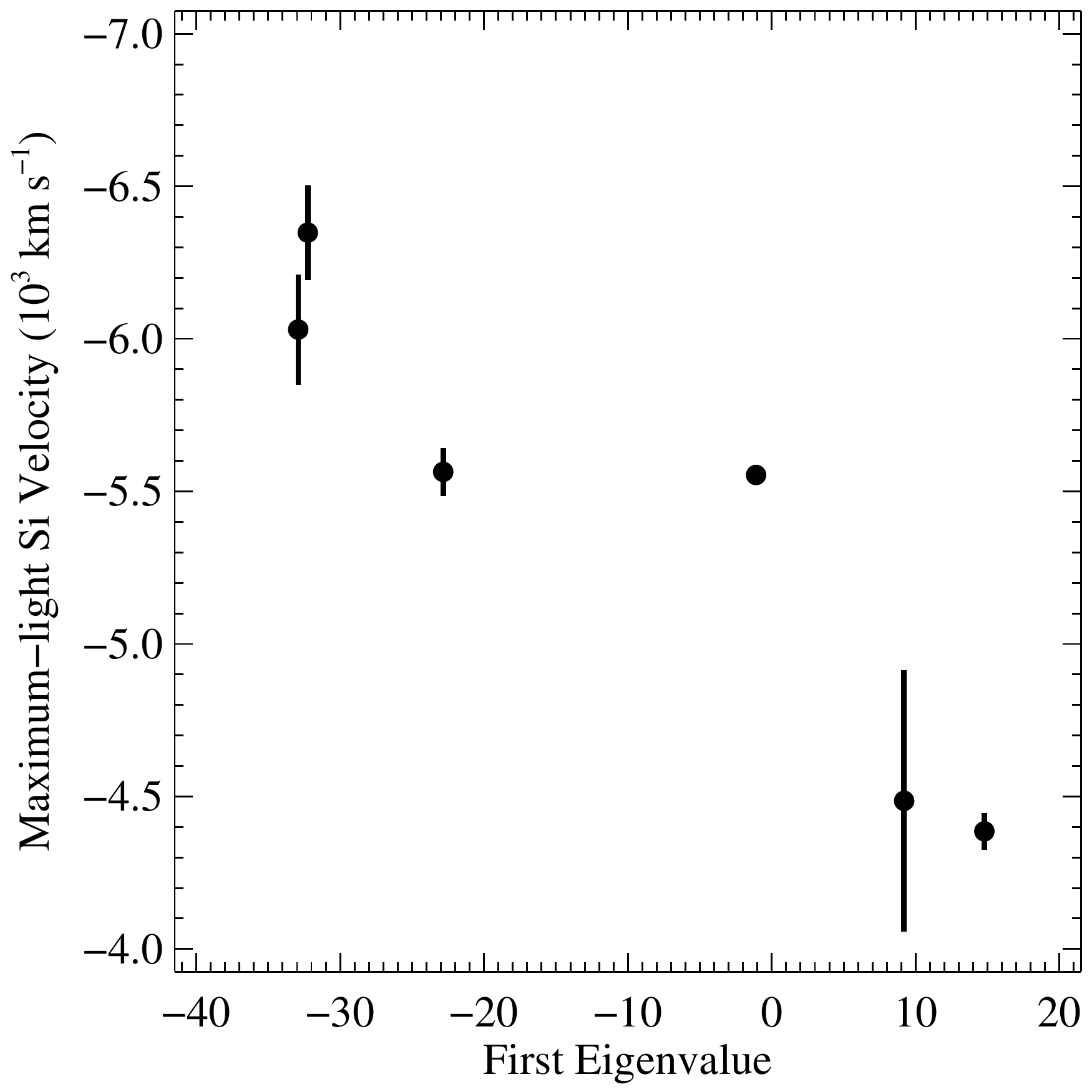}
  \caption{Comparison of maximum-light \ion{Si}{II} $\lambda 6355$
    velocity and first eigenspectrum of the late-time
    spectrum.}\label{f:vph_eigen}
\end{center}
\end{figure}

However, the photospheric velocity is highly correlated ($r = 0.93$)
with the first eigenspectrum (Figure~\ref{f:vph_eigen}).  That is, the
largest portion of the variance in the late-time spectra of SNe~Iax is
physically connected to the ejecta velocity at maximum brightness.
Specifically, SNe with low measured ejecta velocities at maximum
brightness tend to have late-time spectra with weak/absent broad
forbidden lines and strong narrow [\ion{Ca}{II}] lines.

\subsection{Spectral--Photometric Comparisons}

Using the subsample of SNe~Iax that have both late-time spectra and
photometric properties such as $M_{V}$ and $\Delta m_{15}$, we
examined potential correlations between the photometric properties and
those derived from the late-time spectra.

There are no strong correlations between $\Delta m_{15} (V)$, the
decline-rate parameter observed for most SNe~Iax in our sample, and
the appearance of the late-time spectra.

\begin{figure*}
\begin{center}
  \includegraphics[angle=0,width=3.2in]{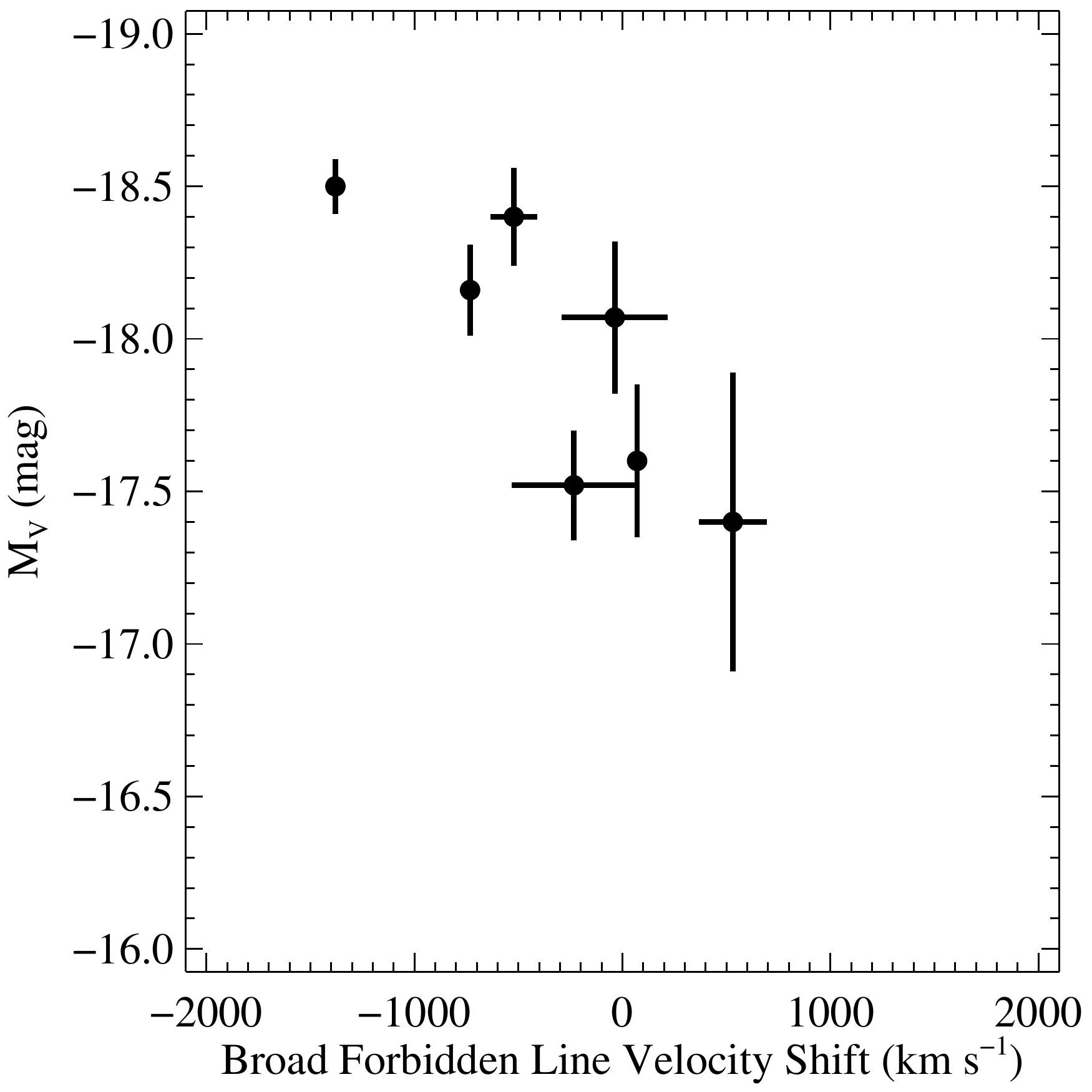}
  \includegraphics[angle=0,width=3.2in]{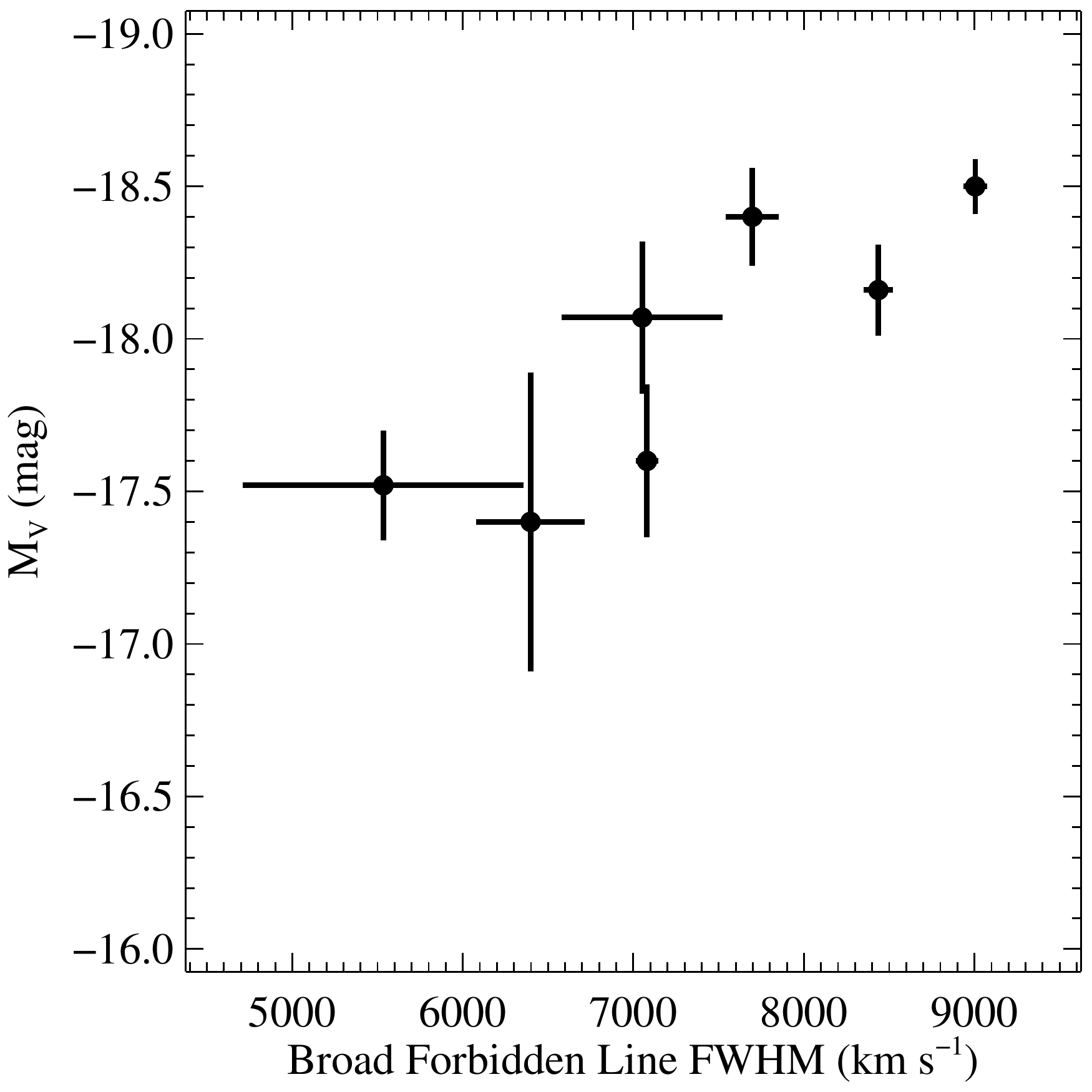}
  \caption{Peak absolute $V$ magnitude as a function of velocity
    shifts (left) and velocity width (right) for the broad
    forbidden-line components as fitted in Section~\ref{ss:for}.  The
    correlation coefficients are 0.83 and $-0.85$,
    respectively.}\label{f:broadmv}
\end{center}
\end{figure*}

There is a strong correlation between the the peak absolute magnitude
in the $V$ band ($M_{V, {\rm peak}}$) and both the velocity shift ($r
= 0.83$; Figure~\ref{f:broadmv}) and width ($r = -0.85$) of the broad
forbidden lines.  While the first eigenvalue is also correlated with
$M_{V, {\rm peak}}$ ($r = 0.68$), it is not as highly correlated as
the direct measurements of the broad forbidden lines.  This is not
caused by the addition of SN~2010ae (which was excluded from the
previous comparisons because of a lack of an identifiable broad
component); excluding SN~2010ae decreases the correlation between the
first eigenvalue and $M_{V, {\rm peak}}$ to $r = 0.51$.

The SNe~Iax having higher peak luminosity tend to have broader, more
blueshifted ``broad'' forbidden lines.  Two possible explanations for
this correlation are either (1) SNe~Iax that produce more $^{56}$Ni
(and are thus more luminous at peak) also produce higher-velocity
ejecta at all layers of the ejecta, or (2) SNe~Iax explosions are
asymmetric and lines of sight pointed along the ``high-velocity axis''
are also more luminous.


\section{Discussion}\label{s:disc}

\subsection{A Nearly Chandrasekhar-Mass Explosion}

Our identification of the [\ion{Ni}{II}] $\lambda 7378$ line in the
late-time spectra of SNe~Iax is a strong indication of the presence of
stable nickel isotopes (e.g., $^{58}$Ni) in the ejecta, as by $>$200~d
after the explosion, radioactive $^{56}$Ni will have decayed to a
fraction \about $10^{-10}$ of its original abundance.

Explosion models which produce a deflagration flame that fails to
unbind the progenitor WD\footnote{These models are sometimes referred
  to as ``failed deflagration'' models, despite the fact that the
  deflagration is successful.} can reproduce the rough spectral and
photometric properties of SNe~Iax \citep{Jordan12, Kromer13,
  Kromer15}.  These models employ a (nearly) $M_{\rm Ch}$ WD
progenitor.  The burning is ignited in the core of this star, which
has a sufficiently high density that electron capture produces
neutronised isotopes such as $^{54}$Fe and $^{58}$Ni
\citep[e.g.,][]{Thielemann86}.

Contrastingly, detonations occurring in (or on) sub-Chandrasekhar WDs
have densities too low for electron capture to occur.  Although a
small amount of $^{58}$Ni may be synthesised using the excess neutrons
from high-neutron species in the progenitor \citep[primarily
$^{22}$Ne;][]{Timmes03}, it is expected that the amount of $^{58}$Ni
is significantly less than that of all Fe species in such explosions.

While detailed nebular spectrum calculations are needed to infer the
nickel and iron abundances, the detection of strong [\ion{Ni}{II}],
especially being much stronger than [\ion{Fe}{II}], can be interpreted
as strong evidence for a (nearly) $M_{\rm Ch}$ progenitor star.
Accordingly, this is further support for the deflagration models which
fail to unbind their progenitor star.

If the progenitor stars at the time of explosion have masses of $1
{\rm M}_{\sun} \lesssim M \lesssim M_{\rm Ch}$, and in particular if
they are close to the Chandrasekhar mass, then the implied ejecta
masses of \about 0.5~M$_{\sun}$ (or less) for most SNe~Iax \citep[see,
e.g.,][]{Foley10:08ha, Foley13:iax, Narayan11, McCully14:iax} require
a bound remnant for nearly all SNe~Iax.

\subsection{Size of the Late-time Photosphere}\label{ss:phot_size}

At late times ($t \gtrsim 200$~d), some SNe~Iax still have permitted
lines with P-Cygni features, indicative of persistent photospheres
\citepalias{Jha06:02cx}.  Although we cannot detect individual P-Cygni
features for some SNe~Iax, the continua of all SNe~Iax are similar.
Additionally, the bluer [\ion{Fe}{II}] and [\ion{Fe}{III}] features
seen in late-time spectra of SNe~Ia are absent in all late-time SN~Iax
spectra.  It is therefore likely that all SNe~Iax have a photosphere
at late times.

We can measure the size of the photosphere in two independent ways.
The first is to measure the luminosity and temperature of a SN~Iax at
late times and determine the radius assuming that the emission comes
from a blackbody.  The second is to assume that the velocity of the
late-time P-Cygni lines is characteristic of the velocity of the
late-time photosphere.  Assuming no acceleration, one can measure the
radius knowing the time between explosion and the time of the
spectrum.

For these measurements, we will emphasise the well-observed SN~2005hk
\citep{Phillips07, Sahu08, McCully14:iax}, focusing on the $+402$~d
spectrum (417 days after explosion) presented by
\citet{Silverman12:bsnip}.  This is the last spectrum of SN~2005hk
which still has a clear continuum.  At this time, the bolometric
luminosity of SN~2005hk was \about $10^{39.9}$~erg~s$^{-1}$
\citep{McCully14:iax}.  The bolometric luminosity was determined from
broad-band photometry, and may overestimate the continuum flux by as
much as 40\% because of line emission (as determined from the optical
spectrum).  Fitting a blackbody spectrum to the continuum of SN~2005hk
at this epoch, we find a best-fitting temperature of \about 4500 --
5500~K, consistent with the presence of both \ion{Fe}{I} and
\ion{Fe}{II} in the spectrum \citep{Hatano99:ion, McCully14:iax}.
Using the above values, we determine that the blackbody radius 417~d
after explosion is
\begin{equation}
  R_{\rm BB} = 1.3 \times 10^{14} \left ( \frac{L}{10^{39.9}
      {\rm ~erg}} \right )^{1/2} \left ( \frac{T}{5000 {\rm ~K}} \right
  )^{-2} {\rm ~cm}.
\end{equation}
The uncertainty in the radius measurement is \about 20\% given the
uncertainties in the luminosity and temperature, and the range in
radius is the result of different assumptions about the continuum
luminosity and blackbody temperature.

The $+402$~d spectrum of SN~2005hk has a photospheric velocity (as
determined from the P-Cygni absorption) of $-410$~\kms, similar to
what was found for SN~2002cx at late times \citep{Jha06:02cx}.  If the
emitting material for SN~2005hk has been in homologous expansion since
explosion, this would place the material at
\begin{equation}
  R_{\rm kin} = 1.5 \times 10^{15} \left ( \frac{v}{410 {\rm ~km~s}^{-1}} \right ) \left ( \frac{t}{417 {\rm ~d}} \right ) {\rm ~cm},
\end{equation}
a radius more than an order of magnitude {\it higher} than the
blackbody radius.  The uncertainty in this measurement is around 2\%
and primarily set by the uncertainty in the measured velocity.

These two discrepant estimates of the photospheric radius cannot be
reconciled by any simple adjustment of the measured quantities.
First, the late-time photospheric velocity would need to be
overestimated such that the true velocity is $v_{\rm ph} \lesssim
40$~\kms, which is much too low to be consistent with the spectrum.
Alternatively, the bolometric luminosity could be \about
$10^{42}$~erg~s$^{-1}$, which is $>$2 orders of magnitude higher than
measured.  Finally, a true temperature of 2500~K would sufficiently
reduce the measured blackbody radius; however, at this temperature, we
would not expect to see any \ion{Fe}{II} emission.  Furthermore, such
a low temperature would require that the continuum seen in the spectra
be caused by a nonblackbody component, making our luminosity
assumption incorrect --- a lower blackbody luminosity with this lower
temperature is similarly inconsistent with the kinematic radius.  In
summary, it does not appear that a poor assumption or measurement
error has resulted in this discrepancy.

Another possible explanation for the different radius estimates is
asymmetry, but this ultimately seems unlikely.  The kinematic radius
describes the radius along the line of sight, while the blackbody
radius describes the (average) radius in the plane of the sky.  A
highly asymmetric explosion could thus have very different
measurements for the radius.  However, this would require an extreme
aspect ratio and very particular viewing angle.  This becomes even
less likely considering the number of other SNe~Iax similar to
SN~2005hk and the lack of strong polarization at early times for
SN~2005hk \citep{Chornock06, Maund10:05hk}.

Alternatively, the photospheric material may not have been expanding
since the time of explosion.  If the material generating the
photosphere were launched at 410~\kms\ 30--40~d before the time of the
spectrum (with a shorter period if there is deceleration), this would
place the kinematic radius at the same radius as the blackbody radius.
However, in this scenario, the material would not be caused by the
initial explosion and must be a wind from either the companion star or
a surviving remnant.

In the case of a wind, the velocity of the photosphere should be
essentially the wind velocity; the SN explosion would have created a
cavity and so there would be minimal deceleration.  (However, if there
is a bound remnant, there may be infalling material even at late times
which could decelerate the wind.)  Assuming that the photosphere seen
at late times for SN~2005hk is the result of a 410~\kms\ wind,
equivalent to the escape speed of a compact remnant, this remnant
would have a radius at late times of
$R = 8 \times 10^{10} \left ( M / 0.5 {\rm M}_{\sun} \right ) {\rm
  ~cm} = 1.2 \left ( M / 0.5 {\rm M}_{\sun} \right ) {\rm R}_{\sun}$.
For this scenario, the mass-loss rate would need to be high enough
such that the wind remains optically thick out to a radius of \about
$10^{14}$~cm, or \about $10^{3} {\rm R}_{\sun}$.

The velocity of the permitted lines seen in the SN~2005hk spectra
decreased by \about 100~\kms\ between 245~d and 417~d after explosion,
corresponding to \about 0.5~\kms~d$^{-1}$.  This incredibly slow
change in the photospheric velocity is also difficult to explain with
a single homologous expansion, thus favouring a wind interpretation.

\subsection{A Two-Component Ejecta Model}

The above radius estimates argue for a late-time wind from the
progenitor system.  Explosion models which match the early-time
spectra and light curves of SNe~Iax also argue for a bound remnant
\citep{Jordan12, Kromer13, Kromer15}.  These models predict that a
significant amount (\about 0.02~M$_{\sun}$) of $^{56}$Ni will remain
in the remnant, providing an energy source that may be able to drive
dynamical outflows.  It is reasonable to think that after a SN~Iax
explosion a bound remnant would expand to $R \approx {\rm R}_{\sun}$
and drive a super-Eddington wind (Bildsten et~al., in prep.).  This
model also solves the problem of having low-velocity material from an
explosion that completely disrupted the star \citep{McCully14:iax}.

However, for our observations, we do not require that the wind comes
from a bound remnant.  Rather, the wind could be caused by the remnant
or the companion star.  While it is not yet possible to distinguish
between these two possibilities, the significant $^{56}$Ni in the
bound remnant would be a natural energy source to drive such a wind.

In the wind scenario, the photosphere is significantly smaller than
the forbidden-line emitting region, which is likely dominated by SN
ejecta.  For the case of SN~2005hk, assuming that nearly all of the
luminosity comes from the photosphere at late times, the photosphere
was at $R \approx 10^{14}$~cm at +402~d, while the forbidden-line
emitting region should be at
\begin{equation}
  R_{\rm ej} \approx 1.3 \times 10^{16} \left ( \frac{v}{3500 {\rm ~km~s}^{-1}} \right ) \left ( \frac{t}{417 {\rm ~d}} \right ) {\rm ~cm},
\end{equation}
where the velocity is given by the half width at half-maximum
intensity (HWHM) of the broad forbidden lines.  At this radius, the
forbidden-line emitting region has a radius about 100 times larger
than the photosphere, and thus the projected area of the
forbidden-line region is \about $10^{4}$ times larger than that of the
photosphere.  Even in the case where the material creating the
photosphere (moving at \about 500~\kms) was ejected at the same time
as the higher-velocity material, the forbidden-line emitting region
will be \about 75 times larger than the photosphere in projected area.
In either case, the photosphere cannot block a significant amount of
the forbidden-line region, and the forbidden lines must track
essentially all of the low-density material, with very little blocked
by the photosphere.

Beyond the optically thick region of the wind, lower-density wind
material may generate the narrow forbidden lines.  For a
constant-velocity wind, unlike a homologous flow, the material at
larger radius would share the same velocity as the photospheric
material.  The HWHM of the narrow forbidden components is typically
\about 500~\kms, although SNe with detected narrow P-Cygni features
(e.g., SNe~2002cx and 2005hk) tend to have slightly narrower forbidden
lines (${\rm HWHM} \approx 400$~\kms).  These velocities are
consistent with a \about 500~\kms\ wind.  The gradual, correlated
change in the velocities for the permitted and narrow forbidden lines
between +230~d and +403~d for SN~2005hk implies that the two are
physically connected, again supporting the idea that the narrow
forbidden lines are linked to a wind that is producing the photosphere
and low-velocity P-Cygni features.  Even the large widths seen in some
spectra could be consistent with a wind if the remnant has a smaller
radius or larger mass (and thus larger escape velocity).

The distribution of narrow-line velocity shifts peaking at zero
velocity and the lack of a correlation between the narrow and broad
components are all consistent with the narrow-line regions being
formed by a wind.  Similarly, the lack of broad Ca lines may be
indicative of different compositions for the broad and narrow
components.  The compositional difference is a natural outcome of the
narrow and broad lines originating from a remnant wind and the SN
ejecta, respectively.  For instance, the model of \citet{Kromer13} has
the remnant composed of 88\% C/O and 3\% iron-group elements (IGEs),
while the ejecta are only 28\% C/O and 59\% IGEs.  While this model
may not perfectly match the relative abundances of the ejecta and wind
(especially since the wind will be composed of surface material), the
model may predict strong, narrow [\ion{O}{I}] $\lambda\lambda 6300$,
6364 emission.  However, strong [\ion{O}{I}] has not yet been observed
in a SN~Iax (\citetalias{Jha06:02cx}; \citealt{McCully14:iax}), and
since both [\ion{Ca}{II}] and \ion{Ca}{II} are seen in the spectra,
the density is sufficiently high to suppress [\ion{O}{I}] emission
\citep{McCully14:iax}.  None the less, this potential compositional
difference may explain the lack of broad [\ion{Ca}{II}] lines.  The
\citet{Kromer13} model has 9.7 and 220 times as much (by mass) IGEs as
Ca in the remnant and SN ejecta, respectively.  This difference is
generally consistent with having relatively strong/absent narrow/broad
Ca lines.

If the narrow forbidden lines are indicative of a wind, then nearly
every SN~Iax must have such a wind.  With the possible exception of
SN~2011ay, which does not have an obviously distinct low-velocity
component, all SNe~Iax in our sample have some narrow lines.

There is significant diversity in the strength, width, and velocity
shifts of the broad component of the forbidden lines.  These
properties are strongly correlated with maximum-light properties such
as peak luminosity, but uncorrelated (or weakly correlated) with the
narrow forbidden lines, indicating two distinct kinematic components.

In the wind model, the narrow forbidden lines, the low-velocity
P-Cygni lines, and the photosphere would be generated by a wind, while
the broad forbidden lines would be related to the SN ejecta, and thus
to early-time SN properties.

\citet{Kromer13} modeled SN~2005hk with such a two-component model and
were able to separate the SN luminosity from the luminosity of the
bound remnant.  In this model, the ejecta of SN~2005hk should have a
luminosity of roughly $10^{39.3}$~erg~s$^{-1}$ at the time of the
late-time spectrum analyzed above (time since explosion of \about
417~d).  This is roughly 20\% of the total bolometric luminosity.
Examining the SN~2005hk spectra, we find that the broad forbidden
lines discussed in this work represent \about 5\% of the optical
luminosity.  For the SN~2012Z spectrum analyzed above, which had the
strongest broad absorption lines of our sample, the broad forbidden
lines represent \about 45\% of the total optical luminosity at late
times.  Therefore, the broad forbidden lines, which one would
naturally associate with the SN emission, have roughly the predicted
fractional flux of the SN ejecta in the \citet{Kromer13} model.

In this two-component ejecta model, where one component is from the SN
explosion and the other is from the wind, there is not necessarily any
direct physical connection between the two components.  That is, the
amount of $^{56}$Ni left in the remnant and the mass of the remnant,
which must be the primary variables for the strength of the wind and
its velocity, may be essentially unrelated to the amount of $^{56}$Ni
ejected and the ejecta mass, which are likely the primary variables
for SN properties.  While a ``stronger'' explosion is expected to
produce more $^{56}$Ni, it is unclear what percentage of the
progenitor star is burned, what the initial mass of the progenitor
star is, or what fraction of $^{56}$Ni is ejected relative to that
left in the remnant.

\citet{Kromer15} produced a model explosion that roughly matches the
observed properties of SN~2008ha \citep{Foley09:08ha, Foley10:08ha,
  Valenti09}, the least-luminous SN~Iax yet discovered.  While the
\citet{Kromer13} model used a C/O WD progenitor, the \citet{Kromer15}
model used a hybrid C/O-Ne WD.  The composition of the WD accounts for
the different outcomes; however, the exact ignition conditions could
affect the mass burned in the deflagration.  While SN~2008ha was not
detected at late times \citep{Foley10:08ha}, this model may provide
insight into the diversity of ejecta for SNe~Iax having different
luminosities. In this model, 46\% of the ejecta is composed of IGEs,
lower than for the \citet{Kromer13} SN~Iax model, indicating a
possible compositional difference for low- and high-luminosity
SNe~Iax.

Using these two models as examples, there is a trend between peak
luminosity and ejecta mass.  The \citet{Kromer13} and \citet{Kromer15}
models have peak $V$-band absolute magnitudes of $-18.2$ and $-14.8$,
respectively (and a factor of 23 in luminosity), while they also
produce 0.372 and 0.014~M$_{\sun}$ of ejecta, respectively (a factor
of 27).  Based on these models, one may expect that peak luminosity is
related to the relative strength of the broad forbidden-line emission,
which would have some dependence on ejecta mass.

\begin{figure}
\begin{center}
  \includegraphics[angle=0,width=3.2in]{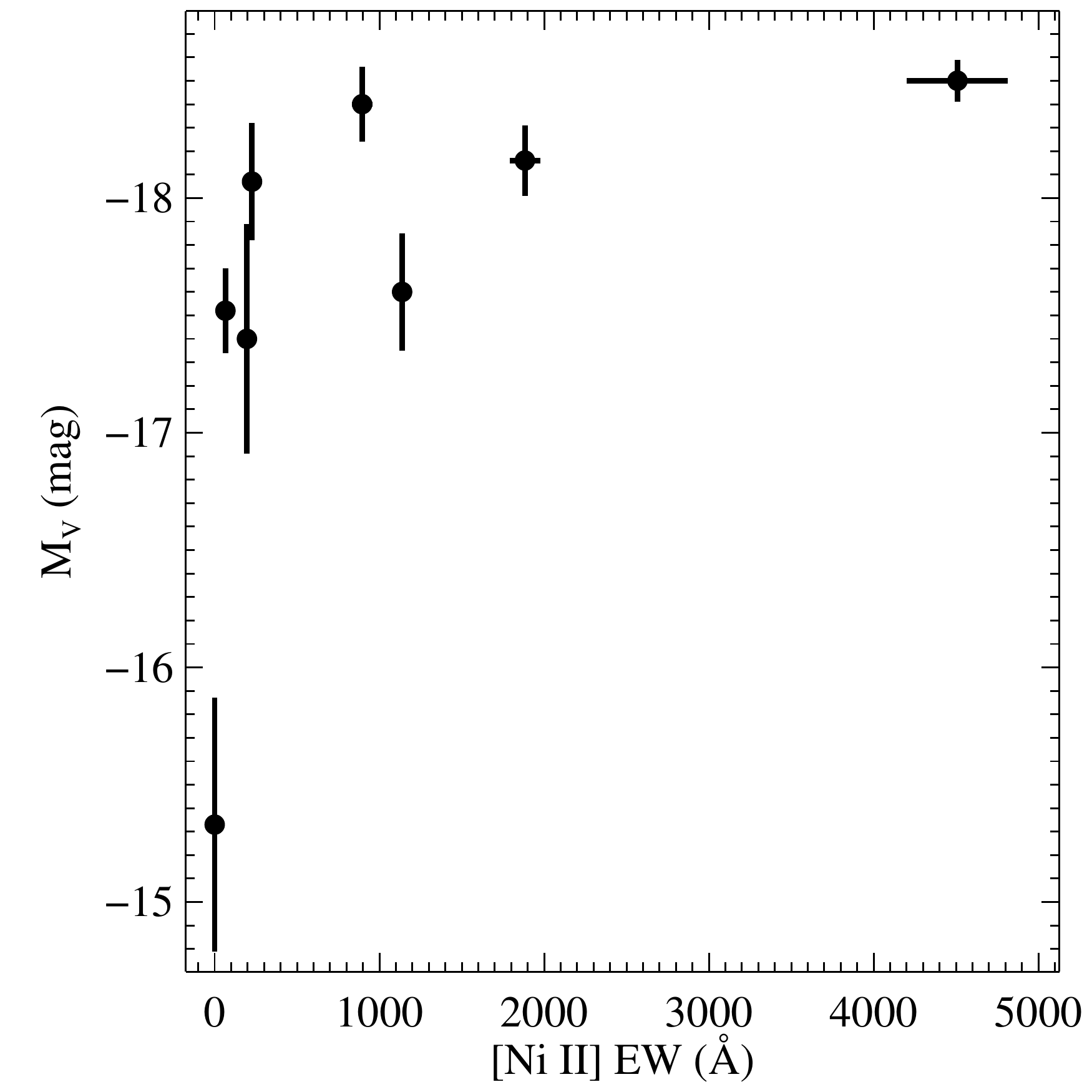}
  \caption{Peak absolute $V$ magnitude as a function of the
    broad-component [\ion{Ni}{II}] $\lambda 7378$ EW.  The correlation
    coefficient is $-0.53$.}\label{f:mv_niew}
\end{center}
\end{figure}

To examine this possibility, Figure~\ref{f:mv_niew} displays the
correlation between $M_{V, {\rm ~peak}}$ and broad-component
[\ion{Ni}{II}] $\lambda 7378$ EW.  There is a modest correlation
(correlation coefficient of $-0.53$), where more-luminous SNe~Iax tend
to have stronger broad emission lines.  However, there is no
correlation between the narrow-component [\ion{Ni}{II}] $\lambda 7378$
EW and peak luminosity ($r = -0.15$), consistent with a wind that is
relatively independent of the SN ejecta.

\subsection{An Asymmetric Explosion?}

The two-component model described above is insufficient for
reproducing the correlation between peak luminosity and velocity
shifts of the broad forbidden-line emitting region (i.e.,
Figure~\ref{f:broadmv}).  That model also does not explain the
predominantly blueshifted broad forbidden lines regardless of any
correlation with peak luminosity.

The latter is difficult to explain with a simple two-component model.
For a predominantly blueshifted population, one would expect that the
redshifted emission be blocked by a photosphere.  However, above we
found that the broad forbidden-line emitting region is \about 10,000
times larger than the photosphere (in projection) at \about 417~d
after explosion.  Even for the typical phases of the spectra analyzed
in this work, the broad forbidden-line emitting region is likely
\about 3000 times larger than the photosphere.  In this scenario, the
photosphere would be unable to block most of the redshifted emission.

If we require a photosphere to block the redshifted emission of the
broad forbidden-line emitting region, it should have a projected area
$\lesssim$10$\times$ that of the area of the photosphere.  One
scenario is that the SNe with blueshifted forbidden lines also have
much larger photospheres ($\gtrsim 2 \times 10^{15}$~cm) than that of
SN~2005hk.  This photosphere, if at \about 5000~K, would have a large
luminosity of \about $10^{42}$~erg~s$^{-1}$.  This is much larger than
any SN~Iax measured at late times \citep[e.g.,][]{McCully14:iax} and
not significantly less than the peak SN luminosity.

Alternatively, the broad emission, which should come from the SN
ejecta, may not be distributed symmetrically.  In fact, reasonable
explosion models expect few plumes, which could result in highly
asymmetric ejecta \citep{Jordan12, Kromer13}.  If, for instance, there
is higher-velocity material ejected primarily along a single axis,
then when we see a large velocity, corresponding to a line of sight
along this axis, a smaller photosphere could block the redshifted
emission.  Correspondingly, looking perpendicular to this axis would
result in no broad lines.  This is an intriguing model to describe the
diversity of SN~Iax late-time spectra, including the transition
objects, which would be viewed at an angle intermediate to the two
examples mentioned above.

A downside of this model is that one would predict extremely large
asymmetries in the SN ejecta and thus large polarization, which is
inconsistent with measurements made for a single SN~Iax, SN~2005hk
\citep{Chornock06, Maund10:05hk}.  However, the photosphere at the
times of polarization measurements might not have been dominated by
this asymmetric material or SN~2005hk may be an atypical object.  In
fact, it may be the case that the blueshifted objects come from a
subpopulation that have strong asymmetries, while other SN~Iax
explosions are more spherical.  Additional spectropolarimetric
observations of SNe~Iax, and comparisons to other spectral and
photometric properties, will test this possibility.

\subsection{The ``Late-time'' SN~2008ha Spectra}

SN~2008ha is an exceptional SN~Iax, being the least luminous member of
the class \citep[$M_{V, {\rm peak}} = -14.2$~mag;][]{Foley09:08ha,
  Valenti09}, fading very quickly \citep[$\Delta m_{15}(B) =
2.2$~mag;][]{Foley09:08ha}, and having very low-velocity ejecta at
peak brightness \citep[$v_{\rm ph} = -3700$~\kms;][]{Foley10:08ha}.
Combined, the data suggest that the SN ejected $M \lesssim
0.3$~M$_{\sun}$ \citep{Foley10:08ha}, significantly less than that
expected for a WD SN that completely unbinds its star.  Intriguingly,
at $t \approx +4$~yr, there is a very red star detected in {\it HST}
images coincident with the position of SN~2008ha \citep{Foley14:08ha}.
While this may be a chance coincidence, it is also possibly the
surviving remnant of the WD.

For all of these reasons, SN~2008ha appears to be an extremely
interesting object for testing the wind model.  Unfortunately, the
only late-time spectrum of SN~2008ha, at $t = +231$~d, did not reveal
any SN emission \citep{Foley10:08ha}.  Therefore, the latest
spectroscopic data for SN~2008ha extend to only \about 2~months after
peak \citep{Foley09:08ha, Valenti09}.  However, even at these early
times, the spectrum exhibited strong [\ion{Ca}{II}] emission
\citep{Foley09:08ha, Valenti09}.

For the first time here, we note that for phases of $t \gtrsim +37$~d,
the spectra of SN~2008ha look remarkably similar to those of other
SNe~Iax at $t > 200$~d.  Specifically, SN~2008ha has emission from
[\ion{Ca}{II}], [\ion{Fe}{II}], and [\ion{Ni}{II}] starting around a
month after peak brightness.  We detect this emission in the +37~d
spectrum, but it is absent in the +22~d spectrum.

\begin{figure}
\begin{center}
  \includegraphics[angle=0,width=3.2in]{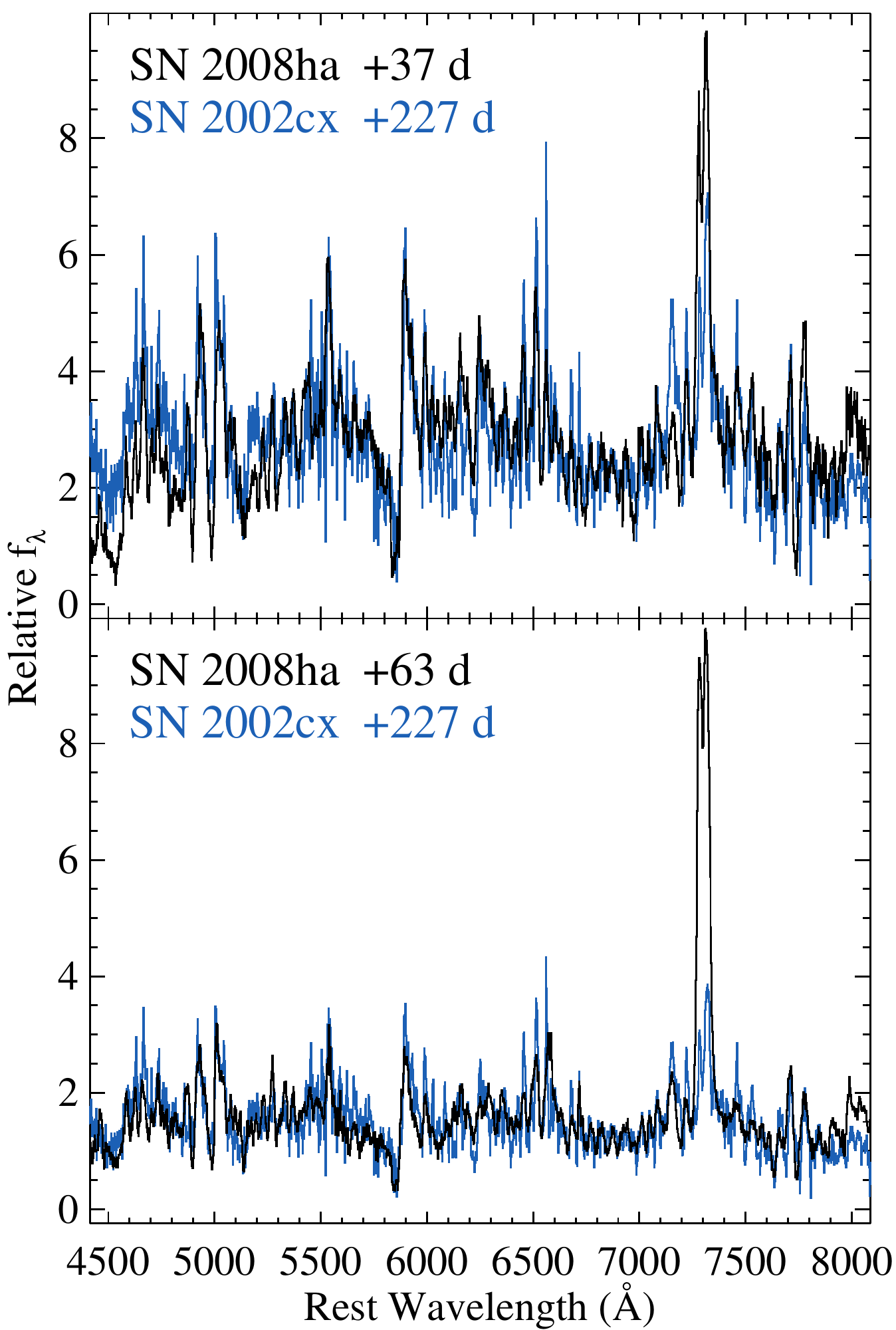}
  \caption{Optical spectra of SN~2008ha (black curves; top panel is at
    $t = +37$~d; bottom panel is at $t = +63$~d) and SN~2002cx (blue
    curves; $t = +227$~d).  The spectra are remarkably similar despite
    their very different phases.}\label{f:08ha}
\end{center}
\end{figure}

We compare the +37~d and +63~d spectra of SN~2008ha to the +227~d
spectrum of SN~2002cx in Figure~\ref{f:08ha}.  The continua and
permitted lines in the spectra of the two SNe are nearly identical.
The primary difference is the strength of the forbidden lines.  At
+37~d, SN~2008ha has relatively stronger [\ion{Ca}{II}] and weaker
[\ion{Fe}{II}] and [\ion{Ni}{II}] than SN~2002cx.  However, at +63~d,
SN~2008ha has [\ion{Ca}{II}] emission that is roughly 10 times as
strong as for SN~2002cx (relative to the continuum emission).  At this
time, the [\ion{Fe}{II}] and [\ion{Ni}{II}] emission is similar in the
two objects (again, relative to the continuum).

From spectra alone, it appears that SN~2008ha has a ``late-time''
appearance starting only \about 1~month after peak.  This is
exceptionally fast evolution.  If our wind model is correct, we would
expect the wind to be launched before \about 50~d after explosion
(since SN~2008ha had a rise time of \about 10~d), placing a strong
lower limit to this condition in at least the lowest-luminosity
SNe~Iax.  Regardless, it appears that SN~2008ha transitioned to have
``late-time'' behaviour at an incredibly early time.

The \citet{Kromer15} model for SN~2008ha resulted in six times as much
$^{56}$Ni in the bound remnant than in the SN ejecta.  As a result,
the instantaneous energy deposition is always larger for the remnant
than in the ejecta.  Depending on the distribution of $^{56}$Ni in the
remnant, one could imagine the wind being the dominant component at
these early times.

We also examined the spectra of SN~2010ae, another low-luminosity,
low-velocity SN~Iax \citep{Stritzinger14:10ae}.  Despite the other
similarities to SN~2008ha, SN~2010ae does not exhibit any forbidden
emission through +57~d.  There are no published spectra of SN~2010ae
between +57~d and +252~d (we examine this late-time spectrum above);
therefore, we cannot assess if SN~2010ae transitioned to having a
``late-time'' spectrum at a relatively early time.  None the less, we
can definitively say that this transition happened later in SN~2010ae
than in SN~2008ha.

Finally, while the $>$1~month SN~2008ha spectra are extremely similar
to those of SNe~Iax at $>$6~months, we have chosen to not include
these spectra in the other analysis presented here so as to examine
only SNe~Iax at late times rather than select objects based solely on
spectral similarities.

\subsection{Lack of Dust in SN~2014dt}

\citet{Fox15} detected SN~2014dt as a relatively strong IR source at
phases of +302 to +329~d after peak brightness.  The IR flux was
interpreted as dust emission either from pre-existing circumstellar
dust or dust newly formed in the SN ejecta.  Fitting the two IR bands,
they infer a dust mass of \about $5 \times 10^{-6}$~M$_{\sun}$ (using
our preferred distance) and a blackbody temperature of 700~K.
\citet{Fox15} measure an increase in IR luminosity during this month
of observations, and while this is only significant at \about
1~$\sigma$, it may be indicative of an increasing luminosity of an
IR-bright component to SN~2014dt.

As noted by \citet{Foley15:14dt} and \citet{Fox15}, there is no
indication of dust reddening for SN~2014dt at early times.
\citet{Fox15} notes that SN~2014dt becomes redder (in $B-V$) at \about
250~d after peak.  This claim is based on a single data point that is
discrepant from other SN~Iax colour curves at \about 3~$\sigma$.
However, there is some indication of SN~2014dt becoming redder at
these times in our late-time spectra.  \citet{Fox15} interpret the
change in colour as coming from additional flux of a redder component
and not new dust reddening.  Using the light-curve data presented by
\citet{Fox15}, including the extrapolated optical light curves, we
find that based on the IR luminosity, a 2000~K blackbody can
contribute at most 0.3\% of the $V$-band flux, with lower temperatures
contributing even less flux.  Therefore, it is unlikely that a single
blackbody can both account for the IR flux and change the $B-V$
colour.

\begin{figure}
\begin{center}
  \includegraphics[angle=0,width=3.2in]{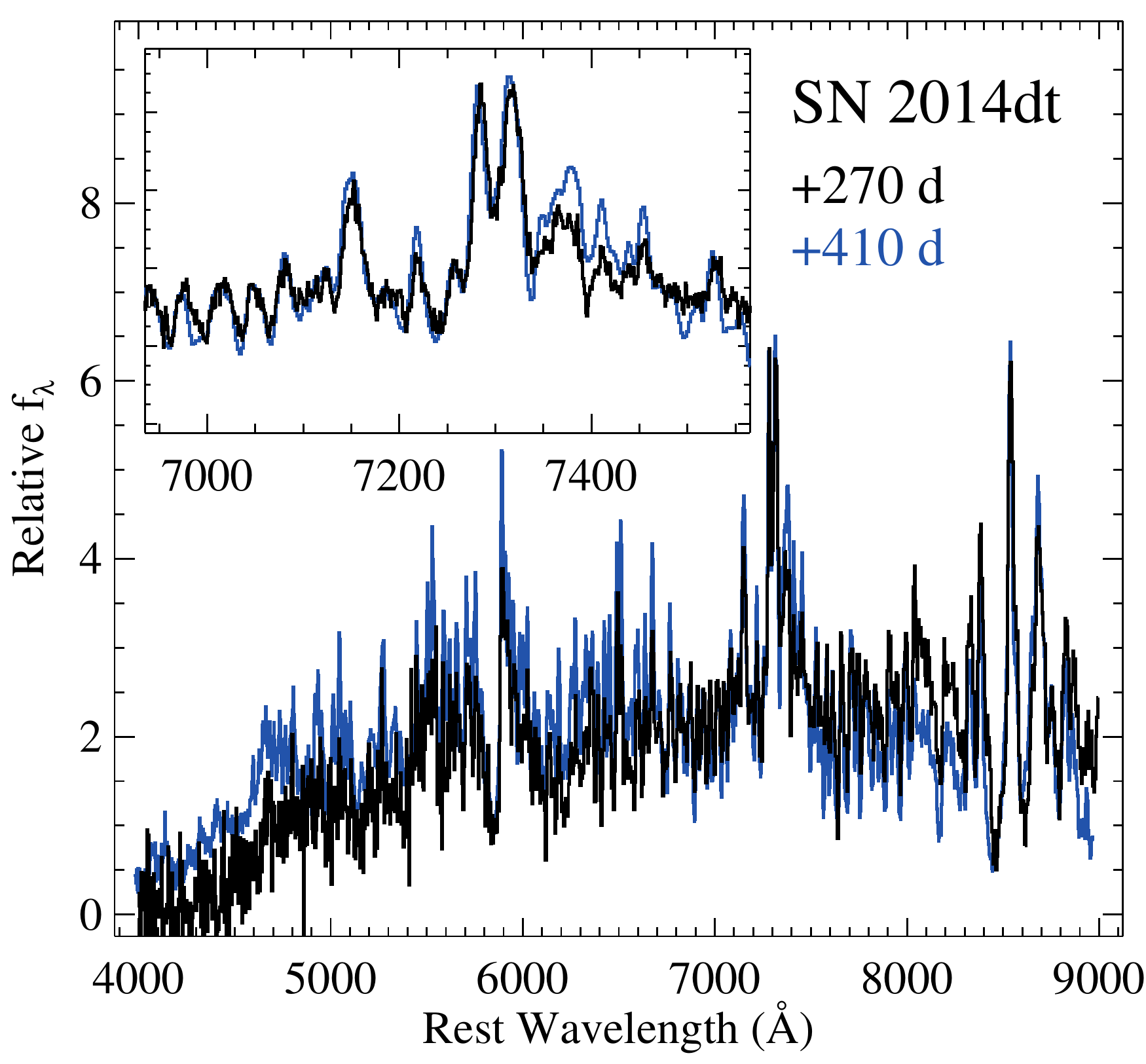}
  \caption{Late-time spectra of SN~2005hk at phases of roughly
    $+270$~d (black curve) and $+410$~d (blue curve).  The spectra are
    very similar.  The slight difference in continuum shape is
    inconsistent with dust reddening, but is consistent with slightly
    different photospheric temperatures.  There is no indication for
    circumstellar interaction in either spectrum.  {\it Inset}:
    Zoom-in of the region near the forbidden emission lines.  There is
    no indication of preferential extinction for the red side of the
    forbidden-line profiles.}\label{f:14dt_dust}
\end{center}
\end{figure}

Given the low reddening at early times, it is unlikely that there was
any significant amount of pre-existing circumstellar dust.  There are
also strong limits on narrow absorption lines in the spectra of
SN~2014dt \citep{Foley15:14dt}, indicating a gas-poor circumstellar
environment.  Furthermore, there is no indication of any circumstellar
interaction in any spectra (Figure~\ref{f:14dt_spec}), including the
+410-day spectrum (Figure~\ref{f:14dt_dust}).  The circumstellar dust
scenario seems unlikely given existing data.

We can also test the possibility of newly formed dust with data
presented here.  Dust can form in the SN ejecta, and this will redden
the SN, produce an IR excess, and preferentially extinguish the
redshifted light in forbidden lines \citep[e.g.,][]{Smith08:06jc}.
With our +410-day spectrum of SN~2014dt, which was obtained after the
IR data, we can rule out any significant newly formed dust.

Most obviously, there are no clear changes to the forbidden-line
shapes.  Figure~\ref{f:14dt_dust} displays the late-time spectra of
SN~2014dt from +270~d and +410~d.  While there is a slight change in
the strength of the [\ion{Ni}{II}] emission (see
Section~\ref{ss:evol}), the forbidden lines are otherwise nearly
identical in strength, width, velocity shift, and profile.  In
particular, there is no preferential extinction of redshifted
material.  This is consistent for all spectra of SN~2014dt, including
the +410~d spectrum, which was taken after the IR observations.

Additionally, there is no obvious emission from a warm, \about 2000~K
(corresponding to the dust deposition temperature) blackbody (see
Figure~\ref{f:14dt_spec}) as was seen for the dust-forming SN~2006jc
\citep{Smith08:06jc}.  However, given the IR luminosity, such a
component is not necessarily expected.  None the less, this is further
indication that the change in $B-V$ colour is intrinsic to the SN and
not a result of an additional blackbody component.

Examining the wavelength regions near hydrogen Balmer, \ion{He}{I},
and \ion{He}{II} lines, we see no indication of narrow emission as
would be expected from circumstellar interaction.  While this does not
rule out the presence of circumstellar material, it does constrain the
amount of such matter and the mechanism for heating potential
circumstellar dust.

Finally, there is no indication of additional reddening.  While the
continuum of SN~2014dt has slowly become redder with time, we are
unable to deredden later spectra to match earlier spectra using normal
reddening laws and reasonable reddening parameters.  While one can
deredden later spectra to roughly match the continuum at either bluer
or redder wavelengths, it is not possible to adequately match all
wavelengths simultaneously.  Moreover, such a manipulation changes the
strengths of the spectral features.  While all observed spectra have
roughly similar line strengths from +172 to +410~d, dereddening
spectra causes bluer lines to become relatively stronger and redder
lines to become relatively weaker.  For the SN to have additional
reddening, it must be a particularly odd reddening law and the SN
spectral features must evolve in a way to perfectly counteract the
effects of reddening.  We find this behaviour to be highly unlikely.

An alternative explanation for the IR emission in SN~2014dt, which was
not explored by \citet{Fox15}, is that it comes from a bound remnant
with a super-Eddington wind.  Such a mechanism is consistent with the
SN~2014dt data, late-time data for all SNe~Iax, and the potential
counterpart seen for SN~2008ha \citep{Foley14:08ha}.  Considering the
lack of any obvious circumstellar material or dust, as well as the
observed long-lasting photosphere and low photospheric velocities, the
best model for the IR emission is that it is somehow related to the
bound remnant, and most likely as an optically thick super-Eddington
wind.


\section{Conclusions}\label{s:conc}

We have presented an analysis of the late-time spectra of a sample of
10 SNe~Iax.  We add 8 spectra of SN~2014dt, the closest SN~Iax yet
discovered, to literature data to form our dataset.  We find that
while there are some subtle changes to the spectra at $t \gtrsim
200$~d after peak brightness, a single late-time spectrum is generally
sufficient to describe the late-time behaviour of a SN~Iax.  In
particular, we find SNe~Iax to be in a continuum between two extremes:
(1) those having low-velocity (\about 500~\kms) permitted P-Cygni
lines and strong/narrow forbidden [\ion{Fe}{II}], [\ion{Ni}{II}], and
[\ion{Ca}{II}], and weak (or absent)/broad [\ion{Fe}{II}] and
[\ion{Ni}{II}]; and (2) those having relatively smooth continuum
emission with a shape similar to that of other SNe~Iax, relatively
weak (perhaps even absent in one case)/narrow forbidden
[\ion{Fe}{II}], [\ion{Ni}{II}], and [\ion{Ca}{II}], and strong/broad
[\ion{Fe}{II}] and [\ion{Ni}{II}].

By fitting the forbidden lines, cross-correlating the spectra, and
performing a PCA, we have quantitatively shown that the spectral
continuum described above is real, with the various correlated
properties listed being significant.  We further note that besides the
relative strength of the narrow/broad forbidden lines, the two
kinematic components appear to be physically disconnected.  That is,
the velocity shifts and widths of the broad and narrow components are
uncorrelated.

We find a strong correlation between the width/strength of the broad
forbidden lines and their blueshift.  We also find no SNe~Iax that
have clearly redshifted broad forbidden lines, while there are several
that are significantly blueshifted.  It is unclear if this trend is
simply a result of a relatively small sample.

We find that SNe~Iax that have higher ejecta velocities measured at
maximum brightness also have stronger broad forbidden lines.  This can
be explained if explosions with higher kinetic energy per unit mass
also eject more material.  We also find that the more luminous (at
peak) SNe~Iax have stronger, broader, and more blueshifted broad
forbidden lines.  This requires that either SNe~Iax that produce more
$^{56}$Ni (in their ejecta) also have higher-velocity ejecta, or that
SNe~Iax are highly asymmetric and lines of sight that see
higher-velocity ejecta are also more luminous.  Such claims can be
tested in the future with additional spectropolarimetry of SNe~Iax.

The strong [\ion{Ni}{II}] lines in SN~Iax spectra at $>$200~d after
explosion must come from stable Ni, presumably $^{58}$Ni.  Producing
such a large amount of stable Ni requires electron capture, which can
only occur at the high densities of a (nearly) $M_{\rm Ch}$ WD.
Although full nebular modeling is necessary to confirm the Ni (and Fe)
abundances, this is further support for the idea that deflagrations of
(nearly) $M_{\rm Ch}$ WDs that fail to unbind their stars produce
SNe~Iax \citep{Jordan12, Kromer13, Kromer15}.

We found that the kinematic radius of SN~2005hk (determined from the
velocity of the photosphere at late times and the time since
explosion) is an order of magnitude higher than the blackbody radius
(determined from the luminosity and temperature).  This discrepancy
along with others point to SN~2005hk and several other SNe~Iax --- and
perhaps all SNe~Iax --- to have a wind component at late times.  A
two-component model consisting of the SN ejecta and a wind, either
driven from a remnant or companion, solves the radius problem, the
slow decline of the late-time light curve, the lack of velocity
evolution of the photosphere from about 200~d to $>$400~d after peak
brightness, and the fact that SNe~Iax have a photosphere even at
extremely late times.

For the two-component model, the photosphere, P-Cygni features, and
narrow forbidden lines are caused by the wind while the broad
forbidden lines are from the SN ejecta.  In this case, the two
components would be relatively decoupled.  However, the details of the
progenitor and explosion likely affect both the SN ejecta (through the
amount of $^{56}$Ni generated, the ejecta mass, the ejecta velocity,
and the ejecta composition) and the wind (through the amount of
$^{56}$Ni trapped in the remnant, the mass of the remnant, and the
composition of the remnant).  Such a model may have compositional
differences, which can be tested with detailed modeling.

We consider if SNe~Iax are primarily asymmetric explosions.  While not
fully explored in current SN~Iax models \citep{Jordan12, Kromer13,
  Kromer15}, it is possible that the explosion is highly asymmetric.
However, the current spectroscopic data for a single SN~Iax disfavour
large asymmetries \citep{Chornock06, Maund10:05hk}.  Although such
extreme asymmetries currently seem unlikely, additional data will test
if SN~Iax explosions are generally symmetric.

We found that the low-luminosity SN~2008ha had a spectrum similar to
the $\gtrsim$200~d spectra of other SNe~Iax only \about 1~month after
peak brightness.  As SN~2008ha likely did not unbind its progenitor
star \citep{Foley09:08ha, Foley10:08ha, Foley13:iax, Foley14:08ha}, it
is an excellent candidate for having a bound remnant and wind.  This
early transition to a ``wind-dominant'' spectrum can possibly be
explained by the relative amounts of $^{56}$Ni in the remnant and
ejecta \citep{Kromer15}.  Detailed spectral sequences, especially for
low-luminosity events are necessary to determine if the timing of this
transition is related to the explosion energetics.

Finally, we examined the spectra of SN~2014dt in detail, focusing on
the possibility of there being dust and/or circumstellar material
\citep[as suggested by][]{Fox15}.  We find no evidence for newly
formed or circumstellar dust, or any other circumstellar material, and
the existing data disfavour dust emission as the source of the IR
flux.  As an alternative, the strong IR flux seen about 315~d after
peak is perhaps caused by an extended optically thick super-Eddington
wind.  Such a scenario is consistent with all existing data.  If this
emission is dominated by the remnant, it may be the second such
detection after SN~2008ha \citep{Foley14:08ha}.

Late-time spectra of future SNe~Iax will continue to constrain their
progenitors and explosions.  Such data are critical for understanding
the potential remnant star and the properties of a possible
remnant-blown wind.

\section*{Acknowledgements}

  {\it Facility:} SOAR (Goodman), Keck:I (LRIS), Shane (Kast Double
  spectrograph), SALT (RSS)

\bigskip

R.J.F.\ gratefully acknowledges support from NSF grant AST-1518052 and
the Alfred P.\ Sloan Foundation.  SN~Iax research at Rutgers
University is supported by NASA/HST grants GO-12913 and GO-12973 to
S.W.J.  This work was supported by the NSF under grants PHY 11-25915
and AST 11-09174.  A.V.F.'s research was funded by NSF grant
AST-1211916, the TABASGO Foundation, and the Christopher R. Redlich
Fund.

We thank the participants of the ``Fast and Furious: Understanding
Exotic Astrophysical Transients'' workshop at the Aspen Center for
Physics, which is supported in part by the NSF under grant
PHY-1066293.  Some of the work presented in this manuscript was
initiated there during discussions with L.\ Bildsten \& D.\ Kasen.
Portions of this manuscript were also written during the Aspen Center
for Physics workshop, ``The Dynamic Universe: Understanding ExaScale
Astronomical Synoptic Surveys.''  We are grateful to the Aspen Center
for Physics for its hospitality during the ``Fast and Furious'' and
``Dynamic Universe'' workshops in June 2014 and May 2015,
respectively.

This research has made use of the NASA/IPAC Extragalactic Database
(NED) which is operated by the Jet Propulsion Laboratory, California
Institute of Technology, under contract with the National Aeronautics
and Space Administration (NASA).
Based in part on observations obtained at the Southern Astrophysical
Research (SOAR) telescope, which is a joint project of the
Minist\'{e}rio da Ci\^{e}ncia, Tecnologia, e Inova\c{c}\~{a}o (MCTI)
da Rep\'{u}blica Federativa do Brasil, the U.S. National Optical
Astronomy Observatory (NOAO), the University of North Carolina at
Chapel Hill (UNC), and Michigan State University (MSU).
KAIT and its ongoing operation were made possible by donations from
Sun Microsystems, Inc., the Hewlett-Packard Company, AutoScope
Corporation, Lick Observatory, the NSF, the University of California,
the Sylvia \& Jim Katzman Foundation, and the TABASGO Foundation.
Research at Lick Observatory is partially supported by a generous gift
from Google.
Some of the data presented herein were obtained at the W. M. Keck
Observatory, which is operated as a scientific partnership among the
California Institute of Technology, the University of California, and
NASA; the observatory was made possible by the generous financial
support of the W. M. Keck Foundation.
This research has made use of the Keck Observatory Archive (KOA),
which is operated by the W. M. Keck Observatory and the NASA Exoplanet
Science Institute (NExScI), under contract with NASA.  We thank the
staffs of the various observatories and telescopes (SOAR, Keck, SALT,
Lick) where data were obtained, as well as observers who helped obtain
some of the data (see Table~\ref{t:spec}).


\appendix

\section{Reassessment of PTF Classifications}\label{a:ptf}

\citet{White15} presented spectra and light curves of several SNe
discovered by the (Intermediate) Palomar Transient Factory [(i)PTF].
The intent of their investigation was to construct a sample of SNe~Iax
and SNe similar to the low-velocity and peculiar Type~I SN~2002es
\citep{Ganeshalingam12}.

To select their sample, \citet{White15} examined SNe~I in the (i)PTF
sample.  They then compared their spectra to those of previously
classified SNe using {\tt superfit} \citep{Howell05}, allowing the
redshift to vary by $\pm$0.02.  If one of the top 15 spectral matches
was a SN~Iax or SN~2002es, it was investigated further.  After
smoothing the spectrum, they measured the number of ``peaks'' seen in
the spectrum between 6000 and 8000~\AA.  SNe with a large number of
measured peaks were retained in the sample; however, the exact number
necessary for inclusion in the final sample is not mentioned and some
spectra do not cover this full spectral range.  SNe that then have a
strong \ion{Ti}{II} $\lambda 4200$ line are considered SN~2002es-like
(although one SN in their sample does not definitely have this
feature), while those lacking this line and having a ``peak'' near
6200~\AA\ are considered SNe~Iax.  In total, \citet{White15} presented
six new SNe classified as Type Iax and three new SNe classified as
SN~2002es-like.

\citet{White15} also re-evaluated literature SNe to determine if they
were SNe~Iax and/or SN~2002es.  Their main conclusion from this
additional investigation is that SNe~2004cs and 2007J, which have
prominent \ion{He}{I} lines in their spectra but are otherwise very
similar to other SNe~Iax \citep{Foley09:08ha, Foley13:iax}, should be
classified as SNe~IIb instead of SNe~Iax.

As part of our study of late-time spectra of SNe~Iax, we examined the
\citet{White15} sample to determine if any members should be included
in the current study.  While only one SN in their sample (PTF09ego)
has a spectrum at $>200$~d after peak brightness, we examined the
entire sample for completeness.  Through this analysis, we found that
four SNe are likely SNe~Iax and two SNe are probably SN~2002es-like
SNe.  However, we show below that two SNe were misclassified and that
one SN has insufficient data for a clear classification.  Below we
examine these misclassified and ambiguous SNe in detail.

Additionally, we re-evaluate the claim that SNe~2004cs and 2007J are
SNe~IIb.  There is no strong evidence that SNe~2004cs and 2007J are
SNe~IIb, but significant evidence against this claim.  In addition to
other data, there is no evidence for hydrogen emission, arguing
against the ``Type II'' designation.  While it is still unclear if
SNe~2004cs and 2007J are physically linked to SNe~Iax (as discussed by
\citetalias{Foley13:iax}), they do not appear to be SNe~IIb.

\subsection{PTF09ego}

PTF09ego was discovered with PTF imaging and reported by
\citet{White15}; however, no discovery information is explicitly
listed.  Using two spectra, obtained on 23 September 2009 (at +13~d)
and 15 May 2010 (at +225~d), \citet{White15} classified PTF09ego as a
SN~Iax.

\begin{figure}
\begin{center}
  \includegraphics[angle=0,width=3.2in]{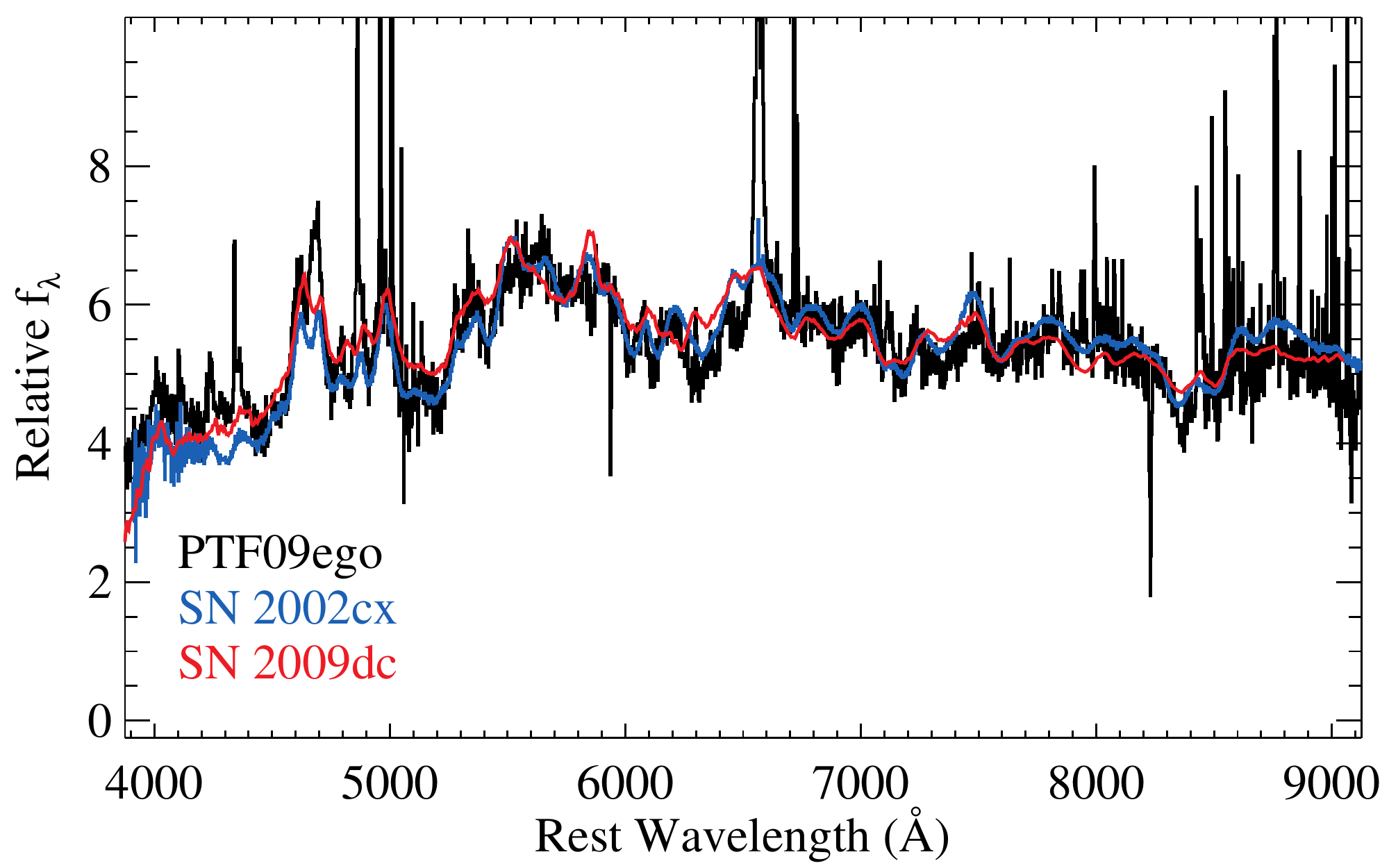}
  \caption{Optical spectrum of PTF09ego (black curve) as presented by
    \citet{White15}.  Also shown are spectra of the SN~Iax~2002cx
    \citep[blue curve;][]{Li03:02cx} and the luminous SN~Ia~2009dc
    \citep[red curve;][]{Taubenberger11} at phases of $+20$ and
    $+31$~d, respectively.  The spectra of SNe~2002cx and 2009dc have
    had a galaxy template spectrum added to roughly match the
    continuum seen in the spectrum of PTF09ego.}\label{f:09ego}

\end{center}
\end{figure}

We retrieved these spectra from WISERep \citep{Yaron12}.  The later
spectrum, although noisy, is consistent with being primarily or all
galaxy light.  Therefore, it cannot be included in the current study.
Furthermore, the early-time spectrum, while similar to that of
SN~2002cx (see Figure~\ref{f:09ego}), is equally similar to that of
the high-luminosity SN~2009dc \citep[e.g.,][]{Silverman11:09dc,
  Taubenberger11}, sometimes referred to as a ``super-Chandrasekhar''
SN~Ia.  While the peak luminosity of PTF09ego reported by
\citet[$M_{R} = -18.6$~mag]{White15} is more consistent with being a
SN~Iax, their reported rise time of \about 21~d is significantly
longer than that of any other SN~Iax ($t_{\rm rise} \approx 15$~d) and
more consistent with SN~2009dc \citep[$t_{\rm rise} \approx 23$~d;
e.g.,][]{Silverman11:09dc}.  Similarly, its relatively slow decline
rate is similar to that of SN~2009dc and related objects.

Although PTF09ego may be a SN~Iax, an alternative explanation is that
it is similar to SN~2009dc with significant host-galaxy dust
reddening.  We therefore consider PTF09ego to have an ambiguous
classification.  Regardless, its low-S/N late-time spectrum is not of
adequate quality to be included in the current study.

\subsection{PTF09eiy}

\begin{figure}
\begin{center}
  \includegraphics[angle=0,width=3.2in]{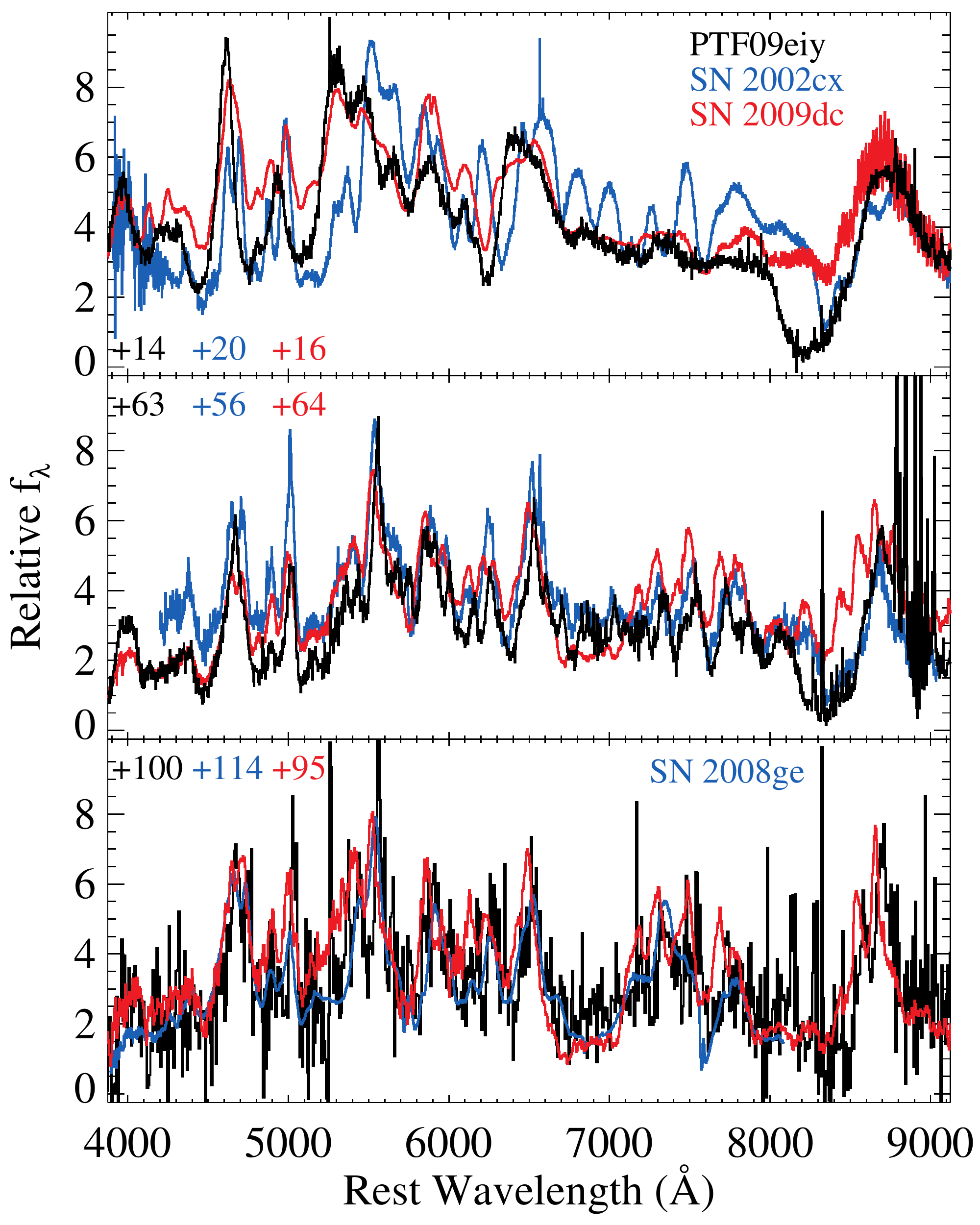}
  \caption{Optical spectra of PTF09eiy (black curve) as presented by
    \citet{White15}.  Also shown are spectra of the SN~Iax~2002cx
    \citep[blue curve;][]{Li03:02cx}, the SN~Iax~2008ge
    \citep{Foley10:08ge}, and the luminous SN~Ia~2009dc \citep[red
    curve;][]{Silverman11:09dc, Taubenberger11}.  The SN~2009dc
    spectra have been reddened by $E(B-V) = 0.45$~mag (corresponding
    to $A_{V} = 1.4$~mag).}\label{f:09eiy}

\end{center}
\end{figure}

PTF09eiy was discovered with PTF imaging and reported by
\citet{White15}; however, no discovery information is explicitly
listed.  They present five spectra of the SN with phases of roughly
(the time of maximum brightness is not well measured) +14, +33, +63,
+100, and +121~d.  We present the three spectra having relatively high
S/N in Figure~\ref{f:09eiy}.

\citet{White15} classify this SN as a SN~Iax despite having high
velocities ($-9600$~\kms) in their first spectrum.  This spectrum
differs from every other SN~Iax spectrum in the
\citetalias{Foley13:iax} and \citet{White15}
samples\footnote{\citet{White15} state that SN~2003gq, a SN~Iax in the
  \citetalias{Foley13:iax} sample, has a velocity of about
  $-10$,000~\kms\ a week before maximum brightness.  However,
  \citetalias{Foley13:iax} used this spectrum to measure a velocity of
  $-5600$~\kms, which we verified during the present analysis.}.  The
classification appears to be mostly based on the later spectra, which
are similar to spectra of SNe~Iax at comparable phases (assuming that
the SN was discovered near peak) and with the assumed redshift of
0.06.

Examining the PTF09eiy spectra in detail, we find that the later
spectra are similar to those of other SNe~Iax.  However, the first
spectrum is very different from any other SN~Iax.  This first spectrum
is similar to those of typical SNe~Ia at similar phases, but the later
spectra are unlike any spectra of typical SNe~Ia.

An alternative scenario is that PTF09eiy is not a SN~Iax, but rather
an atypical SN~Ia similar to the high-luminosity SN~Ia SN~2009dc
\citep[e.g.,][]{Silverman11:09dc, Taubenberger11}.
Figure~\ref{f:09eiy} presents spectral comparisons between PTF09eiy
and both SNe~Iax and SN~2009dc at similar phases.  For this
comparison, SN~2009dc has been artificially extinguished by $A_{V} =
1.4$~mag, corresponding to a reddening of $E(B-V) = 0.45$~mag.  After
applying this reddening, SN~2009dc is similar to PTF09eiy at all
phases.

According to \citet{White15}, PTF09eiy peaked at $M_{R} < -18.0$~mag.
If we correct for an extinction of $A_{V} = 1.4$~mag, this corresponds
to $M_{R} < -19.1$~mag; however, SN~2009dc may itself have $A_{V}
\approx 0.9$~mag \citep{Silverman11:09dc}, for which we did not
correct in the spectral comparisons.  Adopting this additional
extinction, PTF09eiy peaked at $M_{R} < -20.0$~mag, significantly
brighter than most SNe~Ia.

Given the spectral similarity at all available epochs as well as
consistent luminosities, we believe PTF09eiy is more likely to be
similar to SN~2009dc than an atypical SN~Iax.

\subsection{PTF10bvr}

PTF10bvr was discovered with PTF imaging and reported by
\citet{White15}; however, no discovery information is explicitly
listed.  Using a spectrum obtained 7.64 March 2010 (PI Kulkarni;
Program C247LA) with the Low Resolution Imaging Spectrometer
\citep[LRIS;][]{Oke95}, they classify PTF10bvr as a SN~2002es-like SN
at $z = 0.015$.  SN~2002es is a peculiar Type I SN similar to
SN~1991bg in many regards, but having significantly lower ejecta
velocity \citep{Ganeshalingam12}.  Because of their low expansion
velocities, SNe~2002es-like objects may be physically related to
SNe~Iax.

While the nominal host galaxy of PTF10bvr, CGCG~224-067, is an
early-type galaxy at $z = 0.02954$, \citet{White15} claim to detect a
strong Na~D absorption line at $z = 0.015$ and use that redshift to
classify the SN as having low velocities.  There is no other possible
host galaxy detected in any images presented by \citet{White15},
although presumably it would need to have a low luminosity in order to
be closer than CGCG~224-067 yet remain undetected.

We obtained the LRIS data from the Keck Observatory Archive and
rereduced the data using our own data-reduction pipeline.  Standard
CCD processing and spectrum extraction were accomplished with IRAF.
The data were extracted using the optimal algorithm of
\citet{Horne86}.  Low-order polynomial fits to calibration-lamp
spectra were used to establish the wavelength scale, and small
adjustments derived from night-sky lines in the object frames were
applied.  We employed our own IDL routines to flux calibrate the data
and remove telluric lines using the well-exposed continua of
spectrophotometric standards \citep{Wade88, Foley03,
  Silverman12:bsnip}.

There were two standard-star observations obtained during the night:
G191B2B and BD+33$^\circ$2642.  We used the former and latter to
calibrate the blue and red data, respectively.  However, BD+33\,2642
is not an ideal standard as its absorption lines, particularly the
Paschen series, make defining a continuum in regions of the spectrum
affected by telluric absorption difficult.  We carefully removed these
features from our spectrum, but caution that the final result may
still have residual problems.

\begin{figure}
\begin{center}
  \includegraphics[angle=0,width=3.2in]{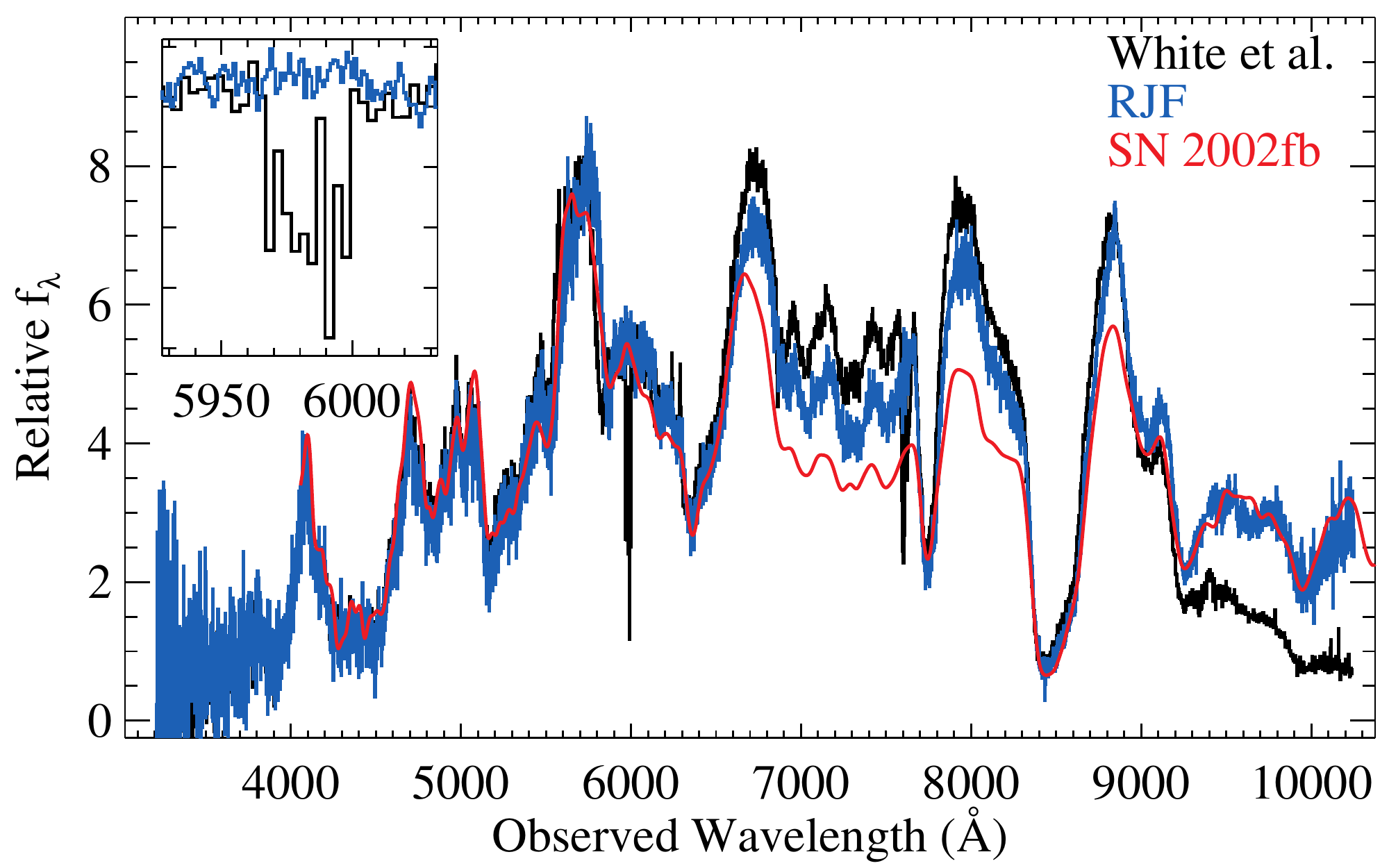}
  \caption{Optical spectrum of PTF10bvr.  The black spectrum was
    reduced by M.\ Kasliwal and presented by \citet{White15}.  The
    blue spectrum is a rereduction of the same data by one of us
    (R.J.F.).  We note that the \citet{White15} version appears to
    have been binned.  The red spectrum is a smoothed spectrum of
    SN~2002fb, a SN~1991bg-like SN, at a phase of $+18$~d
    \citep{Silverman12:bsnip}, and shifted to be at the redshift of
    CGCG 224-067, $z = 0.02954$.  The inset shows the region near the
    claimed $z = 0.015$ Na~D absorption, which is only present in the
    previous reduction and is likely an artifact.}\label{f:10bvr}

\end{center}
\end{figure}

Both our reduction and the \citet{White15} reduction of the spectrum
are presented in Figure~\ref{f:10bvr}.  Examining the two, it is clear
that the \citet{White15} version suffers from several data-quality
issues.  First, the wavelength solution near the dichroic (covering at
least (5500 -- 6150~\AA) is incorrect by up to 40~\AA.  Second, the
flux beyond 9000~\AA\ is likely significantly underestimated; this was
verified by independently reducing other spectra obtained during the
night.  The low flux at these wavelengths is perhaps partially caused
by a lack of telluric correction, which is evident as the telluric ``A
band'' near 7600~\AA\ is uncorrected.

\begin{figure*}
\begin{center}
  \includegraphics[angle=0,width=6.4in]{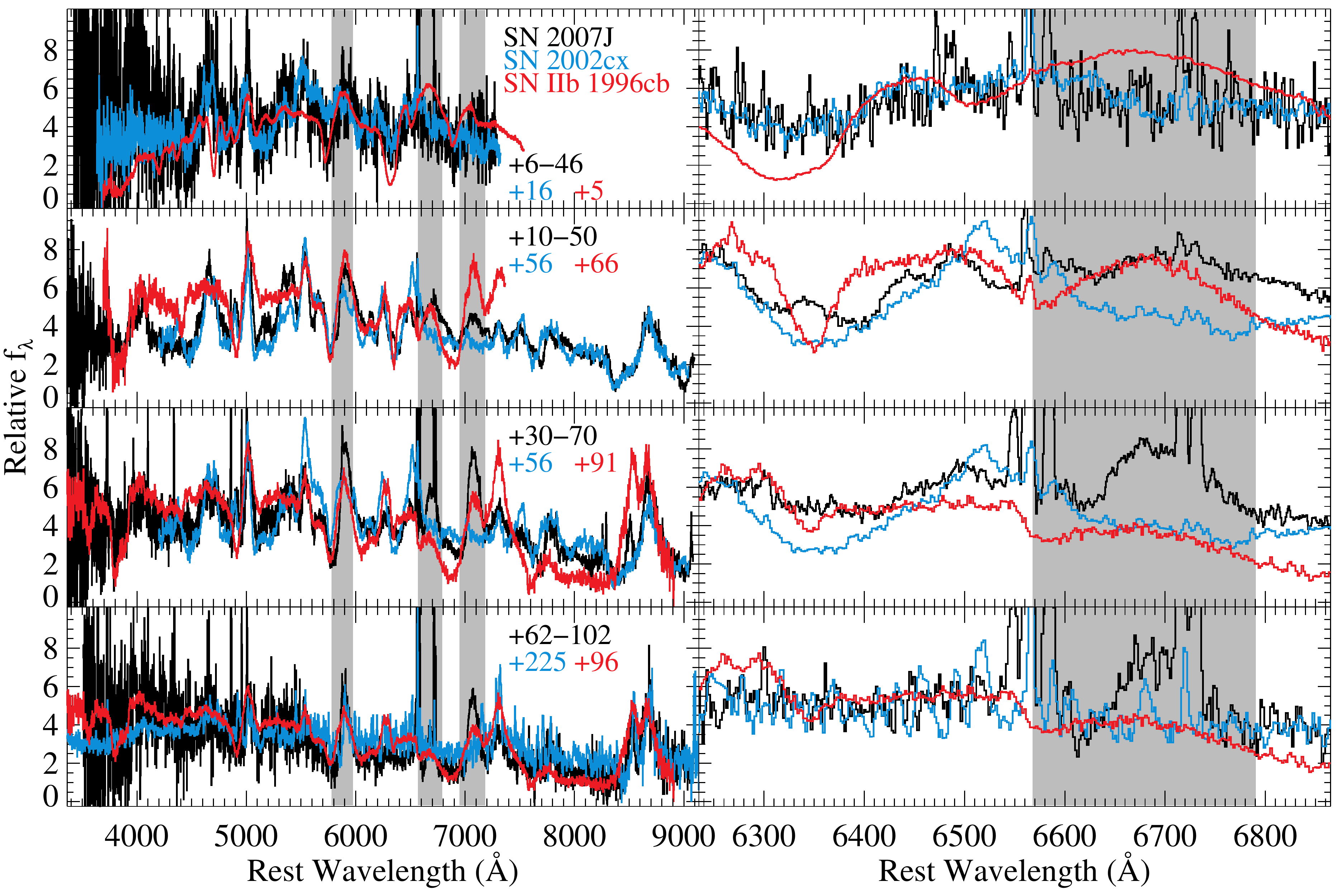}
  \caption{Optical spectra of SN~2007J (black curve).  The phase
    ranges for the spectra are noted in each panel; the exact phase is
    not known, but is constrained by a nondetection and the first
    detection (see \citealt{Foley09:08ha, Foley13:iax}).  The
    left-side panels show the entire optical range, while the
    right-side panels display in detail the area around H$\alpha$.
    There is no broad H$\alpha$ from SN~2007J visible in any spectrum.
    Comparison spectra of SN~Iax~2002cx and SN~IIb~1996cb are shown as
    blue and red curves, respectively, with phases also noted.
    SN~1996cb was specifically chosen as the SN~IIb with the weakest
    H$\alpha$ emission at late times (to match the lack of obvious
    H$\alpha$ in SN~2007J).  Regions corresponding to the strong
    \ion{He}{I} $\lambda\lambda$5876, 6678, 7065 lines are shaded
    grey.}\label{f:07j}

\end{center}
\end{figure*}

We further examine the claimed Na~D absorption.  This feature is
completely absent in the new reduction.  Examining the two-dimensional
image, there is a cosmic ray near the position of the object at
roughly the correct wavelength, although this is in a region where the
\citet{White15} wavelength calibration is highly suspect, so it is
difficult to confirm if that particular cosmic ray is causing the
``absorption.''  Close examination of the profile of this feature
reveals that it does not have a typical shape and that its noise
properties differ significantly from those of the continuum.  We
conclude that there is no Na~D absorption, especially at $z = 0.015$,
in the spectrum of PTF10bvr.  A nearby cosmic ray on the detector is
also a more likely scenario than PTF10bvr being hosted by a very
low-luminosity galaxy in the direct foreground of a more luminous
galaxy.

If we use the redshift of CGCG~224-067 as the redshift of PTF10bvr, it
is clear that PTF10bvr is a SN~1991bg-like SN~Ia, not similar to
SN~2002es (Figure~\ref{f:10bvr}).  Allowing the redshift to be a free
parameter does not change the classification; the spectrum of PTF10bvr
is much more similar to SN~1991bg-like objects at $z = 0.03$ than to
SN~2002es at $z = 0.015$.

\subsection{SN 2004cs and SN 2007J}

\citet{Foley09:08ha} first noticed that SN~2007J was spectroscopically
similar to other SNe~Iax with the exception of strong \ion{He}{I}
lines present in the spectra of SN~2007J.  \citetalias{Foley13:iax}
identified SN~2004cs as a similar object, being spectroscopically
similar to SNe~Iax, but with strong \ion{He}{I} lines.  The possible
physical association of these objects with the SN~Iax class has
far-reaching implications for the progenitors and explosions of
SNe~Iax, and was one of the strongest reasons (but not the only one)
that \citetalias{Foley13:iax} first suggested that SNe~Iax had
WD/He-star progenitor systems.

\citet{White15} disagreed with this classification and claimed the
detection of hydrogen lines in the spectra of both SNe, reclassifying
these SNe as Type IIb.  We re-evaluate this claim here.

SN~2007J was relatively well observed with four spectra at distinct
phases (see Figure~\ref{f:07j}).  Although the exact time of maximum
brightness was not measured, a nondetection was useful in constraining
that time to within 40~d \citep{Foley09:08ha, Foley13:iax}.

Examining these spectra, we do not detect any hydrogen lines in
SN~2007J.  Comparisons to SN~2002cx and SN~IIb~1996cb, the SN~IIb with
the weakest H$\alpha$ emission at late times in the \citet{Modjaz14}
sample and the best-matching SN~IIb found, show that SN~2007J is more
similar to SN~2002cx --- even when considering the \ion{He}{I} lines.
This is especially true at early times, where there are significant
deviations from SN~1996cb at bluer wavelengths.  In particular, there
is a strong H$\beta$ line in the earliest spectrum of SN~1996cb, but
no corresponding feature for SN~2007J.  From this comparison alone,
SN~2007J is highly discrepant with even the most similar SN~IIb known.

In addition to the lack of an H$\beta$ line in SN~2007J, we note that
the feature \citet{White15} identified as H$\alpha$ in SN~2007J is
also present in SN~2002cx and identified as \ion{Fe}{II} by
\citet{Branch04}.  For SN~2007J, this feature is more similar to that
of SN~2002cx at early times and evolves in a similar way (at later
times the feature is relatively weak in all spectra).  As SN~2002cx
(and other SNe~Iax) never show any strong hydrogen emission, including
at late times, it is unlikely that this feature is H$\alpha$ in
SN~2002cx, and similarly unlikely that it is H$\alpha$ in SN~2007J.

In conclusion, there is no evidence that SN~2007J is a SN~IIb.  While
SN~2007J may not be physically related to SNe~Iax, it is most similar
to these objects, and we consider this classification the most prudent
at this time.

\begin{figure}
\begin{center}
  \includegraphics[angle=0,width=3.2in]{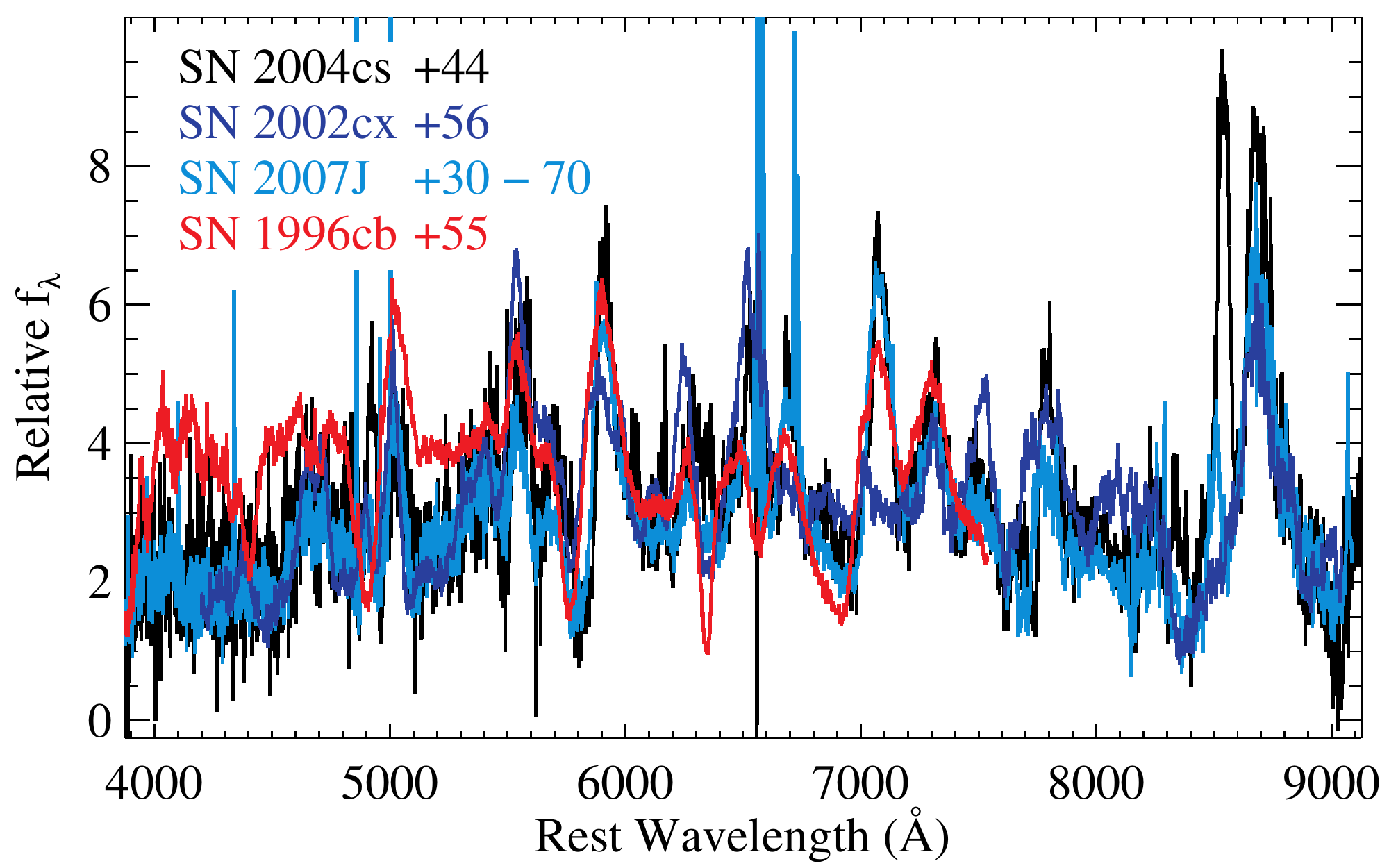}
  \caption{Optical spectrum of SN~2004cs (black curve); there is no
    obvious broad H$\alpha$ in the spectrum.  Similar-phase comparison
    spectra of SN~Iax~2002cx, SN~2007J, and SN~IIb~1996cb are shown as
    dark blue, light blue, and red curves,
    respectively.}\label{f:04cs}

\end{center}
\end{figure}

SN~2004cs does not have as much spectral data as SN~2007J; however, it
does have a very constraining light curve presented by
\citetalias{Foley13:iax}.  As presented by \citetalias{Foley13:iax}
and re-examined here (Figure~\ref{f:04cs}), SN~2004cs is very similar
to SN~2007J as well as SN~2002cx (besides the \ion{He}{I} lines).  As
was seen for SN~2007J, there are no obvious hydrogen lines in the
spectrum of SN~2004cs.  Based on the spectral data available,
SN~2004cs is extremely similar to SN~2007J, and since SN~2007J is not
a SN~IIb, it is unlikely that SN~2004cs is a SN~IIb.

\begin{figure}
\begin{center}
  \includegraphics[angle=0,width=3.2in]{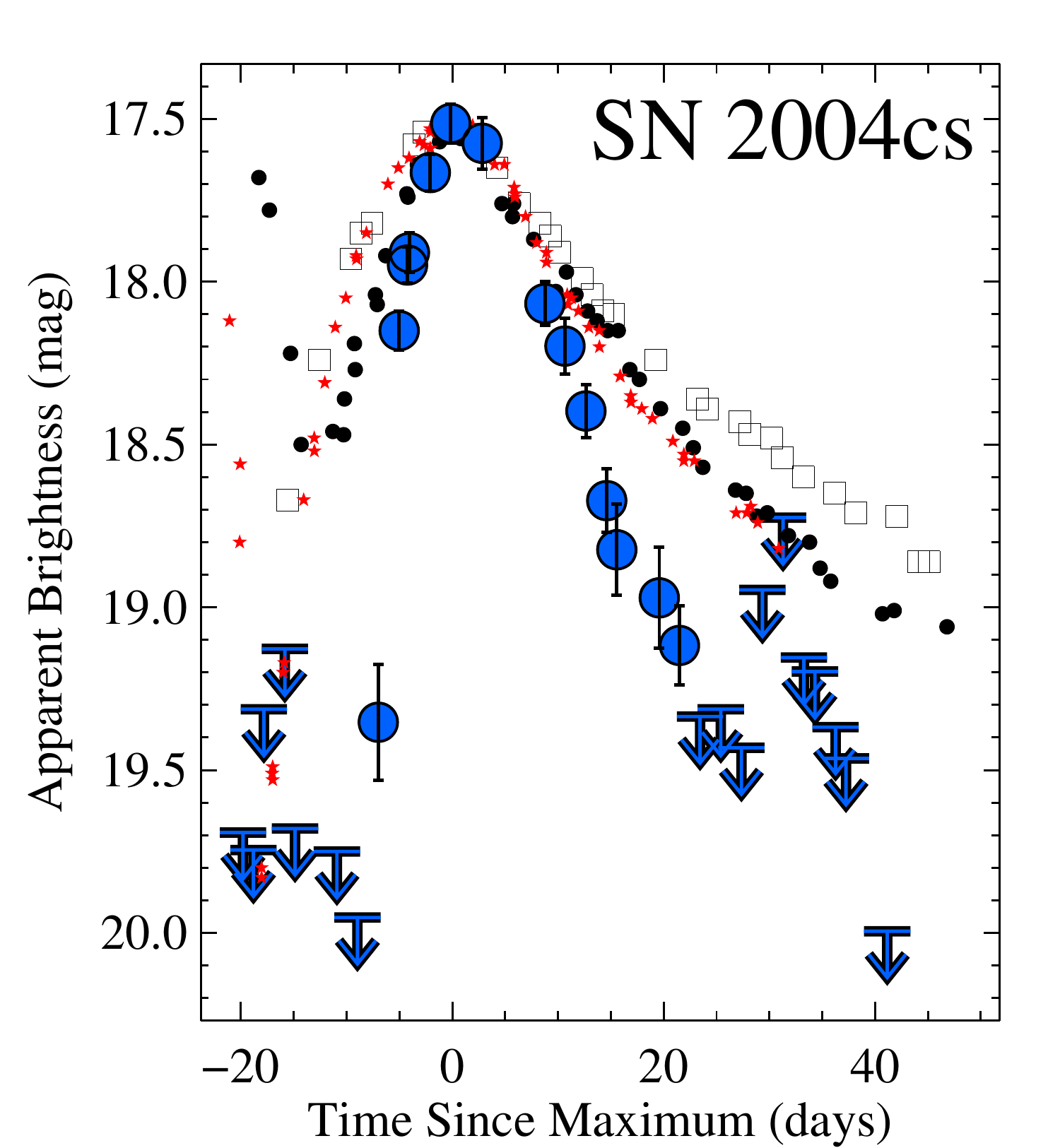}
  \caption{Unfiltered KAIT light curve of SN~2004cs (blue circles, and
    blue arrows indicating upper-limit nondetections).  Also displayed
    are $R$-band (similar to unfiltered) light curves of SNe~1993J
    \citep[black circles;][]{Richmond94}, 1996cb
    \citep[squares;][]{Qiu99}, and 2011dh \citep[red
    stars;][]{Arcavi11, Ergon14}, all shifted to match the peak of
    SN~2004cs.}\label{f:04cs_lc}

\end{center}
\end{figure}

There is stronger evidence against the SN~IIb classification for
SN~2004cs based on its light curve (Figure~\ref{f:04cs_lc}), which is
unlike that of any known SN~IIb.  We reprocessed the unfiltered KAIT
data presented by \citetalias{Foley13:iax} to improve the overall
quality of the photometry and to include several nondetections before
the SN rise and after it declined; see Table~\ref{t:04cs}.

\begin{deluxetable}{llll}
\tabletypesize{\footnotesize}
\tablewidth{0pt}
\tablecaption{KAIT Unfiltered Light Curve of SN~2004cs\label{t:04cs}}
\tablehead{
\colhead{MJD} &
\colhead{Mag} &
\colhead{$\sigma$ (mag)} &
\colhead{Limit (mag)}}

\startdata

53164.412 & \nodata & \nodata & 19.7 \\
53165.426 & \nodata & \nodata & 19.8 \\
53166.423 & \nodata & \nodata & 19.3 \\
53168.436 & \nodata & \nodata & 19.1 \\
53169.412 & \nodata & \nodata & 19.7 \\
53173.431 & \nodata & \nodata & 19.7 \\
53175.397 & \nodata & \nodata & 20.0 \\
53177.399 & 19.35   & 0.18    & \nodata \\
53179.405 & 18.15   & 0.06    & \nodata \\
53180.203 & 17.95   & 0.06    & \nodata \\
53180.388 & 17.91   & 0.06    & \nodata \\
53182.394 & 17.66   & 0.06    & \nodata \\
53184.353 & 17.51   & 0.06    & \nodata \\
53187.371 & 17.57   & 0.08    & \nodata \\
53193.387 & 18.07   & 0.07    & \nodata \\
53195.312 & 18.20   & 0.09    & \nodata \\
53197.323 & 18.40   & 0.08    & \nodata \\
53199.318 & 18.67   & 0.10    & \nodata \\
53200.287 & 18.82   & 0.14    & \nodata \\
53204.359 & 18.97   & 0.16    & \nodata \\
53206.296 & 19.12   & 0.12    & \nodata \\
53208.270 & \nodata & \nodata & 19.3 \\
53210.259 & \nodata & \nodata & 19.3 \\
53212.241 & \nodata & \nodata & 19.4 \\
53214.261 & \nodata & \nodata & 19.0 \\
53216.231 & \nodata & \nodata & 18.7 \\
53218.242 & \nodata & \nodata & 19.2 \\
53219.296 & \nodata & \nodata & 19.2 \\
53221.272 & \nodata & \nodata & 19.4 \\
53222.272 & \nodata & \nodata & 19.5 \\
53226.230 & \nodata & \nodata & 20.0

\enddata

\end{deluxetable}

SN~2004cs rises to maximum light in $<$10 days and declines on a
similar timescale.  The new nondetections, including a relatively deep
limit about 40~d after maximum, rule out a change in the decay rate
right after the last detection.  In contrast, the prototypical
SN~IIb~1993J \citep{Richmond94}, the well-observed SN~IIb~2011dh
\citep{Arcavi11, Ergon14}, and the hydrogen-weak SN~IIb~1996cb
\citep{Qiu99} all have much broader light curves and a change in decay
rate occurring between 20~d and 40~d after peak.  \citet{White15}
found that the light curve of SN~2004cs was consistent with the SN~IIb
template light curve of \citet{Arcavi12}; however, this comparison
indicated that SN~2004cs declined faster than the template.  In
addition, the comparison presented by \citet{White15} is not ideal.
They used the start of the template light curve as the time of
maximum, but the \citet{Arcavi12} template begins roughly 5~d after
maximum, when the light curve is declining much faster than right at
peak.  The template also only covers \about 12 days of the light
curve, and so the comparison ignores all premaximum data and the later
data where SN~2004cs continues to quickly decline while SNe~IIb
decline slower at these phases.  Finally, this template was generated
from only two light curves, with one having only 5 data points over 36
days (and 3 at maximum brightness or later), so this template is not
the best comparison when excellent data, such as those for SNe~1993J
and 2011dh, exist.

In addition, SN~2004cs does not have the initial peak and decline
before rising to maximum that is associated with the cooling of a
shocked stellar envelope and is seen in many SNe~IIb, including
SNe~1993J and 2011dh.  While this has not been detected in all SNe~IIb
(e.g., SN~1996cb; Figure~\ref{f:04cs_lc}), it has for those SNe with
deep, high-cadence observations around the time of explosion, like
what was obtained for SN~2004cs.  Therefore, the deep nondetections
before the rise are highly constraining and indicate that SN~2004cs
was not a SN~IIb.  Furthermore, it is unclear if a SN~IIb progenitor
could produce such a rapidly evolving light curve (requiring very
small ejecta mass).

Finally, \citet{Rajala05} measured a single epoch of multiband
photometry for SN~2004cs at \about 5~d before maximum brightness,
finding colours that were significantly inconsistent with those of
SNe~IIb as well as all core-collapse SNe.  However, the colours were
consistent with those of a young SN~Ia, whose colours are similar to
those of SNe~Iax.

In summary, SN~2004cs has a spectrum similar to that of SN~2007J as
well as SN~2002cx (besides the prominent \ion{He}{I} lines).  While
SN~2004cs is spectroscopically similar to some SNe~IIb (although we
cannot confirm any hydrogen in its spectrum), its light curve is
unlike that of any SN~IIb, rising faster and declining faster than any
known SN~IIb.  Furthermore, SN~2004cs lacks the signature of a cooling
envelope seen in all SNe~IIb with similar-quality data.  From this, we
conclude that it is highly unlikely that SN~2004cs is a SN~IIb, while
being very similar to SNe~Iax.

\subsection{Summary and Discussion}

Above, we showed that PTF09ego may not be a SN~Iax, that PTF09eiy is
unlikely to be a SN~Iax and more likely to be similar to SN~2009dc and
other high-luminosity SNe~Ia, and that PTF10bvr is not a
SN~2002es-like SN.  None the less, 6/9 of the \citet{White15} sample
appear to be likely SNe~Iax or SN~2002es-like objects.

We also re-examine the claims of \citet{Foley09:08ha} and
\citetalias{Foley13:iax} that SNe~2004cs and 2007J are SNe~Iax and not
SNe~IIb as claimed by \citet{White15}.  We find no evidence of these
SNe being SNe~IIb and strong evidence against this classification.  We
also show that other than the presence of \ion{He}{I} lines, they are
very similar to SNe~Iax.  We therefore continue to classify SNe~2004cs
and 2007J as SNe~Iax, although we also caution that observations of
similar SNe in the future may indicate that SNe~2004cs and 2007J are
physically distinct from SNe~Iax.

One of the main goals of \citet{White15} was to measure the relative
rate of SNe~Iax and SNe~Ia, finding 5.6 SNe~Iax (and SN~2002es-like
objects) per 100 SNe~Ia. This value was much smaller than that found
by \citetalias{Foley13:iax}, 31 SNe~Iax per 100 SNe~Ia (and not
counting SN~2002es-like objects).  While the reclassification of up to
1/3 of the \citet{White15} sample may point to an even lower rate, we
caution against this conclusion.

We first note that \citet{White15} did not correct for the photometric
and spectroscopic selection of their survey.  Considering that SNe~Iax
are 1--5~mag fainter than typical SNe~Ia at peak and typically fade
twice as fast as SNe~Ia, there must be some photometric selection
bias.  Additionally, the contrast of relatively faint SNe~Iax compared
to their host galaxies likely makes detecting the SNe more difficult,
and may (partially) explain the large fraction of relatively
low-surface brightness host galaxies in their sample.

Similarly, the \citet{White15} sample almost certainly suffers from
spectroscopic selection bias.  They mention nine SNe that have some
spectroscopic similarities to SNe~Iax and SN~2002es, but most were
rejected from the final sample because of low-S/N spectra.  These
additional objects alone could double the measured rate.

We further note that of the six \citet{White15} SN~Iax candidates, the
earliest spectra were obtained at +13, +14, +23, +25, +26, and +56~d
relative to maximum brightness, with the earliest spectra coming from
PTF09ego (which is perhaps not a SN~Iax) and PTF09eiy (which is
unlikely to be a SN~Iax).  However, 30--55\% of the PTF SN~Ia sample
have at least one spectrum before +5~d \citep{Maguire14}.  Assuming
that the PTF SN~Ia and SN~Iax spectroscopic selection functions are
identical and adopting the most favourable conditions, there is only a
12\% chance of having no SN~Iax in the \citet{White15} sample with a
spectrum before +5~d (0.8\% assuming the higher fraction (55\%) with
early-time spectra, and it is further unlikely to have no spectra
between +5 and +13~d and extremely unlikely to have no spectra before
+23~d, which is the earliest spectrum of a definitive SN~Iax in the
sample).  Considering that SNe~Iax are less luminous than SNe~Iax, one
would expect relatively {\it more} spectra near maximum light for this
class.

One likely reason for this discrepancy is the method for selecting
members of the class.  \citet{White15} make two decisions that
probably bias against selecting SNe~Iax near maximum brightness.  The
first is that they only selected SNe where {\tt superfit} returned a
match with a SN~Iax in its top 15 matches.  However, it is well known
(e.g., \citealt{Li03:02cx}; \citetalias{Jha06:02cx}) that near maximum
brightness, SN~Iax spectra are very similar to those of SNe~Ia except
for their velocity, which is degenerate with redshift when doing a
$\chi^{2}$ fit as {\tt superfit} does.  In fact, 12 of the 25 SNe~Iax
in the \citetalias{Foley13:iax} sample were at some point
misclassified, often because of this effect.

To test this possibility, we classified the $-1$~d spectrum of
SN~2002cx \citep{Li03:02cx} using {\tt SNID} \citep{Blondin07}.
Although {\tt SNID} is a different algorithm than {\tt superfit}, this
experiment is illustrative.  With no prior on the redshift, there were
no SNe~Iax in the top 20 spectral matches.  Putting the correct prior
on the redshift, the SN is still not correctly classified, with no
SNe~Iax in the top 15 spectral matches.  Therefore, SNe without
host-galaxy redshifts are particularly prone to misclassification.
This problem is likely amplified with noisy data.

An additional selection bias is the ``peak counting'' employed by
\citet{White15}.  Starting \about 2~weeks after maximum light, SNe~Iax
have very complex spectra with many distinct spectral features.
However, this is not the case near maximum brightness.  While
\citet{White15} do not explicitly state how many peaks (in the range
6000 -- 8000~\AA) are necessary for inclusion in their final sample,
it appears to be around 7 given the objects that were included.
Although there is no description of exactly how significant a peak
must be to be counted, we attempted this analysis for SN~Iax~2011ay,
which has one of the best spectral sequences between peak and +30~d
\citepalias{Foley13:iax}.  We find that the spectra spanning phases of
$-2$ to +11~d all have $<$7 peaks, while the spectra after +26~d all
have $\ge$7 peaks.  Therefore, if there were only a single
maximum-light spectrum of SN~2011ay, it would likely be excluded from
the \citet{White15} analysis, even if it passes the {\tt superfit}
criterion.

While the smaller sample of genuine SNe~Iax in the \citet{White15}
sample may, at first glance, appear to make the relative rate even
less consistent with the rate of \citetalias{Foley13:iax}, the various
photometric biases, spectroscopic biases, sample-selection biases, and
already identified ambiguous objects are likely the reasons for the
difference.

As further confirmation of the ``high'' rate of
\citetalias{Foley13:iax}, we recalculate the relative rate using only
SNe within $D \lesssim 20$~Mpc.  In the last 10~yr, there have been 5
SNe~Iax discovered within this volume: SNe~2008ge
\citep{Foley10:08ge}, 2008ha \citep{Foley09:08ha, Foley10:08ha,
  Valenti09}, 2010ae \citep{Stritzinger14:10ae}, 2010el, and 2014dt
\citep{Foley15:14dt}.  During this same time, there were \about
25~SNe~Ia discovered in this volume.  Without any additional
corrections, this places a very robust lower limit on the relative
rate of \about 24 SNe~Iax for every 100 SNe~Ia, consistent with the
\citetalias{Foley13:iax} rate and significantly inconsistent with the
\citet{White15} rate.

\bibliographystyle{mnras}
\bibliography{../astro_refs}

\onecolumn

\begin{deluxetable}{rllrl}
\tablewidth{0pt}
\tablecaption{Log of Spectral Observations of SN~2014dt\label{t:spec}}
\tablehead{
\colhead{} &
\colhead{} &
\colhead{Telescope /} &
\colhead{Exposure} &
\colhead{} \\
\colhead{Phase\tablenotemark{a}} &
\colhead{UT Date} &
\colhead{Instrument} &
\colhead{(s)} &
\colhead{Observer\tablenotemark{b}}}

\startdata

$+172$ & 2015 Apr.\ 16.406 & Lick/Kast    & 1800            & IS\\
$+204$ & 2015 May   17.829 & SALT/RSS     & 4 $\times$ 425  & AK\\
$+212$ & 2015 May   26.285 & Lick/Kast    & 1800            & MG\\
$+228$ & 2015 June  11.718 & SALT/RSS     & 4 $\times$ 425  & AK\\
$+233$ & 2015 June  16.068 & SOAR/Goodman & $2 \times 1800$ & RF, SD, YP\\
$+233$ & 2015 June  16.328 & Keck/LRIS    & 600             & AF, MG, WZ\\
$+270$ & 2015 July  24.039 & SOAR/Goodman & $2 \times 1800$ & RF, RH, SD\\
$+410$ & 2015 Dec.\ 11.641 & Keck/LRIS    & $2 \times 1200$ & MG, SV

\enddata

\tablenotetext{a}{Days since $B$ maximum, 2015 Oct.\ 25.2 (JD
  2,456,955.7).}

\tablenotetext{b}{AF = A.\ Filippenko, IS = I.\ Shivvers, MG = M.\
  Graham, AK = A.\ Kniazev, RF = R.\ Foley, RH = R.\ Hounsell, SD =
  S.\ Downing, SV = S.\ Valenti, WZ = W.\ Zheng, YP = Y.-C.\ Pan}

\end{deluxetable}

\newpage
\begin{landscape}
\begin{deluxetable}{@{\extracolsep{2pt}}lrrrrrrrrrrrrrrr@{}}
\tabletypesize{\tiny}
\tablewidth{0pt}
\tablecaption{Forbidden-Line Fit Parameters\label{t:neb_fit}}
\tablehead{
\colhead{} &
\colhead{} &
\multicolumn{8}{c}{Narrow Component} &
\multicolumn{6}{c}{Broad Component} \\
\cline{3-10}
\cline{11-16}
\colhead{} &
\colhead{} &
\colhead{} &
\colhead{} &
\multicolumn{2}{c}{[\ion{Fe}{II}] $\lambda 7155$} &
\multicolumn{2}{c}{[\ion{Ca}{II}] $\lambda \lambda 7291$, 7324} &
\multicolumn{2}{c}{[\ion{Ni}{II}] $\lambda 7378$} &
\colhead{} &
\colhead{} &
\multicolumn{2}{c}{[\ion{Fe}{II}] $\lambda 7155$} &
\multicolumn{2}{c}{[\ion{Ni}{II}] $\lambda 7378$} \\
\cline{5-6}
\cline{7-8}
\cline{9-10}
\cline{13-14}
\cline{15-16}
\colhead{} &
\colhead{Phase} &
\colhead{FWHM} &
\colhead{Shift} &
\colhead{Rel.\ Line} &
\colhead{EW} &
\colhead{Rel.\ Line} &
\colhead{EW } &
\colhead{Rel.\ Line} &
\colhead{EW} &
\colhead{FWHM} &
\colhead{Shift} &
\colhead{Rel.\ Line} &
\colhead{EW } &
\colhead{Rel.\ Line} &
\colhead{EW} \\
\colhead{SN} &
\colhead{(days)} &
\colhead{(\kms)} &
\colhead{(\kms)} &
\colhead{Strength} &
\colhead{(\AA)} &
\colhead{Strength} &
\colhead{(\AA)} &
\colhead{Strength} &
\colhead{(\AA)} &
\colhead{(\kms)} &
\colhead(\kms){} &
\colhead{Strength} &
\colhead{(\AA)} &
\colhead{Strength} &
\colhead{(\AA)}}

\startdata

2002cx &    +227 & 1430 (110) & $   31$ (41) & 1 &  37  (7) & 1.13 (0.21) &  83 (15) & 0.11 (0.10) &   3.9  (3.5) & 7870 (910) & $ 1130$ (430) & 0.34 (0.10) &  69 (23) &  0.46  (0.10) &   90  (20) \\
2005P  & $>$+109 & 1570  (60) & $  530$ (27) & 1 &  77 (10) & 1.87 (0.24) & 288 (28) & 0.78 (0.12) &  59.9  (7.8) & 7950 (180) & $  840$ (110) & 0.95 (0.11) & 373 (28) &  1.27  (0.14) &  500  (30) \\
2005hk &    +224 &  680  (30) & $ -295$ (13) & 1 &  62  (9) & 2.22 (0.26) & 273 (34) & 0.41 (0.09) &  25.2  (5.7) & 7050 (470) & $  -40$ (260) & 0.14 (0.03) &  90 (19) &  0.35  (0.05) &  230  (40) \\
2008A  &    +220 & 1490  (60) & $  488$ (27) & 1 & 126 (15) & 1.65 (0.19) & 416 (49) & 0.40 (0.10) &  50.0 (12.3) & 8440  (90) & $ -730$  (40) & 0.83 (0.06) & 593 (28) &  2.64  (0.19) & 1880  (90) \\
2008ge &    +225 & 2680 (100) & $  898$ (38) & 1 & 151 (13) & \nodata     & \nodata  & 0.78 (0.10) & 117.5 (15.8) & 7080  (70) & $   70$  (20) & 0.83 (0.06) & 331 (16) &  2.84  (0.16) & 1140  (40) \\
2010ae &    +252 &  770  (50) & $   54$ (25) & 1 &  25  (5) & 6.93 (1.37) & 351 (62) & 0.16 (0.11) &   4.1  (2.7) & \nodata    & \nodata       & \nodata     & \nodata  & \nodata       & \nodata  \\
2011ay &    +176 & 3320 (270) & $-1144$ (72) & 1 &  18 (10) & 3.26 (1.95) & 116 (29) & 0.51 (0.97) &   9.1 (14.2) & 7700 (160) & $ -520$ (110) & 7.43 (3.83) & 308 (32) & 21.55 (10.34) &  900  (60) \\
2011ce &    +371 &  780  (60) & $   87$ (26) & 1 &  12  (3) & 2.12 (0.43) &  51  (8) & 0.32 (0.13) &   3.8  (1.4) & \nodata    & \nodata       & \nodata     & \nodata  & \nodata       & \nodata  \\
2012Z  &    +248 & 1790 (120) & $ -107$ (45) & 1 & 115 (20) & 0.36 (0.10) &  84 (29) & 0.11 (0.12) &  12.6 (14.8) & 9000  (70) & $-1380$  (30) & 1.33 (0.11) & 773 (63) &  7.74  (0.52) & 4510 (310) \\
2014dt &    +233 &  950  (60) & $ -333$ (25) & 1 &  30  (5) & 1.18 (0.20) &  72 (12) & 0.59 (0.14) &  17.9  (4.2) & 6400 (320) & $  530$ (160) & 0.32 (0.07) &  65 (11) &  0.95  (0.12) &  200  (20)

\enddata

\end{deluxetable}
\end{landscape}

\end{document}